\pdfoutput=1
\documentclass[11pt,a4paper]{article}
\usepackage{color,slashed}
\usepackage[colorlinks=true,linkcolor=blue,citecolor=red]{hyperref}
\usepackage{verbatim}
\usepackage{lscape}
\usepackage[utf8]{inputenc} 
\input{epsf}
\usepackage{epsfig}
\usepackage{mathrsfs}
\usepackage[all]{xy}
\usepackage{dsfont}
\usepackage{soul}
\usepackage{comment}
\usepackage{cite}
\usepackage{amssymb}
\usepackage{amsmath}
\usepackage{amsthm}
\usepackage{amsfonts}
\usepackage{mathtools}

\usepackage{tikz}
\usetikzlibrary{arrows}
\usetikzlibrary{arrows.meta}
\usetikzlibrary{positioning}
\usetikzlibrary{shapes,decorations}
\usetikzlibrary{fit}

\tikzset{
  ->-/.style={decoration={markings, mark=at position 0.5 with {\arrow{to}}},
              postaction={decorate}},
}

\tikzset{-<-/.style={decoration={markings,mark=at position 0.5 with %
    {\arrow[scale=1.5,>=stealth]{<}}},postaction={decorate}}}

\numberwithin{equation}{section}

\def\fig#1{Fig.\ #1}

\def\be{ \begin{eqnarray} }
\def\ee{ \end{eqnarray}}

\def\bea#1\eea{\begin{align}#1\end{align}}

\def\CA{{\cal A}}

\def\calD {{\cal D}}
\def\CE {{\cal E}}
\def\CF {{\cal F}}
\def\CG {{\cal G}}
\def\CH {{\cal H}}
\def\CI {{\cal I}}
\def\CJ {{\cal J}}
\def\CK {{\cal K}}
\def\CL {{\cal L}}
\def\CM {{\cal M}}
\def\CN {{\cal N}}
\def\CO {{\cal O}}

\def\CR {{\cal R}}
\def\CV {{\cal V}}
\def\CW {{\cal W}}

\def\CO {{\cal O}}

\def\CE {{\cal E}}
\def\CG {{\cal G}}
\def\CH {{\cal H}}
\def\CI {{{\cal I}}}

\def\CQ {{\cal Q}}
\def\CS {{\cal S}}

\def\CU {{\cal U}}

\def\IC{\mathbb{C}}

\def\IP{\mathbb{P}}
\def\IQ{\mathbb{Q}} % % % % MODIFIED % % % %
\def\IR{{\mathbb{R}}}

\def\IZ{{\mathbb{Z}}}

\def\fg{\mathfrak{g}}

\def\fH{\mathfrak{H}}

\def\fl{\mathfrak{l}}

\def\fu{\mathfrak{u}}

\def\fQ{\mathfrak{Q}}

\def\i{\mathsf{i}}
\def\Tr{{\mathrm{Tr}}}

\def\W{W}

\def\i{{\mathrm i}}

\newcommand{\GL}{\mathrm{GL}}
\newcommand{\U}{\mathrm{U}}
\newcommand{\SO}{\mathrm{SO}}
\newcommand{\SU}{\mathrm{SU}}

\def\bphi{{\boldsymbol{\phi}}}

\def\nn{\nonumber}
\def\lm{\limits}

\def\p{\partial}
\def\Tr{{\rm Tr}}
\def\I{{\rm i}}

\newcommand{\hatE}{\hat{\mathbf{e}}}
\newcommand{\hatF}{\hat{\mathbf{f}}}

\begin{document}
%%%%%%%%%%%%%%%%%%%%%%%%%%%%%%%%%%%
%%%%%%%%%%%%%%%%%%%%%%%%%%%%%%%%%%%

\pagenumbering{Alph} %trick to avoid warning
\begin{titlepage}

\vskip 1.5in
\begin{center}
{\bf\Large{Quiver Yangian\\ \vskip 0.2in  and\\ \vskip 0.2in  Supersymmetric Quantum Mechanics}}
\vskip 1cm 
\renewcommand{\thefootnote}{\fnsymbol{footnote}}
{Dmitry Galakhov$^{1,2,}$\footnote[2]{e-mail: dmitrii.galakhov@ipmu.jp; galakhov@itep.ru} and Masahito Yamazaki$^{1,}$\footnote[3]{e-mail: masahito.yamazaki@ipmu.jp}} 
\vskip 0.2in 
\renewcommand{\thefootnote}{\roman{footnote}}
{\small{ 
\textit{$^1$Kavli Institute for the Physics and Mathematics of the Universe (WPI), }\vskip -.4cm
\textit{University of Tokyo, Kashiwa, Chiba 277-8583, Japan}
\vskip 0 cm 
\textit{$^2$Institute for Information Transmission Problems,}\vskip -.4cm
\textit{ Moscow, 127994, Russia}
}}
\end{center}

\vskip 0.5in
\baselineskip 16pt

\begin{abstract}
The statistical model of crystal melting represents BPS configurations of D-branes on a toric Calabi-Yau three-fold. Recently it has been noticed that an infinite-dimensional algebra, the quiver Yangian, acts consistently on the crystal-melting configurations. We physically derive the algebra and its action on the BPS states, starting with the effective supersymmetric quiver quantum mechanics on the D-brane worldvolume. This leads to remarkable combinatorial identities involving equivariant integrations on the moduli space of the quantum mechanics, which can be checked by numerical computations.
\end{abstract}

\date{\today}
\end{titlepage}
\pagenumbering{arabic} % trick

\newpage
\tableofcontents
\newpage

%%%%%%%%%%%%%%%%%%%%%%%%%%%%%%%%%%%%%%%%%%%%%%%%%%%%%%%%%%%%%%%%%%%%%%%%%%%%%%%%%%%%
\section{Introduction}
%%%%%%%%%%%%%%%%%%%%%%%%%%%%%%%%%%%%%%%%%%%%%%%%%%%%%%%%%%%%%%%%%%%%%%%%%%%%%%%%%%%%

In Calabi-Yau compactifications of type IIA string theory,
D-branes wrapping holomorphic cycles represent Bogomol'nyi-Prasad-Sommerfield (BPS) particles in the four remaining dimensions.
The counting of the degeneracies of these BPS particles has been an important problem both in string theory and mathematics.

While the counting problem of BPS degeneracies of these BPS particles
has been a challenging problem in general, the problem has been solved beautifully for 
non-compact toric Calabi-Yau three-folds \cite{Szendroi:2007nu,Young:2008hn,MR2836398,MR2592501,Ooguri:2008yb,MR2999994,Jafferis:2008uf,Chuang:2008aw,Nagao:2009ky,Nagao:2009rq,Sulkowski:2009rw,Aganagic:2010qr} (see \cite{Yamazaki:2010fz} for a summary).
Here the counting problem of BPS states is 
simplified with the help of the localization with respect to the torus action,
and the result gives the counting of configurations of the statistical-mechanical model of crystal melting (generalizing the results \cite{Okounkov:2003sp,Iqbal:2003ds} for $\mathbb{C}^3$).

Recently, a new infinite-dimensional algebra associated with an arbitrary toric Calabi-Yau manifold was introduced in \cite{Li:2020rij}.
The algebra, called the BPS quiver Yangian, acts consistently on the configurations of the crystal melting counting BPS states.
This seems to have solved problems posed in \cite{Rapcak:2018nsl}, and generalizes previous discussions 
for $\mathbb{C}^3$, where the  affine Yangian of $\mathfrak{gl}_1$ 
(which is equivalent with the universal enveloping algebra of the $\mathcal{W}_{1+\infty}$ algebra  \cite{SV,MO,Tsymbaliuk,Tsymbaliuk:2014fvq,Prochazka:2015deb,Gaberdiel:2017dbk})
acts on plane partitions \cite{FJMM,MR2793271,Tsymbaliuk:2014fvq,Prochazka:2015deb} (see \cite{Gaberdiel:2018nbs,Li:2019lgd,Li:2019nna,RSYZ} for further examples of toric Calabi-Yau geometries).

While explicit algebras and their representations were already constructed in \cite{Li:2020rij}, 
one might still ask if it is possible to revisit them starting directly with the physics and geometry of effective field theories for BPS state counting.
The goal of this paper is to answer this question.

Our approach is based on the supersymmetric quantum mechanics (SQM) on the 
D-brane worldvolume \cite{Denef:2002ru}. We define ``raising''/``lowering'' operators on the BPS states
(adding/removing atoms from the crystal melting configuration), 
which fit nicely with the intuition of the crystal-melting configuration
as a ``bound state'' of atoms.
The operators are defined as versions of the Hecke shift operator, and we obtain explicit expressions for the
matrix elements of the operators in terms of equivariant integration on the moduli space of SQM.

With these ingredients we identify the generators and relations of the BPS state algebra,
and verify that the resulting representation of the algebra coincides with
the representation \cite{Li:2020rij} of the quiver Yangian.
Our approach is based on localization onto the Higgs branch of the SQM moduli space (albeit with the $\Omega$-deformation \cite{Nekrasov:2002qd}),
and hence is complementary to the mathematical discussions (see e.g.\ \cite{Rapcak:2018nsl,RSYZ}),
which seem to deal with localization onto the Coulomb branch. We include further comments on the relation with the Coulomb-branch localization
in section \ref{s:Coulomb}.  

It is worth mentioning how the BPS algebra for Calabi-Yau three-folds enters a wider sequence of examples of BPS algebras for Calabi-Yau $n$-folds. An interest to BPS algebras of D-brane systems wrapping Calabi-Yau $n$-folds was getting a second breath in the physics community after a discovery of the Alday-Gaiotto-Tachikawa relation \cite{Alday:2009aq}. A majority of proof techniques for this relation (see e.g. \cite{Alba:2010qc,Mironov:2011dk}) refers to the norms of specialized Whittaker vectors \cite{kostant1978whittaker} in two related algebras on both sides of the relation. In the case of the instanton partition function this is the Heisenberg representation of the Yangian of affine $\fg\fl_1$---the BPS algebra of instanton impurities, whereas in the case of conformal blocks in the 2d conformal field theory this algebra is the $\CW_N$-subalgebra of $\CW_{1+\infty}$. A later  construction of the Coulomb branch algebra for 3d $\CN=4$ theories by Braverman-Finkelberg-Nakajima \cite{Braverman:2010ef} (see also \cite{Braverman:2014xca,Braverman:2016wma}), and its physical counterpart in \cite{Bullimore:2016hdc}, translates this approach into a similar construction for the vortex moduli space, and appears to be the first (despite historically the second) example in the sequence. Indeed, 3d BPS ``monopole" operators according to \cite{Kapustin:2006pk} correspond to Hecke shift operators on the vortex moduli space, and its brane description gives rise to the BPS algebra of one-folds. The next example  in this sequence refers to original brane description of the moduli space of instantons on ALE space \cite{kronheimer1990yang}, so that the affine Yangian corresponds to the BPS algebra of two-folds. Construction we are interested in this paper appears to be the third in this line, and, finally, there are indications \cite{Nekrasov:2017cih} that some similar construction is possible for four-folds. 
%however after four-fold this sequence is expected to be terminated. 

Spectacularly, as abstract algebras the BPS algebras of one-, two- and three-folds are very similar. For canonical sets of generalized conifolds they will correspond to affine Yangians of Lie superalgebras. What differs drastically is a representation that is delivered automatically in this construction as a BPS Hilbert space. In the cases of $\IC$, $\IC^2$ and $\IC^3$ one derives vector, Fock and MacMahon modules correspondingly of the same $Y\left(\widehat{\fg\fl}_1\right)$.

The rest of this paper is organized as follows. 
In Section \ref{s:derivation} we include our general derivation of the 
BPS algebra from SQM. We will discuss many concrete examples in Section \ref{s:Examples}.
We also include several appendices for some technical materials, which include
examples of explicit computational details.

%%%%%%%%%%%%%%%%%%%%%%%%%%%%%%%%%%%%%%%%%%%%%%%%%%%%%%%%%%%%%%%%%%%%%%%%%%%%%%%%%%%%
\section{Derivation of BPS Algebras}\label{s:derivation}
%%%%%%%%%%%%%%%%%%%%%%%%%%%%%%%%%%%%%%%%%%%%%%%%%%%%%%%%%%%%%%%%%%%%%%%%%%%%%%%%%%%%

%%%%%%%---------------------------------------------------------------------------------------------------------------------------------------------------------------------------------------------
\subsection{Toric Calabi-Yau Threefolds, Quivers and Crystals}\label{s:CY3}

A standard way to probe physically the geometry of a Calabi-Yau threefold is to put on it a system of D6-D4-D2-D0 branes wrapping the Calabi-Yau manifold itself
and holomorphic cycles inside. In the classic paper \cite{Douglas:1996sw} Douglas and Moore showed that the effective dynamics of such systems of D-branes is described by quiver gauge theories. 
Details of this association for toric Calabi-Yau threefolds, in the formalism needed for this paper, 
can be found in \cite{Ooguri:2008yb,Yamazaki:2010fz,Li:2020rij} and references therein. Here we would be content with quoting the results.

A quiver diagram (an oriented graph) $\CQ$ consists of a set of vertices $\CV$ and a set of arrows $\CA$ connecting the vertices. To quiver arrows one associates maps we will denote by variables $a\in\CA$; when we will need to specify its head $w\in \CV$ and tail $v\in\CV$ vertices we will denote such an arrow as $a \colon v\to w$. The path algebra $\IC \CQ$ is generated over $\IC$ by oriented arrow paths inside the quiver with a multiplication defined by the natural path concatenation. One in addition specifies a quiver superpotential $\W$---a holomorphic map from $\CA$ to a subspace $\CL\subset \IC\IQ$ formed by closed loops modulo cyclic permutations of arrows in a loop.

For a toric Calabi-Yau three-fold there is a procedure to produce a pair $(\CQ,\W)$. This procedure defines a periodic quiver $\hat{\CQ}$ drawn on a torus,
where the original quiver diagram $\CQ$ is the same as $\hat\CQ$ considered as an abstract graph. And one constructs the superpotential $\W$ according to the following rule:
\be\label{W}
\W\coloneqq \sum\lm_{\CF\in\left(\mbox{faces of }\hat\CQ\right)} (\pm 1)\; \Tr \prod\lm_{a\in \p \CF}a  \;,
\ee
where the sign in front of the product is defined by the orientation of the face boundary $\p\CF$, and the product is a cyclically-ordered product of arrows in $\p \CF$.

Consider a two-sided ideal $\CI_{\W}$ in $\IC\CQ$ generated by all derivatives $\p_a \W$ for all $a\in \CA$. The Jacobian ring is defined as:
\be
\CJ(\CQ,\W)\coloneqq \IC\CQ/\CI_{\W} \;.
\ee

\begin{figure}
	\begin{center}
		\begin{tikzpicture}
			% 3d
			\draw[->] (0,0) -- (8,0);
			\draw[->] (0,0) -- (0,-2);
			\draw[->] (0,0) -- (-3,-3);
			\begin{scope}[shift={(1,-0.5)}]
				\foreach \x / \y in
				{0/0, 1.5/0, 3/0, 4.5/0, 0.5/-1.5, 2/-2, 3.5/-1, 3.5/-2, -0.5/-3, 1/-3}
				{
					\shade[ball color=gray] (\x,\y) circle (0.12);
				}
				\draw[thick,<->] (3.5, -1.12) -- (3.5, -1.88);
				\draw[red,thick,->] (0.0848528, 0.0848528) to[out=30,in=150] (1.41515, 0.0848528);
				\draw[red,thick,->] (1.62, 0) -- (2.88, 0);
				\draw[red,thick,->] (3.12, 0) -- (4.38, 0);
				\draw[red,thick,->] (4.41515, -0.0848528) -- (3.58485, -0.915147);
				\draw[blue,thick,->] (0.0848528, -0.0848528) to[out=330,in=210] (1.41515, -0.0848528);
				\draw[blue,thick,->] (1.44, -0.103923) -- (0.56, -1.39608);
				\draw[blue,thick,->] (0.44, -1.60392) -- (-0.44, -2.89608);
				\draw[blue,thick,->] (-0.38, -3) -- (0.88, -3);
				\draw[blue,thick,->] (1.08485, -2.91515) -- (1.91515, -2.08485);
				\draw[blue,thick,->] (2.12, -2) -- (3.38, -2);
			\end{scope}
			% 2d
			\begin{scope}[shift={(0,-4)}]
			\draw[fill=white!90!black] (0,0) -- (8,0) -- (5,-3) -- (-3,-3) -- cycle;
			\begin{scope}[shift={(1,-0.5)}]
				\draw[thick,->,red] (0.0424264, 0.0424264) .. controls (1.05,0.3) .. (1.45757, 0.0424264);
				\draw[thick,->,red] (1.56,0) -- (2.94,0);
				\draw[thick,->,red] (3.06,0) -- (4.44,0);
				\draw[thick,->,red] (4.45757, -0.0424264) -- (3.54243, -0.957574);
				\draw[thick,->,blue] (0.0424264, -0.0424264) .. controls (0.45,-0.3) ..  (1.45757, -0.0424264);
				\draw[thick,->,blue] (1.45757, -0.0424264) -- (0.542426, -0.957574);
				\draw[thick,->,blue] (0.457574, -1.04243) -- (-0.457574, -1.95757);
				\draw[thick,->,blue] (-0.44, -2) -- (0.94, -2);
				\draw[thick,->,blue] (1.04243, -1.95757) -- (1.95757, -1.04243);
				\draw[thick,->,blue] (2.06, -1) -- (3.44, -1);
				\draw[fill=black] (0,0) circle (0.1) (1.5,0) circle (0.06) (3,0) circle (0.06) (4.5,0) circle (0.06) (0.5,-1) circle (0.06) (2,-1) circle (0.06) (3.5,-1) circle (0.1) (-0.5,-2) circle (0.06) (1,-2) circle (0.06); 
			\end{scope}
			\end{scope}
			\draw[dashed] (0,-2) -- (0,-4);
			\foreach \x/\y/\z/\w in {0/0/0/0, 1.5/0/1.5/0, 3/0/3/0, 4.5/0/4.5/0, 0.5/-1.5/0.5/-1, 2/-2/2/-1, 3.5/-2/3.5/-1, -0.5/-3/-0.5/-2, 1/-3/1/-2}
			{
				\draw[dashed,thin] (\x+1,\y-0.62) -- (\z+1,\w-4.5);
			}
			\node[right] at (8,0) {3d crystal};
			\node[right] at (8,-4) {2d crystal};
			\node[above left] at (0,0) {$\IR^3$};
			\node[above left] at (0,-4) {$\IR^2$};
			\node[below left] at (1,-4.5) {$\vec 0$};
			\node[below right] at (4.5,-5.5) {$\vec r$};
			\node[right] at (4.5, -2) {$n$};
			\node[red,above] at (4.5, -1.3) {$\wp_{m,0}$};
			\node[blue, below right] at (4.5, -2.62) {$\wp_{m}$};
		\end{tikzpicture}
	\end{center}
	\caption{The 3d crystal and its 2d projection.} \label{f:cry_paths}
\end{figure}
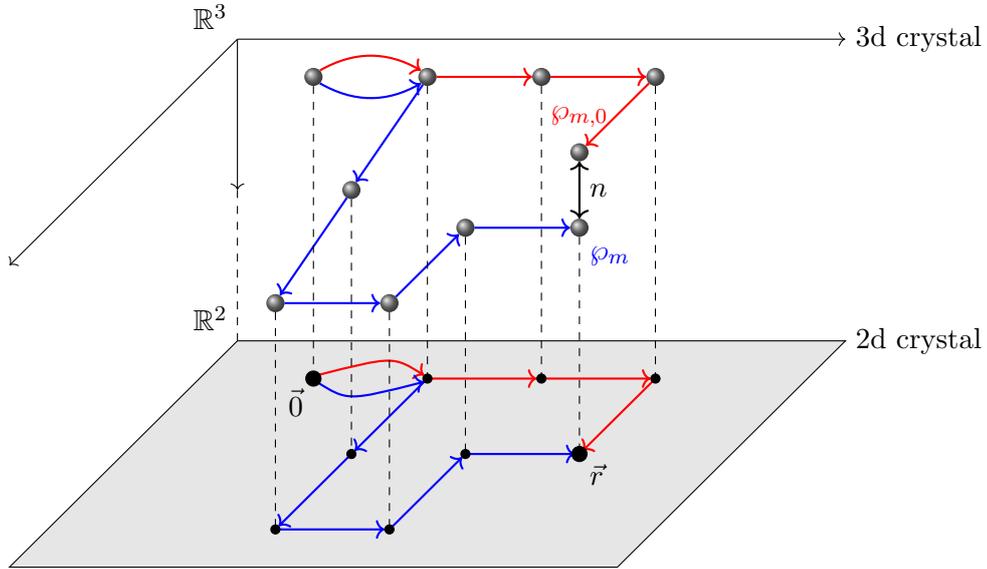

For toric Calabi-Yau three-folds $\CJ(\CQ,\W)$ has a nice visualization in the form of a crystal lattice \cite{Ooguri:2008yb,Yamazaki:2010fz,Li:2020rij}. Consider a lift $\fQ$ of $\hat\CQ$ to $\IR^2$ universally covering the torus. $\fQ$ forms a periodic crystal lattice in 2d. 
In what follows we will consider a single D6-brane wrapping the Calabi-Yau three-fold that defines a quiver framing. It singles out a specific quiver vertex which we will call the framed node. A map from the framing node to the framed node adds an element $\iota$ to $\IC\CQ$. Let us break the translational symmetry of the 2d crystal $\fQ$ by choosing a root atom position in either node lifted from the framed node. Consider monomials in $m\in\CJ(\CQ,\W)\cdot\iota$. Each such monomial defines a path in the quiver path algebra starting at the framing node. We lift this path to a path $\wp_m$ in $\fQ$ starting with the root atom. The end point of this path is an atom position $\vec r$ in $\fQ$. Consider one shortest path $\wp_{m,0}$ connecting the root atom and an atom at position $\vec r$. The difference 
$$
\wp_{m}-\wp_{m,0}
$$
contains $n$ closed loops of $\fQ$ modulo the ideal $\CI_{\W}$. One associates to the monomial $m$ an atom in 3d crystal lattice with coordinates (see Fig.~\ref{f:cry_paths}):
$$
m\rightsquigarrow (\vec r,\;n) \;,
$$
where the integer $n$ specifies the coordinate along the third direction, i.e.\ the ``depth'' from the surface of the crystal.
We will present some examples of crystal lattices for some choices of pairs $(\CQ,\W)$ in Section \ref{s:Examples}.

%%%%%%%%%%%%%%%%%%%%%%%%%%%%%%%%%%%%%%%%%%%%%%%%%%%%%%%%%%%%%%%%%%%%%%%%%
%Crystal figure
%%%%%%%%%%%%%%%%%%%%%%%%%%%%%%%%%%%%%%%%%%%%%%%%%%%%%%%%%%%%%%%%%%%%%%%%%

Considering all the possible paths starting at the framed node we will get a basic crystal $\Lambda_0$ growing from the root atom.

The crystal admits a {\it coloring}. We identify a set of colors with that of the quiver vertices $\CV$. For an atom $\Box$ we denote its color as $\hat\Box$. For an atom $\Box$ identified with a monomial
$$
(a_n\colon v_{n-1}\to v_{n})\cdot (a_n \colon v_{n-2}\to v_{n-1})\cdot\ldots \cdot (a_1\colon v_{0}\to v_{1})\cdot\iota \;,
$$
we define the color $\hat\Box$ of the atom $\Box$ by the endpoint $v_n \in \mathcal{V}$:
$$
\hat \Box=v_n \;.
$$

%%%%%%%---------------------------------------------------------------------------------------------------------------------------------------------------------------------------------------------
\subsection{Quiver Quantum Mechanics and BPS States}
We will consider an effective theory emerging in the system of D-branes probing Calabi-Yau three-fold from the point of view of D-brane worldvolume. The effective  field theory is a quiver SQM with four supercharges \cite{Denef:2002ru}.

In principle, the SQM setup is well defined for an arbitrary pair $(\CQ,\W)$. To specify it completely one needs some extra information which we call the quiver data.

The D-branes wrap holomorphic cycles and hence can be regarded as fractional D0 branes.
The charges of the fractional D0 branes are labeled by quiver vertices $\CV$. By specifying our system we choose our system to contain $n_v\in\IZ_{\geq 0}$ D0 branes for each $v\in \CV$. We incorporate these numbers in a quiver dimension vector $\gamma$:
$$
\gamma=\{n_v \}_{v\in\CV} \;.
$$

To a quiver node $v\in \CV$ one associates a gauge vector multiplet of gauge group $\U(n_v)$, and to an arrow $a \colon v\to w\in \CA$ one associates a chiral multiplet bi-fundamentally charged in $\U(n_v)\times \overline{\U(n_w)}$. We will in addition consider the framing due to the presence of a  D6 brane wrapping the Calabi-Yau manifold. In general, a framing corresponds to a collection of framing nodes $\CF$ and arrows connecting framing and ordinary vertices. The gauge groups in framing nodes correspond to flavor symmetries. The corresponding gauge fields in the vector multiplet associated to framing nodes take constant expectation values equal to flavor masses and all other SUSY partner fields are set to zero. Summarizing, we consider a quantum field theory with the following gauge and flavor group:
$$
\mathcal{G}=\prod\lm_{v\in \CV} \U(n_v)\quad \mbox{and}\quad \mathcal{G}_F=\prod\lm_{f\in\CF}\U(n_f)  \;.
$$

For each gauge group factor $\U(n_v)$ one associates a Fayet-Iliopoulos (FI) term defined by a stability parameter:
$$
\theta_v=2|Z_0|^{-1}\;{\rm Im}(\bar Z_0 Z_v) \;,
$$
where $Z_v$ are the central charges of D0-branes,
 and $Z_0$ is the central charge of the whole D-brane configuration.

These data specify the effective D0-brane theory as a supersymmetric gauged linear sigma-model (GLSM) in 1d. See Appendix \ref{s:SQM} for details.

We have the following set of operators  acting on the Hilbert space of the gauged supersymmetric quantum mechanics: 
Hamiltonian $H$, gauge rotations $\hat G$ and 4 supercharges $Q_{\alpha}$, $\bar Q_{\dot \alpha}$, $\alpha,\dot\alpha=1,2$ satisfying corresponding superalgebra \eqref{superalgebra}.

One defines the BPS states as physical ground states. Using superalgebra relations \eqref{superalgebra} one observes that gauge-invariant BPS states are annihilated by all the supercharges and a gauge transformation generator:
\be
Q_{\alpha}|{\rm BPS}\rangle=\bar Q_{\dot\alpha}|{\rm BPS}\rangle =\hat G|{\rm BPS}\rangle=0 \;.
\ee

The BPS states span a subspace of the Hilbert space of all physical states, which subspace we call the BPS Hilbert space
$$
\CH_{\rm BPS} \;.
$$

We will try to describe the BPS Hilbert space in geometrical terms as it appears in the mathematics literature, see, for example, \cite{Kontsevich:2010px}. The target space of SQM,
which we denote by $\mathcal{M}_{\rm SQM}$, is spanned by the scalar fields of SQM. 
This includes adjoint scalars $X^{1,2,3}_v$ associated with vector multiplets, and bifundamental scalars $q_{(a:\; v\to w)}$ associated with 
chiral multiplets  (see Appendix \ref{s:SQM} for more details): 
\bea
\mathcal{M}_{\rm SQM}: \quad 
\begin{split}
X^3_v\in \fu(n_v),\; \Phi_v\in \fg\fl(n_v,\IC)\;, \quad&v \in\CV\;, \\
q_{(a:\; v\to w)}\in{\rm Hom}(\IC^{n_v},\IC^{n_w})\;,  \quad & a\in\CA \;,
\end{split}
\eea
where we introduced a complex combination $\Phi_v \coloneqq X^1_v+i X^2_v$.

The supercharge has the form of  a differential on the target space (see Appendix \ref{s:SQM}):\footnote{A version of this formula is well-known
since the classic paper \cite{Witten:1982im}, however we are not aware if this particular expression has appeared previously in the literature.}
\be\label{diff}
\bar Q_{\dot 1}=e^{-\fH}\left(d_{X^3}+\bar\p_{\Phi,q}+\iota_V+dW\wedge\right)e^{\fH} \;,
\ee
where we define the height function:
\be
\fH \coloneqq \sum\lm_{v\in\CV}\Tr \; X^3_v\left(\frac{1}{2}\left[\Phi_v,\bar \Phi_v\right]-\mu_{\IR,v}\right) \;,
\ee
a vector field: 
\be\label{vf_before}
V\coloneqq \sum\lm_{(a:\; v\to w)\in\CA}\left(\Phi_wq_a-q_a\Phi_v\right)\frac{\p}{\p q_a} \;,
\ee
and the moment maps define stability conditions:
\be
\label{stability}
\mu_{\IR,v}\coloneqq \theta_v {\mathds{1}}_{n_v\times n_v}-\sum\lm_{x\in\CV}\sum\lm_{(a:v\to x)\in\CA}q_aq_a^{\dagger}+\sum\lm_{y\in\CV}\sum\lm_{(b:y\to v)\in\CA}q_b^{\dagger}q_b \;.
\ee

Using the standard reasoning \cite{Witten:1982im} we can identify the BPS Hilbert space with the $\hat G$-equivariant cohomology of one of the four supercharges,
which we take to be $\bar Q_{\dot 1}$:
\be\label{HH}
\CH_{\rm BPS}\cong H^*_{\hat G}(\bar Q_{\dot 1}) \;.
\ee
Under such geometrical identification the cohomological degree is identified with the fermion number of a physical state.

One subtlety in the identification \eqref{HH} is that  the
target space in the SQM is singular in general. To regularize it 
we introduce the $\Omega$-background \cite{Nekrasov:2002qd}, and 
we consider the cohomology \eqref{HH} after the $\Omega$-deformation.

To introduce the $\Omega$-deformation,  
we introduce an additional $\U(1)$ gauge multiplet  $\U(1)_{F,a}$  for each arrow $a\colon v\to w$,
so that the bifundamental chiral field labeled by an arrow $a\colon v\to w$ has charge $-1$
with respect to this symmetry $\U(1)_{F,a}$.
We then freeze these additional degrees of freedom by setting up expectation values to the  complex adjoint scalar $\Phi_a$
 of the $\U(1)_{F,a}$ vector multiplet: $\langle \Phi_a\rangle=\epsilon_a$.
As a result one finds that this procedure leads to a modification of the vector field action \eqref{vf_before}:
\be\label{vf}
V(q_a)=\sum\lm_{(a:\; v\to w)\in\CA}\left(\Phi_wq_a-q_a\Phi_v-\epsilon_a q_a\right)\frac{\p}{\p q_a} \;.
\ee
In addition, a requirement of supercharge nilpotency (see superalgebra relations \eqref{superalgebra}) leads to a constraint that the superpotential is invariant with respect to the equivariant torus action:
\be\label{W-inv}
\bar Q_{\dot 1}^2=-4\sum\lm_{a\in\CA} \epsilon_a \; \Tr\left(q_a\frac{\p}{\p q_a}\right)W=0 \;.
\ee
Since the superpotential $W$ is a sum of monomials \eqref{W},
\eqref{W-inv} means that $\epsilon_a$ is a flavor symmetry charge assignment to the bifundamental chiral multiplets
consistent with the superpotential. On the crystal the charge assignment 
satisfies the  ``loop constraint''  \cite{Li:2020rij}, i.e.\ the sum of the all the charges around any face of the 
periodic quiver is zero. This charge assignment is parametrized by $|\mathcal{V}|+1$
parameters \cite[Section 4.1]{Li:2020rij}, where the $|\mathcal{V}|$ is the number of the vertices of the quiver $\mathcal{Q}$.
We can use gauge degrees of freedom to shift the charge assignment, and this causes
the reshuffling of the algebra as discussed in \cite[Section 4.3.2]{Li:2020rij}.
We can fix this ambiguity e.g.\ by imposing the ``vertex constraint'',
which reduces the number of parameters to two  \cite{Li:2020rij}---these are nothing but the 
two mesonic flavor symmetries of the SQM, which when combined with an
 R-symmetry (the R-symmetry for the parent four-dimensional $\mathcal{N}=1$ theory) correspond to the three isometries of the toric Calabi-Yau three-fold.
The two parameters are precisely the two parameters for
the 2d projection of the crystal \cite{Li:2020rij}. 

%%%%%%%---------------------------------------------------------------------------------------------------------------------------------------------------------------------------------------------
\subsection{Localization and RG Flow}\label{s:RG}

Let us briefly review an application of the localization technique to a SQM. As we shall see, this is  tightly related to the renormalization group (RG) flow.

Consider a one-parameter family of differentials
\be\label{family}
\bar Q_{\dot 1}^{({\bf s})}=e^{-{\bf s} \fH}\left(d_{X^3}+\bar\p_{\Phi,q}+\iota_{{\bf s}V}+{\bf s}\, dW\wedge\right)e^{{\bf s}\fH} \;.
\ee

The supercharge of the SQM we considered before in equation \eqref{diff} is a member of this family for ${\bf s}=1$. 
Since $\bar Q_{\dot 1}^{({\bf s})}$ with different values of ${\bf s}$ are related by conjugation,
the corresponding cohomology is an invariant of this one-parameter family.
The major idea of the localization technique is to evaluate this cohomology in a special limit ${\bf s}\to\infty$.

The supercharge $\bar Q_{\dot 1}^{({\bf s})}$ leads to a family of corresponding Hamiltonians:
\be
H^{({\bf s})}=\left\{\bar Q_{\dot 1}^{({\bf s})},\left(\bar Q_{\dot 1}^{({\bf s})}\right)^{\dagger} \right\}\sim \Delta+{\bf s}^2\left|\vec\nabla\fH\right|^2+{\bf s}^2\left|\vec V\right|^2+{\bf s}^2\left|\vec\nabla W\right|^2+\ldots \;.
\ee
In the limit ${\bf s}\to\infty$ the contributions of potential terms will grow, therefore trajectories represented by particles sitting in classical vacua---zeroes of the potential---will give the dominant contribution to the path integral.

Let us reparameterize the degrees of freedom as:
$$
X_i=\langle x_i\rangle+{\bf s}^{-\frac{1}{2}}x_i \;.
$$
We observe that the Hamiltonian and the supercharge decompose as:
\be
H={\bf s} H_0+O\left({\bf s}^\frac{1}{2}\right),\quad \bar Q_{\dot 1}^{(s)}={\bf s}^{\frac{1}{2}} \bar Q_{\dot 1}^{(0)}+ \bar Q_{\dot 1}^{(1)}+O({\bf s}^{-\frac{1}{2}}) \;.
\ee
Here $H_0$ and $\bar Q_{\dot 1}^{(0)}$ are simple expressions representing a free particle: 
$$
H_0\sim \sum\lm_i \left(-\p_{x_i}^2+\omega_i^2x_i^2\right)+\sum\lm_i \omega_i\left(\psi_i\psi_i^{\dagger}-\psi_i^{\dagger}\psi_i\right),\quad\bar Q_{\dot 1}^{(0)}\sim\sum\lm_i\psi_i\left(\p_{x_i}+\omega_i x_i\right)\;. 
$$

Let us choose a cutoff $\Lambda_{\rm cf}$ for frequencies $\omega_i$. Then the leading contribution to the supercharge is of order ${\bf s}^{\frac{1}{2}}\Lambda_{\rm cf}^{\frac{1}{2}}$. We can split the wave-function in two parts, fast modes $x_{|\omega|>\Lambda_{\rm cf}}$ and slow ones $x_{|\omega|<\Lambda_{\rm cf}}$:
$$
\Psi=\Psi_{|\omega|<\Lambda_{\rm cf}}\left(x_{|\omega|<\Lambda_{\rm cf}}\right)\Psi_{|\omega|>\Lambda_{\rm cf}}\left(x_{|\omega|<\Lambda_{\rm cf}},x_{|\omega|>\Lambda_{\rm cf}}\right)+O({\bf s}^{-1}) \;.
$$

The first order BPS equation reads:
$$
\left(\bar Q_{\dot 1}^{(0)}\right)^{\dagger}\Psi_{|\omega|>\Lambda_{\rm cf}}=\bar Q_{\dot 1}^{(0)}\Psi_{|\omega|>\Lambda_{\rm cf}}=0 \;.
$$
The leading order of the supercharge $\bar Q_{\dot 1}^{(0)}$ will not have derivatives with respect to slow modes $x_{|\omega|<\Lambda_{\rm cf}}$, which will enter the corresponding expression for the wave-function only as parameters. Therefore this equation is not enough to define $\Psi_{|\omega|<\Lambda_{\rm cf}}$. To derive a defining equation we incorporate contributions to the supercharge up to the first order and multiply this equation by a bra-vector $\Psi_{|\omega|<\Lambda_{\rm cf}}^{\dagger}$:
$$
\Psi_{|\omega|>\Lambda_{\rm cf}}^{\dagger}\left({\bf s}^{\frac{1}{2}} \bar Q_{\dot 1}^{(0)}+ \bar Q_{\dot 1}^{(1)}\right)\Psi_{|\omega|>\Lambda_{\rm cf}} \Psi_{|\omega|<\Lambda_{\rm cf}} =0\;.
$$
The first term in this sum cancels out.

Splitting modes into slow and fast modes is a familiar procedure for the Wilsonian RG flow: the unknown part $\Psi_{|\omega|<\Lambda_{\rm cf}}$ of the first-order approximation to the wave-function is annihilated by 1-loop corrected effective supercharge:
\be
Q_{\rm eff}^{\dagger}\Psi_{|\omega|<\Lambda_{\rm cf}}=0,\quad Q_{\rm eff}^{\dagger}\coloneqq\left\langle\Psi_{|\omega|>\Lambda_{\rm cf}}\Big|\bar Q_{\dot 1}^{(1)}\Big|\Psi_{|\omega|>\Lambda_{\rm cf}}\right\rangle \;.
\ee

%%%%%%%---------------------------------------------------------------------------------------------------------------------------------------------------------------------------------------------
\subsection{Higgs Branch Localization and Crystal Melting}\label{s:Higgs}

In our setup  we choose a Higgs branch localization. In other words, we assume that $\theta_v$-parameters are large and the vacuum expectation values are given to the chiral fields---we will choose the following orders in the size of the parameters in question:
\be \label{hierarchy}
|\epsilon|\ll \Lambda_{\rm cf} \ll |\theta|^{\frac{1}{2}} \;.
\ee
Actually, as we will see in what follows, the vacuum 
expectation values of the vector-multiplet scalars are also non-zero being resolved by the $\Omega$-background parameters.

Following the procedure of the previous section we associate wave-functions with vacuum field values---critical points of the height function $\fH$ and the superpotential $\W$ fixed with respect to the action of complexified gauge field $V$ introduced in \eqref{vf}.

Critical points of $\fH$ define a zero of the real moment map \eqref{stability}.
This equation is an analog of the constant curvature equation in the Narashiman-Shishadri-Hitchin-Kobayashi correspondence \cite{Donaldson}, and can be traded for a stability condition if we complexify the gauge group.  

Consider a complexification 
$$
\CG_{\IC}=\prod\lm_{v\in\CV} \GL(n_v,\IC) 
$$
of the quiver gauge group $\CG$. In general a quiver representation $\CR$ is a $\CG_{\IC}$-orbit of a collection of vector spaces $$\CR=\bigoplus\lm_{v\in\CV}\IC^{n_v}$$
associated to quiver nodes, equipped with the action of morphisms associated to quiver arrows:
$$
q_{(a:\; v\to w)}\in {\rm Hom}(\IC^{n_v},\IC^{n_w}) \;.
$$
For a quiver representation with a dimension vector $\gamma$ we define a function
\be\label{stb}
\theta(\gamma)\coloneqq \sum\lm_{v\in\CV}n_v\theta_v, \quad \theta_v\in \IR \;.
\ee
For a representation $\CR$ with dimension vector $\gamma$, the FI parameters $\theta_v$ satisfy $\theta(\gamma)=0$. This constraint follows naturally if one adds up traces of all the moment maps \eqref{stability}.
A quiver representation $\CR$ is called semi-stable (stable) if for all proper subrepresentations $\CR'$ we have $\theta(\gamma')>0$ ($\theta(\gamma')\geq 0$). A theorem of King \cite{King} (see also \cite{Denef:2002ru} and examples in \cite{Alim:2011kw}) states that each stable quiver representation $\CR$ contains a single solution to \eqref{stability} up to complexified gauge transformations. And all the solutions to \eqref{stability} are contained in orbits of  semi-stable representations. In our consideration all the semi-stable representations will be stable, therefore we can establish an equivalence between solutions to \eqref{stability} and stable quiver representations.

The notion of algebraic stability is translated to a notion of physical stability \cite{Douglas:2000ah,Harvey:1995fq}. Indeed a physical vacuum represents a D-brane configuration, and possible quiver subrepresentations are considered to be constituting more elementary D-branes in the initial composite D-brane. The stability constraint for \eqref{stb} defines a stability chamber in the space of D-brane central charges, where a composite D-brane is stable since its decay to more elementary D-branes is forbidden by conservation laws.

Throughout this paper we will consider a single type of framing---a single framing node with dimension 1 and a framing map $\iota$ connecting this node to any other quiver node. It is simple to generalize King's theorem to framed quivers. We only need to consider the framing node as a gauge node and assign to it some fictitious stability parameter $\theta_f$. Then the stability function is modified as
$$ 
\theta(\gamma)=\theta_f n_f+\sum\lm_{v\in\CV}n_v\theta_v \;, \quad \theta_v\in \IR \;.
$$
A stable representation is, obviously, indecomposable. For us this implies that whole $\CR$ is generated as a module $\IC\CQ\cdot \iota$. For such a representation $n_f=1$, and all subrepresentations are also modules of this form, therefore to be at least non-zero for a subrepresentation $\CR'$ we have $n_f'=1$. Then choosing $\theta_v<0$ and $\theta_f=-\sum\lm_{v\in\CV}n_v\theta_v$ we will find that all the framed  representations are stable. An alternative proof could be derived using a stability condition for framed quivers as in \cite{Reineke}.

Therefore we construct all the solutions to the moment map \eqref{stability} condition $\mu_{\IR}=0$ in the quiver SQM as indecomposable modules $\IC\CQ\cdot \iota$. Then the vacuum manifold $\mu_{\IR}^{-1}(0)\cap d\W^{-1}(0)$ corresponds to modules of the Jacobian ring $\CJ(\CQ,\W)\cdot\iota$. Since we would like to consider modules of finite dimension, we construct such modules as complements of vector spaces 
$$
(\CJ(\CQ,\W)\cdot \iota)/ (\CS\cdot \iota) \;,
$$
where $\CS$ is an ideal in $\CJ(\CQ,\W)$. The equivariant fixed point on $\mu_{\IR}^{-1}(0)\cap d\W^{-1}(0)$ corresponds to $\CS$ generated by monomials.

In Section \ref{s:CY3} we have identified basis elements of $\CJ(\CQ,\W)\cdot \iota$ with atoms of the crystal growing from the root atom. Then the vector space $\CS\cdot \iota$ is naturally identified with a molten crystal. Here the melting rule \cite{Ooguri:2008yb} is a translation of the fact that $\CS$ is an ideal: a subset of atoms of a basic crystal growing from the root atom is a molten crystal $\CK$ if for any atom $\Box\in \CK$ and all maps $(a:\; \hat\Box\to \circ)\in\CA$ all atoms $a\cdot\Box$ are also in $\CK$.

The classical vacua on the Higgs branch for our quiver SQM are in one-to-one correspondence with finite crystals growing from the root atom  and obtained  as a complement of the basic crystal $\Lambda_0$ to some molten crystal $\CK$. We will refer to such a finite crystal $\Lambda$ as just a crystal for the sake of brevity. 

Having such a crystal $\Lambda$ we easily restore a vacuum quiver representation. In the representation we choose a distinguished basis of vectors of $\CR$ labeled by crystal atoms $\Box$ as monomials in $\CJ(\CQ,\W)\cdot\iota$. In particular, we have
\be\label{box_rep}
\IC^{n_v}\cong {\rm Span}\; \{|\Box\rangle \}_{\hat \Box=v} \;.
\ee 

For quiver maps we have the following representation:
\be
q_{(a \colon v\to w)}|\Box\rangle=\left\{\begin{array}{ll}
	|a\cdot\Box\rangle,&\mbox{if }\hat\Box=v,\mbox{ and } a\cdot\Box\in\Lambda \;,\\
	0, & \mbox{otherwise} \;.\\
	\end{array}
	\right. 
\ee
Expectation values of the complex fields $\langle\Phi_v\rangle$ in the gauge multiplets are also simple to calculate in the chosen basis: they are nonzero only if $v$ coincides with the color $\hat \Box$ of the atom $\Box$, and take a diagonal form with eigenvalues $\bphi_{\Box}$:
\be \label{phi_sum}
{\bphi}_{a_1\cdot a_2\cdot\ldots \cdot a_n\cdot\iota}=\sum\lm_{i=1}^n\epsilon_a \;.
\ee

Our aim is to define expressions for the effective wave-functions  $\Psi_{|\omega|<\Lambda_{\rm cf}}$ (which we will denote as $\Psi$ in what follows for brevity) in the vicinity of corresponding fixed points. First notice that the frequencies of the fields contributing to the gauge multiplet are of order $\sim\sqrt{\theta}$, and they do not contribute to the wave-function. The vector field term \eqref{vf} in the supercharge mixes Goldstone modes of the chiral fields with respect to the action of $\CG_{\IC}$ with complex scalars of the vector multiplets. This mixing forces chiral field Goldstone modes to acquire frequencies as well of order $\sim\sqrt{\theta}$. An explicit example of such mixing could be found in Appendix \ref{s:exapmle}.

The effective wave-function $\Psi$ is supported on the meson manifold:
\be
\CM_{\rm meson}=\left(\bigoplus\lm_{a\in\CA} \IC \, \delta q_a\right)/\fg_{\IC} \;,
\ee
where we have decomposed chiral fields into vacuum expectation values and fluctuations: $q_a=\langle q_a\rangle+\delta q_a$. The action of the complexified gauge algebra on the fluctuation degrees of freedom reads:
$$
\{g_v\}\in\fg_{\IC}:\quad \delta q_{(a\colon v\to w)}\mapsto \delta q_{(a\colon v\to w)}+g_w\langle q_{(a\colon v\to w)}\rangle-\langle q_{(a\colon v\to w)}\rangle g_v \;.
$$
The effective supercharge on the meson manifold takes a form of the equivariantly extended Dolbeault differential. The equivariant torus action of the field \eqref{vf} introduces a natural grading on $\CM_{\rm meson}$ where a matrix element $\langle \Box'|\delta q_a|\Box\rangle$ has an equivariant weight equal to
$$
\bphi_{\Box'}-\bphi_{\Box}-\epsilon_a \;.
$$
The effective wave-function $\Psi$ is therefore given by a Thom representative of the Euler class \cite{Cordes:1994fc} associated with the critical point $\Lambda$. We could write down a simplified version of this expression in a specific basis $\{m_i\}$ for $\CM_{\rm meson}$ where the equivariant vector field action diagonalizes 
$$
V=\sum\lm_{i}w_i\; m_i\frac{\p}{\p m_i} \;,
$$
Notice that restricting $\fg_{\IC}$ back to its uncomplexified version $\fg$ for vector field $V$ we produce generators of flavour group transformations. Even if the effective metric on $\CM_{\rm meson}$ is corrected by the RG flow, the resulting metric should remain flavour invariant. This, in turn, implies that different weight spaces are orthogonal to each other. We can hence assume in addition that in chosen basis the metric takes the following simple form:
$$
ds^2=\sum\lm_i d\bar m_i \;dm_i\;.
$$
Effective supercharges have the form of equivariant Dolbeault differential that is rather simple in the chosen basis (compare to \cite[eq.(23)]{Lillywhite2002FormalityIA}):
$$
Q_{\rm eff}^{\dagger} = \sum\lm_i\left(d\bar m_i\;\p_{\bar m_i}+ w_i m_i\; \iota_{\p/\p_{m_i}}\right)\;.
$$
Using dictionary \eqref{dict} we translate them to operators:
\be
Q_{\rm eff}\sim \sum\lm_i\left(-\psi_{1, i}\p_{m_i}+\psi_{ 2,i}\bar w_i\bar m_i\right),\quad Q_{\rm eff}^{\dagger}\sim \sum\lm_i\left(\bar\psi_{\dot 1, i}\p_{\bar m_i}+\bar\psi_{\dot 2,i}w_im_i\right)\;.
\ee
It is easy to single out a harmonic $Q_{\rm eff}^{\dagger}$-cohomology representative by the following condition:
\be
Q_{\rm eff}\Psi_{\Lambda}=Q_{\rm eff}^{\dagger}\Psi_{\Lambda}=0\;.
\ee
A solution to this system of equations reads:
\be\label{wf_dec}
\Psi_{\Lambda}=\left(\prod\lm_{i} \left(w_i-|w_i|\; \bar\psi_{\dot 1,i}\psi_{2,i} \right)e^{-|w_i||m_i|^2}\right)\prod\lm_i\bar\psi_{\dot 2,i}|0\rangle \;.
\ee
This wave-function describes simply Gaussian fluctuations around the vacuum labeled by the crystal $\Lambda$.
One can translate this expression into a differential form using the dictionary \eqref{dict}, the result is precisely the Thom representative of the Euler class:
$$
\Psi_{\Lambda}\sim \bigwedge\lm_i e^{-|w_i||m_i|^2}\left(w_i-|w_i|\; dm_i\wedge d\bar m_i \right)\;.
$$
Indeed, for \eqref{wf_dec} we have:
\begin{align}
\begin{split}
\Psi_{\Lambda}=\left(\prod\lm_i w_i\right)\; e^{-\left\{Q_{\rm eff}^{\dagger},\sum\lm_i \frac{|w_i|}{w_i}\bar m_i \psi_{2,i} \right\}}\prod\lm_i\bar\psi_{\dot 2,i}|0\rangle = \\
= \left(\prod\lm_i w_i\right)\prod\lm_i\bar\psi_{\dot 2,i}|0\rangle+\left(\mbox{$Q_{\rm eff}^{\dagger}$-exact term}\right)\;.
\end{split}
\end{align}

Thus the wave-function is cohomologically equivalent to the Pfaffian of the curvature (compare to \cite[Sections 11.1.2 and 11.6]{Cordes:1994fc}):
\be\label{wf_norm}
\Psi_{\Lambda}\sim {\rm Eul}_{\Lambda}\coloneqq \prod\lm_{i}w_i \;.
\ee

The Euler class satisfies the following normalization conditions following from equivariant integration: 
\be\label{norm}
\int\Psi_{\Lambda}=1 \;, \quad \int\Psi_{\Lambda}\wedge \Psi_{\Lambda'}={\rm Eul}_{\Lambda}\; \delta_{\Lambda,\Lambda'} \;.
\ee
Here $\Psi_{\Lambda}$ and $\Psi_{\Lambda'}$ in these integrals are treated as forms---elements of equivariant cohomology, integration goes over the quiver representation moduli space.

This normalization condition is rather unusual from the physics point of view; it would be more conventional to use the unitary Hermitian norm descending form the Hermitian structure on the Hilbert space. Nevertheless, as it was pointed out in \cite[Section 3.3]{Bullimore:2016hdc} the very transition form the harmonic forms to the equivariant Dolbeault cohomologies we made in Section \ref{s:RG} to apply localization techniques prioritizes the complex structure on $\CG_{\IC}$ over unitarity. As we will see this choice of the norm give rise to a BPS algebra resembling the desired properties of affine Yangians. For a comparison, we could mention a similar phenomenon occurs for a basis of orthogonal Jack polynomials \cite{Macdonald}. Jack polynomials are known to deliver a fixed point basis representation in the BPS Hilbert space for Hilbert scheme on $\IC^2$, see Section \ref{s:gl2} for details. The vectors of this basis are orthogonal simultaneously with respect to two norm choices: a Hermitian one and a ``holomorphic" one. However raising and lowering operators resembled by multiplications by time variables $p_n$ and $p_{-n}$ correspondingly are conjugated to each other only for the holomorphic norm.

%%%%%%%---------------------------------------------------------------------------------------------------------------------------------------------------------------------------------------------
\subsection{Hecke Shift Generators}\label{s:Hecke}
We have considered the effective theory from the point of view of the D0-brane worldvolume.
Let us call this description I. On the other hand, we could have started with an effective theory on the worldvolume of the non-compact D6-brane wrapping the Calabi-Yau manifold. Let us call this alternative description II.

Description II has an interpretation as ``stringy K\"ahler gravity" \cite{Bershadsky:1994sr,Iqbal:2003ds,Cirafici:2008sn},
an effective description of K\"ahler quantum foam. D0-branes represent point-like gravitational sources deforming the initial Calabi-Yau geometry in such a way that the K\"aler form $\omega$ takes the following form:
\be
\omega=\omega_0+g_s F_A \;,
\ee
where $\omega_0$ is an unperturbed K\"ahler form of Calabi-Yau three-fold $X$, $g_s$ is the string coupling constant, and $F_A$ is a curvature of a $\U(1)$ connection $A$. Then, effectively, description II is in terms of $\CN=2$ supersymmetric six-dimensional Yang-Mills theory, whose vacua define constraints on the gauge connection $A$. These equations can be reduced to Donaldson-Uhlenbeck-Yau (DUY) equations for $A$: 
\be\label{DUY}
{\arraycolsep=1.4pt\def\arraystretch{2}
\begin{array}{lcc}
	F_A^{(2,0)}=F_A^{(0,2)}=0 \;, \quad
	\omega_0\wedge \omega_0\wedge F_A^{(1,1)}=0 \;. 
\end{array}}
\ee
These equations define a natural six-dimensional generalization of the instanton self-duality equations in four dimensions.

Descriptions I and II are equivalent to each other by construction as both are applied to the same system of D-branes. One could track down this equivalence to IR degrees of freedom. In the cases of systems of D0-D4 branes and D0-D2 branes this equivalence gives rise to ADHM-like description of instanton \cite{KN} and vortex \cite{Eto:2005yh} moduli spaces. And the very equivalence relation is known as a Nahm transform. However, unfortunately, the Nahm transform is inapplicable to 6d situation since it uses the fact that 4d $2\times 2$ chiral gamma-matrices form a quaternion representation. Nevertheless algebro-geometric construction of Beilinson spectral sequences \cite{Beilinson} remains applicable.

A conventional analysis \cite{Nakajima_book,Donaldson_Kronheimer_book} of DUY equations \eqref{DUY} identifies instanton solutions with stable holomorphic connections on $X$ that are promoted to torsion free sheaves. It is natural to expect that all such sheaves can be constructed as cohomologies of monads. Examples of such analysis for a certain class of Calabi-Yau three-folds is given in \cite{Cirafici:2008sn}. For example, for the case of $\IC^3$ one considers torsion free sheaves on $\IC\IP^3$ with prescribed framing at infinite lines. Each such sheaf $\CE$ is a cohomology of a three-term monad. In this case the quiver is just a trefoil quiver:
\be\label{trefoil_early}
\begin{array}{c}
	\begin{tikzpicture}
	\draw[->] (0.173,-0.1) to[out=330,in=270] (1,0) to[out=90,in=30] (0.173,0.1);
	\node[right] at (1,0) {$B_1$};
	\begin{scope}[rotate=90]
	\draw[->] (0.173,-0.1) to[out=330,in=270] (1,0) to[out=90,in=30] (0.173,0.1);
	\node[above] at (1,0) {$B_2$};
	\end{scope}
	\begin{scope}[rotate=180]
	\draw[->] (0.173,-0.1) to[out=330,in=270] (1,0) to[out=90,in=30] (0.173,0.1);
	\node[left] at (1,0) {$B_3$};
	\end{scope}
	\draw (0,0) circle (0.2);
	%\draw[->] (0,-1) --  (0,-0.2);
	%\draw (-0.1,-1) -- (0.1,-1) -- (0.1,-1.2) -- (-0.1,-1.2) -- (-0.1,-1);
	\end{tikzpicture}
\end{array} \;,\quad W=\Tr\left(\left[B_1,B_2\right]B_3\right) \;.
\ee 
 The quiver representation space is given by:
$$
\IC^k\cong H^2\left(\IC\IP^3,\CE\otimes \CO_{\IC\IP^3}\left(-3\right)\right) \;.
$$

Consider a holomorphic bundle $\CE$ over $X$. We will call $\CE'$ a \emph{Hecke modification} of $\CE$ in a point $x\in X$ if $c_3(\CE')=c_3(\CE)\pm 1$ and there is an isomorphism:
\be 
\mu:\quad \CE'\big|_{X\setminus x}\longrightarrow \CE\big|_{X\setminus x} \;.
\ee
In other words we can say there exists a gauge transform $G_{\mu}$ that is in general singular, however it is smooth in an open subset $X \setminus x$ and satisfies the following relation for connection:
\be\label{Hecke}
\nabla_AG_{\mu}=G_{\mu}\nabla_{A'} \;.
\ee

A singular homomorphism of holomorphic bundles induces corresponding homomorphism of quiver representations. Following \cite{King} we define a homomorphism $\tau$ of quiver representations $\CR$ and $\CR'$:
$$
\tau:\quad \CR'\longrightarrow \CR\;,
$$
to be a collection of linear maps $\{\tau_v\}_{v\in\CV}$:
$$
\tau_v:\quad \IC^{n_v'}\longrightarrow \IC^{n_v}\;,
$$
satisfying the following relations:
\be\label{hmm}
q_{(a\colon v\to w)}\cdot\tau_v=\tau_w\cdot q'_{(a\colon v\to w)} \;.
\ee
Note that the quiver description of the morphism \eqref{hmm}
works for a general Calabi-Yau manifold other than $\mathbb{C}^3$.

The generalization of ADHMN construction to 6d instantons maps a self-dual gauge connection to morphisms $q_a$ of the quiver representation. Then we can treat equation \eqref{hmm} as an image of \eqref{Hecke} under this equivalence relation. 

We define the \emph{raising} Hecke operator $\hatE_v$ as a BPS operator performing a Hecke modification on a bundle associated to description II and \emph{increasing} the number of D0 branes of charge $v\in\CV$ by 1. Similarly, \emph{lowering} Hecke operator $\hatF_v$ \emph{decreases} the number of $v$-colored D0-branes by 1. Correspondingly, we have
$$
n_v'=n_v\pm 1,\quad\mbox{and} \quad n_w'=n_w,\;{\rm for}\; w\neq v \;.
$$

In the case, say, $n_v'=n_v+1$ homomorphism $\tau$ describes $\CR$ as a subrepresentation of $\CR'$. A physical interpretation of this fact is \cite{Douglas:2000ah} that $\CR$ can appear among products of $\CR'$'s decay. Unfortunately, we are unable  to give an immediate description of the decay process in the current framework for the following reason. A decay of a bound state occurs establishing a wall-crossing phenomenon at the boundary of the marginal stability chamber where the stability constraint for \eqref{stb} is not fulfilled. At this boundary some part of FI parameters $\theta_v$ change the sign going through the zero value. However, the Higgs branch description  fails down in a vicinity of $\theta_v=0$ where the Coulomb branch or a mixed branch opens. A non-perturbative parallel transport of branes through such regions is available for some models \cite{Herbst:2008jq} and represents a physical description of braiding for brane categories through a Fourier--Mukai transform we will mention later. 

Here let us present some physical arguments to derive matrix elements for operators $\hatE, \hatF$. A picture of a molten crystal suggests a natural physical intuition behind the decay (recombination) processes that an atom is taken to (brought from) infinity being detached from (attached to) the crystal body. Eigenvalues of the complex scalar field in the gauge multiplet $\bphi_{\Box}$ are in general complex numbers. Let us assign to atoms $\Box\in\Lambda$ positions in $\IC$ corresponding to values $\bphi_{\Box}$. Notice that for a pair of atoms connected by an arrow, $\Box$ and $a\cdot\Box$, their positions differ as (recall \eqref{phi_sum}):
$$
\bphi_{a\cdot\Box}-\bphi_{\Box}=\epsilon_a \;.
$$
Equivariant weights satisfy the gauge-invariance condition for the superpotential $\W$ \eqref{W}. If  arrows $a_1,\ldots,a_n$ form a face of the periodic graph $\hat\CQ$ then
$$
\sum\lm_k\epsilon_{a_k}=0 \;.
$$
This condition implies that for any pair of atoms $\Box$ and $\omega\cdot \Box$,  where $\omega$ is a closed path in the quiver, $\bphi_{\Box}=\bphi_{\omega\cdot\Box}$. Therefore, a collection of points $\bphi_{\Box}\in\IC$ is precisely the 2d projection of the 3d crystal $\Lambda$ (Fig.~\ref{f:cry_paths}). Then a process of taking away an atom $\Box$ corresponds to a limit $\bphi_{\Box}\to\infty$.

In principle we are unable to make this limit adiabatic since not all the intermediate states are BPS. Nevertheless let us parallel transport the wave-function $\Psi$ to the limit $\bphi_{\Box}\to\infty$ and observe how the RG flow modifies it. We will consider a process of crystal decay when one atom $\Box$ is detached from crystal $\Lambda$ resulting in another crystal $\Lambda-\Box$. If $\Lambda-\Box\subset\Lambda$ one easily constructs the expectation value of morphism $\tau$ for representations of critical points $\Lambda$ and $\Lambda-\Box$:
\be\label{tau_in}
\langle\tau\rangle|\Box'\rangle=\left\{\begin{array}{ll}
	|\Box'\rangle \:,&\Box'\neq\Box \;;\\
	0 \;,&\mbox{otherwise} \;.
\end{array} \right. 
\ee
Then for tangent directions we have a linearized version of \eqref{hmm}
\be\label{lin_hmm}
\langle q_{(a\colon v\to w)}\rangle\cdot\delta\tau_v+ \delta q_{(a\colon v\to w)}\cdot\langle\tau_v\rangle=\delta\tau_w\cdot\langle q'_{(a\colon v\to w)}\rangle+\langle\tau_w\rangle\cdot \delta q'_{(a\colon v\to w)} \;.
\ee

Let us denote linearized meson space around critical point $\Lambda-\Box$ as $\CM_1$ and around critical point $\Lambda$ as $\CM_2$. Linear equations \eqref{lin_hmm} cut out a hyperplane $\Sigma$ inside $\CM_1\oplus\CM_2$. Let us denote coordinates in $\CM_1$ as $m_i^{(1)}$ and coordinates in $\CM_2$ as $m_i^{(2)}$. Then in $\CM_1\oplus \CM_2$ we can choose a specific basis so that the equation system defining an arrangement of $\Sigma$ inside $\CM_1\oplus\CM_2$ has the following form:
\be\label{Sigma_eq}
\begin{array}{ll}
	m_i^{(1)} = m_i^{(2)}\;,& i=1,\ldots,n_1\;,\\
	m_i^{(1)} = 0\;, &i=n_1+1,\ldots,n_2\;,\\
	m_i^{(2)} = 0\;, &i=n_1+1,\ldots,n_3\;,\\
\end{array}
\ee
where $n_k$ are some dimensions.

We split our spaces and denote subspaces accordingly:
\be
\begin{split}
&U\coloneqq \;{\rm Span}\left\{m_i^{(1)} \right\}_{i=1}^{n_1}\;\cong\;{\rm Span}\left\{m_i^{(2)} \right\}_{i=1}^{n_1}\;,\\
&V_1\coloneqq \;{\rm Span}\left\{m_i^{(1)} \right\}_{i=n_1+1}^{n_2},\quad
V_2\coloneqq \;{\rm Span}\left\{m_i^{(2)} \right\}_{i=n_1+1}^{n_3}\;,\\
&W_1\coloneqq \;{\rm Span}\left\{m_i^{(1)} \right\}_{i=n_2+1}^{{\rm dim}\;\CM_1},\quad W_2\coloneqq \;{\rm Span}\left\{m_i^{(2)} \right\}_{i=n_3+1}^{{\rm dim}\;\CM_2}\;.
\end{split}
\ee

Schematically this decomposition may be depicted in the following way:
\be\label{decomp}
\begin{array}{c}
\begin{tikzpicture}
\begin{scope}[shift={(-0.6,0.6)}]
\draw (-0.3,-0.3) -- (-0.3,0.3) -- (0.3,0.3) -- (0.3,-0.3) -- cycle;
\foreach \i in {1,...,4}
{
	\draw[blue] (-0.3+ 0.6*\i/4,0.3) -- (-0.3,0.3- 0.6*\i/4);
}
\foreach \i in {1,...,3}
{
	\draw[blue] (0.3- 0.6*\i/4,-0.3) -- (0.3,-0.3+ 0.6*\i/4);
}
\end{scope}
\begin{scope}[shift={(0,0.6)}]
\draw[fill=green] (-0.3,-0.3) -- (-0.3,0.3) -- (0.3,0.3) -- (0.3,-0.3) -- cycle;
\end{scope}
\begin{scope}[shift={(0.6,0.6)}]
\draw[fill=red] (-0.3,-0.3) -- (-0.3,0.3) -- (0.3,0.3) -- (0.3,-0.3) -- cycle;
\end{scope}
\begin{scope}[shift={(-0.6,-0.6)}]
\draw (-0.3,-0.3) -- (-0.3,0.3) -- (0.3,0.3) -- (0.3,-0.3) -- cycle;
\foreach \i in {1,...,4}
{
	\draw[red] (-0.3+ 0.6*\i/4,0.3) -- (-0.3,0.3- 0.6*\i/4);
}
\foreach \i in {1,...,3}
{
	\draw[red] (0.3- 0.6*\i/4,-0.3) -- (0.3,-0.3+ 0.6*\i/4);
}
\end{scope}
\begin{scope}[shift={(0,-0.6)}]
\draw[fill=green] (-0.3,-0.3) -- (-0.3,0.3) -- (0.3,0.3) -- (0.3,-0.3) -- cycle;
\end{scope}
\begin{scope}[shift={(0.6,-0.6)}]
\draw[fill=blue] (-0.3,-0.3) -- (-0.3,0.3) -- (0.3,0.3) -- (0.3,-0.3) -- cycle;
\end{scope}
\node at (0,0) {\rotatebox{90}{$\cong$}};
\draw[thick, dashed] (-0.4, 0.9) to[out = 90,in=180] (-0.3,1.0) -- (0.9,1.0) to[out=0, in=180] (1.0,0.9) -- (1.0,-0.9) to[out=270,in=0] (0.9, -1.0) -- (0.3,-1.0) to[out=180,in=270] (0.2,-0.9) -- (0.2, 0.1) to[out=90,in=0] (0.1,0.2) -- (-0.3,0.2) to[out=180,in=270] (-0.4,0.3) -- cycle;
\node[left] at (-0.9,0.6) {$\CM_1$};
\node[left] at (-0.9,-0.6) {$\CM_2$};
\node[right] at (1.0,0) {$\Sigma$};
\node[above] at (-0.6,1.0) {$V_1$};
\node[above] at (0,1.0) {$U$};
\node[above] at (0.6,1.0) {$W_1$};
\node[below] at (-0.6,-1.0) {$V_2$};
\node[below] at (0,-1.0) {$U$};
\node[below] at (0.6,-1.0) {$W_2$};
\end{tikzpicture}
\end{array}\quad\quad
\begin{array}{l}
\CM_1=V_1\oplus U\oplus W_1\;,\\
\CM_2=V_2\oplus U\oplus W_2\;,\\
\Sigma=U\oplus W_1\oplus W_2\;.
\end{array}
\ee
A simple example of such decomposition is presented in Appendix \ref{s:ideals}.

Having solved \eqref{lin_hmm} we represent morphism $\tau_v$ in the form:
$$
\tau_v = \langle\tau_v\rangle+\delta\tau_v\;.
$$
Gauge transforms for $\fg_{\IC}$ acting on both systems with generators $g_v^{(1)}$ and $g_v^{(2)}$ act on morphisms as well:
$$
\{g_v^{(i)}\}\in\fg_{\IC}^{(i)}:\quad \delta \tau_v\mapsto \delta \tau_v+g_v^{(1)}\langle \tau_v\rangle-\langle \tau_v\rangle g_v^{(2)} \;.
$$
We can use gauge transform $g_v^{(2)}$ to ``kill" $\delta\tau_v$ so that the gauge transformed morphism $\tilde\tau_v$ becomes again a simple projection $\langle \tau_v\rangle$. 

Since $\tilde \tau$ is a  simple projection we can propose a block decomposition of chiral maps $q$: 
\be\label{block}
\tilde q^{(2)}(V_2,U,W_2)\sim\left(\begin{array}{c|c}
q^{(1)}(V_1=0,U,W_1) & f(V_2)\\
\hline
\multicolumn{2}{c}{h(W_2)}
\end{array}\right)\;,
\ee
where $f$ and $h$ are some matrix valued functions, so that different blocks are responsible for maps between bases spanned by subcrystals $\Box$ and $\Lambda-\Box$ inside $\Lambda$:
\be\nn
\begin{split}
&\tilde q^{(2)}:\; \Lambda\to\Lambda \;, \quad q^{(1)}:\; \Lambda-\Box\to\Lambda-\Box\;,\\
&f:\; \Box\to \Lambda-\Box \;,\quad h:\; \Lambda\to\Box\;.
\end{split}
\ee
An explicit example of such block decomposition for the case of Hilbert schemes on $\IC^2$ is given in \eqref{block_decomp}.

This block decomposition allows us to stress the physical meaning of our strategy for splitting $\CM_1$, $\CM_2$ and $\Sigma$ into subspaces. Under morphism $\tilde\tau$ the block $q^{(1)}$ inside $\tilde q^{(2)}$ is directly projected to $q^{(1)}$ in a representation corresponding to $\Lambda-\Box$. 
$W_2$ defines a block corresponding to a kernel of morphism $\tilde\tau$, therefore equations for $\Sigma$ leave this subspace unconstrained. $V_2$ is not in the kernel of $\tilde\tau$ however it is mapped to 0, therefore equations for $\Sigma$ impose a condition $V_2=0$ (see \eqref{Sigma_eq}). On the block $q^{(1)}$ morphism $\tilde\tau$ is invertible, however subspace $V_1$ of $\CM_1$ is not presented in the block $q^{(1)}$ inside $\tilde q^{(2)}$ at all, therefore we set it to zero, and derive another portion of equations in \eqref{Sigma_eq} defining $\Sigma$. Subspaces $U$ and $W_1$ are mapped into the block $q^{(1)}$ under $\tilde\tau^{-1}$, this defines remaining equations in \eqref{Sigma_eq} for $U$. Finally, $W_1$ remains unconstrained since those are not mesonic degrees of freedom in $\CM_2$: we  have brought them in by gauge transform $g^{(2)}$, and in $\CM_2$ they can be gauged away.

Based on these speculations for the wave function of the BPS state in the $\Lambda$ vacuum one can propose the following form:
$$
\Psi_{\omega<|\Lambda_{\rm cf}|}(V_2,U,W_2)\times \Psi_{\omega>|\Lambda_{\rm cf}|}(W_1,{\bf hfm})\;,
$$
here $\bf hfm$ denotes other high frequency modes. $W_1$ belongs to the high frequency modes since we defined it as non-mesonic.

Let us follow how this function varies as we parallel transport it from crystal $\Lambda$ to $\Lambda-\Box$. First we split all the wave functions of subsystems based on decomposition \eqref{decomp}. In general these subspaces have different weight subspaces, and equations for $\Sigma$ are equivariant, so we may work with these subspaces as orthogonal elements:
$$
\Psi_{\omega<|\Lambda_{\rm cf}|}(V_2)\Psi_{\omega<|\Lambda_{\rm cf}|}(U)\Psi_{\omega<|\Lambda_{\rm cf}|}(W_2)\times \Psi_{\omega>|\Lambda_{\rm cf}|}(W_1)\Psi_{\omega>|\Lambda_{\rm cf}|}({\bf hfm})\;.
$$
During parallel transport $W_1$ belonging to a bulk of high frequency modes, becomes low frequency in $\Lambda-\Box$ and contributes to the effective wave function. In particular, the mixing between $W_1$ and $\fg_{\IC}$ is defined by expectation values of some chiral fields $\langle q\rangle$(see, for example, \eqref{mixing}), schematically we could write for field $x\in W_1$:
$$
w_x\sim \sqrt{\langle q\rangle^2+|\epsilon_x|^2}\;.
$$  
As we take an atom $\Box$ away by inflating value of $\bphi_{\Box}$ expectation value $\langle q\rangle$ that contributes to the potential by a term $\left|\bphi_{\Box}\langle q\rangle\right|^2$ should go to zero, and mixing between gauge and chiral degrees of freedom disappears:
\be
{\rm high\;}w: \; w_x\sim \sqrt{\langle q\rangle^2+|\epsilon_x|^2}\sim \langle q\rangle\; \to\; {\rm low\;}w: \; w_x\sim \epsilon_x \; .
\ee
In our consideration we keep track of normalization of wave functions only up to a holomorphic factor in equivariant weights $\epsilon$, therefore for wave functions this transition generates a coefficient as in \eqref{wf_norm} given by an inverse product over all equivariant weights of $W_1$:
\be
\Psi_{\omega>|\Lambda_{\rm cf}|}(W_1) \to \frac{1}{{\rm Eul}(W_1)}\Psi_{\omega<|\Lambda_{\rm cf}|}(W_1)\;.
\ee

$V_2$ establish an opposite behavior, so those degrees of freedom become ``heavy", corresponding weights $w$ being just linear functions in $\epsilon$ get inflated. Therefore for those wave functions we derive:
\bea
\begin{split}
\Psi_{\omega<|\Lambda_{\rm cf}|}(V_2) \to {\rm Eul}(V_2)\Psi_{\omega>|\Lambda_{\rm cf}|}(V_2)\;.
\end{split}
\eea
To derive the wave function for $\Lambda-\Box$, one has to implement also degrees of freedom corresponding to $V_1$, since they are absent in the initial system for crystal $\Lambda$. Gathering all contributions one arrives to the following relation:
\bea
\begin{split}
\Psi_{\omega<|\Lambda_{\rm cf}|}(V_1)& \times \Psi_{\omega<|\Lambda_{\rm cf}|}(\CM_2)\times \Psi_{\omega>|\Lambda_{\rm cf}|}=\\
\quad\quad \quad \quad \quad  &=\frac{{\rm Eul}(V_2)}{{\rm Eul}(W_1)}\times \Psi_{\omega<|\Lambda_{\rm cf}|}(\CM_1)\times\Psi_{\omega<|\Lambda_{\rm cf}|}(W_2)\times \Psi_{\omega>|\Lambda_{\rm cf}|}'\;.
\end{split}
\eea

The wave function $\Psi_{\omega<|\Lambda_{\rm cf}|}(W_2)$ contains degrees of freedom $W_2$ that are not projected to $\CM_1$. Since these degrees of freedom do not become high frequency modes, it is natural to associate this wave function with the D0 brane and $W_2$ with the degrees of freedom carried away. 

We associate to the action of $\hatF$ a matrix coefficient given by a numerical coefficient in the above expression:
\be
\frac{{\rm Eul}(V_2)}{{\rm Eul}(W_1)} = \frac{{\rm Eul}(\CM_2)}{{\rm Eul}(\Sigma)}\;,
\ee
where we used \eqref{decomp}. Let us introduce the following notations:
\be
{\rm Eul}_{\Lambda}\coloneqq{\rm Eul}(\CM_2),\quad {\rm Eul}_{\Lambda-\Box,\Lambda}\coloneqq{\rm Eul}(\Sigma)\;.
\ee

A complete expression for $\hatF$ is given by contributions from all possible atom subtractions from the crystal $\Lambda$.
Denote $\Lambda^-(\Lambda^+)$ a set
of atoms that can be removed from/added to crystal $\Lambda$ and the result will be a crystal
again, then we have:
\be\label{M^-}
\hatF\; \Psi_{\Lambda} =\sum\lm_{\Box\in\Lambda^-}\frac{{\rm Eul}_{\Lambda}}{{\rm Eul}_{\Lambda-\Box,\Lambda}}\Psi_{\Lambda-\Box} \;.
\ee

Expression for $\hatE$ is defined from the requirement that $\hatE$ and $\hatF$ are conjugate with respect to the norm \eqref{norm}:
\be\label{M^+}
\hatE\; \Psi_{\Lambda} =\sum\lm_{\Box\in\Lambda^+}\frac{{\rm Eul}_{\Lambda}}{{\rm Eul}_{\Lambda,\Lambda+\Box}}\Psi_{\Lambda+\Box} \;.
\ee
 
These expressions coincide with ones derived by geometric methods in \cite{Nakajima_book}, where Hecke modification corresponds to a Fourier--Mukai transform of $\Psi_{\Lambda}\in H^*_{\CG_{\IC}}(\CM_2)$ on a product manifold $\CM_1\times\CM_2$ with a kernel given by a structure sheaf of the equivariant incidence locus \eqref{hmm}. Indeed, using orthogonality of norm \eqref{norm} we can calculate corresponding coefficients in expansion \eqref{M^+} as
$$
c_{\Box}\coloneqq \frac{\int\lm_{\CM_1}(\hatE\; \Psi_{\Lambda})\wedge\Psi_{\Lambda+\Box}}{\int\lm_{\CM_1} \Psi_{\Lambda+\Box}\wedge \Psi_{\Lambda+\Box}} \;.
$$
Then pulling back the numerator integral to $\CM_1\times \CM_2$ we can calculate it using standard localization techniques:
$$
c_{\Box}=\frac{\int\lm_{\Sigma}\Psi_{\Lambda}\otimes \Psi_{\Lambda+\Box}}{\int\lm_{\CM_1} \Psi_{\Lambda+\Box}\wedge \Psi_{\Lambda+\Box}}=\frac{{\rm Eul}_{\Lambda}{\rm Eul}_{\Lambda+\Box}}{{\rm Eul}_{\Lambda,\Lambda+\Box}}\cdot\frac{1}{{\rm Eul}_{\Lambda+\Box}}=\frac{{\rm Eul}_{\Lambda}}{{\rm Eul}_{\Lambda,\Lambda+\Box}} \;.
$$

%%%%%%%---------------------------------------------------------------------------------------------------------------------------------------------------------------------------------------------
\subsection{From BPS Algebra to Quiver Yangian} \label{s:construction}

Hecke shift operators we derived in the previous section are BPS operators; they map BPS states into BPS states. They have no perturbative interpretation in general---BPS algebra contains non-perturbative ``monopole-like" operators \cite{Bullimore:2016hdc} 
in the gauge-invariant cohomologies of the supercharge $\bar Q_{\dot 1}$.

% as perturbative operators \cite{Bullimore:2016hdc} that are gauge invariant cohomologies of the supercharge $\bar Q_{\dot 1}$.

There are also another set of perturbative BPS operators of the form $\Tr (\Phi_v^k)$ with $k \in \mathbb{N}$ and $v\in \mathcal{V}$.
Clearly, the supercharge in the form \eqref{diff} commutes with $\Tr (\Phi_v^k)$, and 
gauge-invariance of this operator is guaranteed by the trace. These operators 
may be combined into a generating series with the help of a spectral parameter $z\in\IC$:
\begin{align} \label{z_Phi}
\Tr\; (z-\Phi_v)^{-1} \;.
\end{align}

In the crystal basis representation we have constructed so far these operators become diagonal, the corresponding eigenvalues are given just by expectation value $\langle\Phi_v\rangle$ in the corresponding vacuum. Using atom representation \eqref{box_rep} we easily derive eigen values of these operators:
\be
\left[\Tr\; (z-\Phi_v)^{-1}\right]|\Lambda\rangle=\left(\sum\lm_{\Box\in\Lambda^{(v)}}\frac{1}{z-\bphi_{\Box}}\right)|\Lambda\rangle\;,
\ee
where $\Lambda^{(v)}\subset\Lambda$ is a subset of crystal atoms of color $v$.

In the case of the Hecke modification on a complex plane \cite{Bullimore:2016hdc} the corresponding BPS algebra representation is known to give rise to a vector module of the corresponding quiver Yangian. The vector module is labeled by a one-dimensional crystal that is just a one-dimensional array of atoms growing from a root atom, where new atoms can be added and removed just at the tip end of this string. Corresponding BPS algebra of non-perturbative Hecke modifications closes to the BPS subalgebra of perturbative operators:
$$
\hatF\, \hatE\sim f_1(\Phi),\quad \hatE \, \hatF\sim f_2(\Phi)\;.
$$
Unfortunately, this is not the case for two-dimensional and three-dimensional crystals where vacant atom positions are scattered all over the crystal boundary.

Momentarily, we will discuss relations in our BPS algebra. Let us implement first another set of notations adopted form \cite{Li:2020rij}. Define matrix elements of operators $\hatE, \hatF$:
\begin{align}
\begin{split}
\hat E(\Lambda\to\Lambda+\Box)\coloneqq \frac{\langle\Psi_{\Lambda+\Box}|\hatE|\Psi_{\Lambda}\rangle}{\langle\Psi_{\Lambda+\Box}|\Psi_{\Lambda+\Box}\rangle}=\frac{{\rm Eul}_{\Lambda}}{{\rm Eul}_{\Lambda,\Lambda+\Box}} \;,\\
\hat F(\Lambda\to\Lambda-\Box)\coloneqq\frac{\langle\Psi_{\Lambda-\Box}|\hatF|\Psi_{\Lambda}\rangle}{\langle\Psi_{\Lambda-\Box}|\Psi_{\Lambda-\Box}\rangle}= \frac{{\rm Eul}_{\Lambda}}{{\rm Eul}_{\Lambda-\Box,\Lambda}} \;.
\end{split}
\end{align}

Calculation of these matrix coefficients for each concrete crystal $\Lambda$ and atom $\Box$ is a well-posed linear algebra problem, however dimensions of involved vector spaces grow quite rapidly with the crystal size. Currently, we are unable to present generic combinatorial expressions for matrix coefficients $E$ and $F$. Using programming tools, however, we are able to predict relations between matrix coefficients and to check them in a vast variety of quiver examples and for various crystals. We will concentrate on these examples in section \ref{s:Examples} and put some explicit calculations in Appendix \ref{s:calc}.

We find that the matrix elements satisfy the following set of relations (cf.\  \cite{Li:2020rij}):
\bea\label{EF0}
\begin{split}
	&\hat E(\Lambda\to\Lambda+\Box)\hat F(\Lambda+\Box\to \Lambda)=\mathop{\rm res}\lm_{z=p_{\Box}}\hat\psi_{\Lambda}^{(\hat \Box)}(z)=-\mathop{\rm res}\lm_{z=p_{\Box}}\hat\psi_{\Lambda+\Box}^{(\hat \Box)}(z) \;, \\
	&\frac{\hat E(\Lambda+\Box_1\to\Lambda+\Box_1+\Box_2)\hat F(\Lambda+\Box_1+\Box_2\to\Lambda+\Box_1)}{\hat F(\Lambda+\Box_1\to \Lambda)\hat E(\Lambda\to\Lambda+\Box_2)}=1 \;, \\
	&\frac{\hat E(\Lambda\to\Lambda+\Box_1)\hat E(\Lambda+\Box_1\to\Lambda+\Box_1+\Box_2)}{\hat E(\Lambda\to\Lambda+\Box_2)\hat E(\Lambda+\Box_2\to\Lambda+\Box_1+\Box_2)}\hat \varphi_{\hat\Box_1,\hat\Box_2}\left(\bphi_{\Box_1}-\bphi_{\Box_2}\right)=1 \;, \\
	&\frac{\hat F(\Lambda+\Box_1+\Box_2\to\Lambda+\Box_1)\hat F(\Lambda+\Box_1\to\Lambda)}{\hat F(\Lambda+\Box_1+\Box_2\to\Lambda+\Box_2)\hat F(\Lambda+\Box_2\to\Lambda)}\hat \varphi_{\hat\Box_1,\hat\Box_2}\left(\bphi_{\Box_1}-\bphi_{\Box_2}\right)=1 \;,
\end{split}
\eea
where we introduced the following functions (cf.\  \cite{Li:2020rij}):
\begin{align}\label{psihat}
\begin{split}
&\hat\varphi_{v,w}(z)\coloneqq \boxed{(-1)^{\delta_{v,w}+|a:\; v\to w|}}\frac{\prod\lm_{(a\colon v\to w)\in\CA}\left(z+\epsilon_a\right)}{\prod\lm_{(b:\;w\to v)\in\CA}\left(z-\epsilon_b\right)} \;,\\
&\hat\psi^{(v)}_{\Lambda}(z)\coloneqq \left(\frac{1}{z}\right)^{\delta_{{\bf f.n.},v}}\left(\prod\lm_{(a:\;v\to v)\in\CA}-\frac{1}{\epsilon_a}\right)\prod\lm_{\Box\in\Lambda}\hat \varphi_{v,\hat\Box}(z-\bphi_{\Box}) \;.
\end{split}
\end{align}
Here $\bf f.n.$ denotes the color of the root atom corresponding to the node that is ``framed", i.e. it is the target of the map $\iota$ from the framing node.

The BPS algebra is often compactly represented by the introduction of the spectral parameter.
In addition to \eqref{z_Phi} we can also introduce spectral-parameter dependence to the generators $\hatE, \hatF$
by the commutator:
\bea\label{gen_e_f}
\begin{split}
\hat e^{(v)}(z) & \coloneqq \left[\Tr\; (z-\Phi_v)^{-1},\hatE\right] \;,\\
\hat f^{(v)}(z) & \coloneqq -\left[\Tr\; (z-\Phi_v)^{-1},\hatF\right] \;.
\end{split}
\eea
This modification helps to split the action of operators $\hatE, \hatF$ on the vacant atom positions in $\Lambda^{+}$ and $\Lambda^{-}$, so that the matrix elements of $\hat e^{(v)}$ and $\hat f^{(v)}$ will have poles in vacant atom positions projected to the $\Phi$-plane \cite{Li:2020rij}:
\bea
\begin{split}
	\hat e^{(v)}(z)|\Lambda\rangle=\sum\lm_{\substack{\Box\in\Lambda^+\\ \hat \Box=v}}\frac{1}{z-\bphi_{\Box}}\times \hat E(\Lambda\to\Lambda+\Box)|\Lambda+\Box\rangle \;,\\
	\hat f^{(v)}(z)|\Lambda\rangle=\sum\lm_{\substack{\Box\in\Lambda^-\\ \hat \Box=v}}\frac{1}{z-\bphi_{\Box}}\times \hat F(\Lambda\to\Lambda-\Box)|\Lambda-\Box\rangle \;.
\end{split}
\eea
Also we introduce an operator $\hat \psi^{(v)}(z)$ through its matrix elements on the crystal representation:
\be
\hat\psi^{(v)}(z)|\Lambda\rangle=\hat\psi_{\Lambda}^{(v)}(z)\times |\Lambda\rangle\; .
\ee

We would like to consider a BPS algebra generated by the following set of generators:
\be\nn
\hat e^{(v)}(z)  \;, \,\, \hat f^{(v)}(z)   \;, \,\, \hat \psi^{(v)}(z)  \;, \quad v\in \mathcal{V}\;.
\ee

Using relations \eqref{EF0} we find that these generators satisfy the following closed set of OPE (cf.\ \cite[section 6]{Li:2020rij}):
\bea\label{OPE0}
\begin{split}
	\left[\hat e^{(v)}(x),\hat f^{(w)}(y)\right]&\sim\delta_{vw}\frac{\hat \psi^{(v)}(x)-\hat\psi^{(w)}(y)}{x-y} \;,\\
	\hat\psi^{(v)}(x)\hat e^{(w)}(y)&\simeq\hat\varphi_{v,w}(x-y)\hat e^{(w)}(y)\hat \psi^{(v)}(x)\;,\\
	\hat \psi^{(v)}(x)\hat f^{(w)}(y)&\simeq\left[\hat\varphi_{v,w}(x-y)\right]^{-1}\hat f^{(w)}(y)\hat \psi^{(v)}(x)\;,\\
	\hat e^{(v)}(x)\hat e^{(w)}(y)&\sim\hat\varphi_{v,w}(x-y)\hat e^{(w)}(y)\hat e^{(v)}(x)\;,\\
	\hat f^{(v)}(x)\hat f^{(w)}(y)&\sim\left[\hat\varphi_{v,w}(x-y)\right]^{-1}\hat f^{(w)}(y)\hat f^{(v)}(x) \;,\\
\end{split}
\eea
where $\sim$ ($\simeq$) imply that both sides coincide in expansion in $x$ and $y$ up to monomials $x^j y^{k\ge 0}$ and $x^{j\ge 0} y^k$ ($x^{j\ge 0} y^{k\ge 0})$.

The OPE relations are slightly different from those expected for the BPS quiver Yangian $\mathsf{Y}_{(\mathcal{Q},\W)}$ of \cite{Li:2020rij}. In particular, when the Calabi-Yau 3-fold in question is a generalized conifold $\mathsf{Y}_{(\mathcal{Q},\W)}$ becomes a Yangian of affine Lie superalgebra $Y(\widehat{\fg\fl}_{m|n})$, and generators $e$ and $f$ acquire parity. In this case the $e$--$f$ commutator in \eqref{OPE0} is expected to be substituted by a supercommutator taking into account generator parity, for example.  

We could assign a parity to atoms of color $v\in \CV$ as it appears in the cohomological Hall algebra construction \cite{Kontsevich:2010px}. Similarly, according to the boxed contribution in \eqref{psihat} a permutation of two $\hat e^{(v)}$ gives an extra minus sign if the number $|a\colon v\to v|$ is even and a plus sign if this number is odd. So we define a parity for node $v\in\CV$ as:
\be
|v|=\frac{1+(-1)^{|a:\; v\to v|}}{2}\;.
\ee

For \emph{non-chiral} quivers with 
$|a\colon v\to w|=|a\colon w\to v|$
one can define a canonical binding factor $\varphi$ as  \cite{Li:2020rij}: 
\be
\varphi_{v,w}(z)\coloneqq\frac{\prod\lm_{(a:\, v\to w)\in\CA}\left(z+\epsilon_a\right)}{\prod\lm_{(b:\;w\to v)\in\CA}\left(z-\epsilon_b\right)} \;.
\ee
If quiver $\CQ$ is non-chiral this binding factor satisfies the following relation under node permutation:
\be
\varphi_{v,w}(z)=\varphi_{w,v}(-z)^{-1}\;.
\ee
It allows one to establish explicitly how the mutual parity of generators affects OPE. So for generators, say, $e^{(v)}$ with parity defined by $|v|$ we expect to observe the following OPE relation:
\be
e^{(v)}(x) e^{(w)}(y)\sim (-1)^{|v||w|}\varphi_{v,w}(x-y)e^{(w)}(y) e^{(v)}(x)\;.
\ee

Let us modify matrix elements by mere sign shifts:
\begin{align}
\begin{split}
E(\Lambda\to\Lambda+\Box)= \sigma_+(\Lambda,\Box) \hat{E}(\Lambda\to\Lambda+\Box)\;,\\
F(\Lambda+\Box\to\Lambda)= \sigma_-(\Lambda,\Box) \hat{F}(\Lambda+\Box\to\Lambda)\;,
\end{split}
\end{align}
where $\sigma_+$ and $\sigma_-$ take only values $\pm 1$. It is simple to calculate, that if one defines these sign shifts as a result of pairwise ``interactions" of a new atom with each crystal atom:
\be
\sigma_\pm(\Lambda,\Box)=\prod\lm_{\Box'\in\Lambda}\nu_\pm(\Box',\Box)\;,
\ee
where functions $\nu_{\pm}$ satisfy the following defining relations:
\begin{equation}\label{stat}
\begin{split}
\nu_+(\Box_1,\Box_2)&=(-1)^{|\hat \Box_1||\hat \Box_2|+\delta_{\hat \Box_1,\hat \Box_2}+|a:\;\hat\Box_2\to \hat\Box_1|}\nu_+(\Box_2,\Box_1) \;,\\
\nu_-(\Box_1,\Box_2)&=(-1)^{|\hat \Box_1||\hat \Box_2|}\nu_+(\Box_2,\Box_1)  \;,
\end{split}
\end{equation} 
then new matrix coefficients satisfy the following set of relations  \cite[section 6]{Li:2020rij}:
\bea\label{EF1}
\begin{split}
&	E(\Lambda\to\Lambda+\Box)F(\Lambda+\Box\to \Lambda)=\mathop{\rm res}\lm_{z=p_{\Box}}\psi_{\Lambda}^{(\hat \Box)}(z)=-(-1)^{|\hat \Box|}\mathop{\rm res}\lm_{z=p_{\Box}}\psi_{\Lambda+\Box}^{(\hat \Box)}(z) \;, \\
&	\frac{E(\Lambda+\Box_1\to\Lambda+\Box_1+\Box_2)F(\Lambda+\Box_1+\Box_2\to\Lambda+\Box_1)}{F(\Lambda+\Box_1\to \Lambda)E(\Lambda\to\Lambda+\Box_2)}=(-1)^{|\hat\Box_1||\hat\Box_2|} \;, \\
&	\frac{E(\Lambda\to\Lambda+\Box_1)E(\Lambda+\Box_1\to\Lambda+\Box_1+\Box_2)}{E(\Lambda\to\Lambda+\Box_2)E(\Lambda+\Box_2\to\Lambda+\Box_1+\Box_2)}\varphi_{\hat\Box_1,\hat\Box_2}\left(p_{\Box_1}-p_{\Box_2}\right)=(-1)^{|\hat\Box_1||\hat\Box_2|} \;, \\
&	\frac{F(\Lambda+\Box_1+\Box_2\to\Lambda+\Box_1)F(\Lambda+\Box_1\to\Lambda)}{F(\Lambda+\Box_1+\Box_2\to\Lambda+\Box_2)F(\Lambda+\Box_2\to\Lambda)}\varphi_{\hat\Box_1,\hat\Box_2}\left(p_{\Box_1}-p_{\Box_2}\right)=(-1)^{|\hat\Box_1||\hat\Box_2|} \;,
\end{split}
\eea
where $\psi$ is related to $\varphi$ in the same way as $\hat \psi$ is related to $\hat \varphi$ in \eqref{psihat}.

Then generators defined by the following matrix elements:
\bea
\begin{split}
	e^{(v)}(z)|\Lambda\rangle& \coloneqq\sum\lm_{\substack{\Box\in\Lambda^+\\ \hat \Box=v}}\frac{1}{z-\bphi_{\Box}}\times  E(\Lambda\to\Lambda+\Box)|\Lambda+\Box\rangle \;,\\
	f^{(v)}(z)|\Lambda\rangle& \coloneqq\sum\lm_{\substack{\Box\in\Lambda^-\\ \hat \Box=v}}\frac{1}{z-\bphi_{\Box}}\times F(\Lambda\to\Lambda-\Box)|\Lambda-\Box\rangle \;,\\
	\psi^{(v)}(z)|\Lambda\rangle& \coloneqq\psi_{\Lambda}^{(v)}(z)\times |\Lambda\rangle\;, 
\end{split}
\eea
satisfy the following set of OPE relations \cite[section 6]{Li:2020rij}:\footnote{We define a supercommutator as a bilinear form on the algebra generators:
$$
\left[x,y\right\}\coloneqq x y-(-1)^{|x||y|}y x \;.
$$}
\bea\label{q_OPE}
\begin{split}
	\left[ e^{(v)}(x),f^{(w)}(y) \right\}&\sim\delta_{v,w}\frac{\psi^{(v)}(x)-\psi^{(v)}(y)}{x-y}  \;,\\
	\psi^{(v)}(x)e^{(w)}(y)&\simeq\varphi_{v,w}(x-y)e^{(w)}(y)\psi^{(v)}(x)\;,\\\
	\psi^{(v)}(x)f^{(w)}(y)&\simeq \left[\varphi_{v,w}(x-y)\right]^{-1}f^{(w)}(y)\psi^{(v)}(x)\;,\\\
	e^{(v)}(x)e^{(w)}(y)&\sim(-1)^{|v||w|}\varphi_{v,w}(x-y)e^{(w)}(y) e^{(v)}(x)\;,\\\
	f^{(v)}(x)f^{(w)}(y)&\sim(-1)^{|v||w|}\left[\varphi_{v,w}(x-y)\right]^{-1}f^{(w)}(y)f^{(v)}(x) \;.\\
\end{split}
\eea
%where $\sim$ ($\simeq$) denotes equality up to regular terms $y^{k\ge 0}$ ($x^{j\ge 0} y^{k\ge 0})$.

We will be interested in studying the properties of this BPS algebra for a quiver $\mathcal{Q}$ and a superpotential $\W$ generated for a Calabi-Yau three-fold.  The OPE relations \eqref{q_OPE} define the quiver Yangian $\mathsf{Y}_{(\mathcal{Q},\W)}$ of \cite{Li:2020rij}.\footnote{Note 
	that the precise expressions for the coefficients
	$E(\Lambda\to \Lambda+\Box)$ and $F(\Lambda\to \Lambda-\Box)$
	are different between this paper and \cite{Li:2020rij};
	we can change expressions of $E$ and $F$
	by changing relative normalizations of the basis $|\Lambda\rangle$.
	The algebra relations \eqref{q_OPE} themselves do not depend on the choice of these normalizations.}
	
In Section \ref{s:Examples} we will show for a large class of examples that the resulting algebra coincides with known examples of affine Yangian algebras with $e$ and $f$ generators acting like raising/lowering operators. %This justifies the name the BPS quiver Yangian. 
We also verify extra higher order Serre relations, to obtain
the reduced quiver Yangian $\underline{\mathsf{Y}}_{(\mathcal{Q},\W)}$ of \cite{Li:2020rij}. Presently, these Serre relations are known for quiver Yangians resembling Yangians of affine Lie superalgebras.

Let us emphasize here that the quiver SQM construction delivers not only the BPS algebra as an abstract algebra rather it constructs a concrete representation module---crystal representation. 

%%%%%%%---------------------------------------------------------------------------------------------------------------------------------------------------------------------------------------------
\subsection{\texorpdfstring{Coulomb Branch Localization, Anyon Statistics, Shuffle Algebras and $R$-Matrix}{Coulomb Branch Localization, Anyon Statistics, Shuffle Algebras and R-Matrix}}
\label{s:Coulomb}

In \cite{Kontsevich:2010px} the cohomological Hall algebra (CoHA) was defined as a shuffle algebra of polynomials. To a quiver with a dimension vector $\gamma=\{n_v \}_{v\in\CV}$ one associates a ``state" $\Psi$. The wave-function $\Psi$ is a polynomial in variables $\phi_{v,i}\in\IC$, $v\in\CV$, $i=1,\ldots, n_v$, symmetric under group $\prod\lm_{v\in\CV}\mathfrak{S}_{n_v}$ permuting subscripts $i$ and preserving subscripts $v$. One can multiply states $\Psi^{(1)}$ and $\Psi^{(2)}$ with dimension vectors $\gamma_1$ and $\gamma_2$ corresponding to a chosen quiver $\CQ$, the result is a state with dimension vector $\gamma_1+\gamma_2$:
\begin{equation}\label{COHA_mul}
\begin{split}
&(\Psi^{(1)}*\Psi^{(2)})\left[\left\{\phi_{v,i}^{(1)} \right\}_{v\in\CV,i={1,\dots, n^{(1)}_v}}\cup \left\{\phi_{w,j}^{(2)} \right\}_{w\in\CV,j={1,\dots, n_w^{(2)}}}\right] \\
&\qquad \coloneqq \sum\lm_{\rm shuffles} \Psi^{(1)}\left(\phi^{(1)}_{v,i} \right)\Psi^{(2)}\left(\phi^{(2)}_{w,j} \right)\frac{\prod\lm_{v,w\in\CV}\prod\lm_{i=1}^{n^{(1)}_v}\prod\lm_{j=1}^{n^{(2)}_w}(\phi^{(2)}_{w,j}-\phi^{(1)}_{v,i})^{|a\colon v\to w|}}{\prod\lm_{v\in \CV}\prod\lm_{i=1}^{n^{(1)}_v}\prod\lm_{j=1}^{n^{(2)}_v}(\phi^{(2)}_{v,j}-\phi^{(1)}_{v,i})} \;,
\end{split}
\end{equation}
where shuffles are taken inside each group $\{\phi^{(1)}_{v,i} \}_{i={1,\dots, n^{(1)}_v}}\cup \{\phi^{(2)}_{v,j} \}_{j={1,\dots, n^{(2)}_v}}$ for each $v\in\CV$.

In \cite{Galakhov:2018lta} a proposition was made to identify the CoHA with an algebra of effective wave-functions of scattering states on the Coulomb branch \cite{Manschot:2010qz, Manschot:2013sya}, for a specific class of quivers without loops. A localization to the Coulomb branch sets expectation values of chiral fields to zero. Expectation values of the vector fields are undefined at the zeroth loop order. At the first loop order the effective wave-function describes dynamics of particles on
$$
\bigotimes\lm_{v\in \CV}((\IR\otimes \IC)^{n_v}/\mathfrak{S}_{n_v}) \;,
$$
where $\IR$ components correspond to eigenvalues of $X^3_v$ and $\IC$ components to eigenvalues of $\Phi_v=X^1_v+ i X^2_v$, and the action of the permutation group is an action of gauge Weyl subgroup remaining unbroken in the RG flow. Further one could consider the effective wave-functions solely as holomorphic functions of eigenvalues $\phi_{v,i}$, $i=1,\ldots,n_v$ of $\Phi_v$ subjected to $\prod\lm_{v\in\CV}\mathfrak{S}_{n_v}$ symmetry. 

We would like to draw reader's attention to the following observation. The wave-functions in both Higgs and Coulomb branch localizations incorporate holomorphic functions of eigenvalues of vector multiplet complex field $\Phi_v$. Moreover, it is rather  suggestive to treat $\Phi_v$ eigenvalues $\phi_{v,i}\in\IC$ as coordinates of $\sum_{v\in\CV}n_v$ particles in $\IC$ in the Higgs branch localization scheme as well. Indeed, in this case expectation values $\langle \phi_{v,i}\rangle$ are localized in positions corresponding to 2d projections of atom positions in the 3d crystal  as we stated in Section \ref{s:Hecke}. Holomorphicity implies that these ``atoms" localized in the $\Phi$-plane acquire properties of \emph{anyons}, and that the algebraic structure we derived is due to anyon permutations. Let us in the rest of this section investigate such a \emph{hypothesis} that the quiver SQM flows to a theory of anyons in the $\Phi$-plane and the BPS algebra follows from the anyonic factors. In principle, the Coulomb branch localization could have justified this conjecture, however the very Coulomb localization for more complicated quivers containing loops reveals its own complications such as the appearance of scaling states \cite{Bena:2012hf, Manschot:2013sya, Denef:2007vg}, so we will not follow this route and choose a different path.

An argument for such a simple identification of contributions of the Coulomb branch and those of the Higgs branch is the following. The anyonic properties of quiver SQM wave-functions are ``sealed" by a non-perturbative operator (see \eqref{angular})
$$
\CJ_+= J_3+R
$$
commuting with both $Q_{1}$ and $\bar Q_{\dot 1}$. Here $J_3$ is the angular momentum operator  corresponding to  $\SO(3)$ rotations of vector-multiplet scalars $X^i$, and $R$ is an R-symmetry generator of the $\CN=4$ 1d SUSY algebra (see Appendix \ref{s:SQM}).  On the Coulomb branch this operator rotates
the $\Phi$-plane, while on the Higgs branch this operator becomes a half form degree operator \cite[Section 4.3]{Denef:2002ru}. The expectation values $\langle\phi\rangle$ are linear functions in equivariant weights $\epsilon_a$ and have therefore homological degree 2 and eigenvalue of $\CJ_+$ is equal to 1. This implies that the Coulomb branch wave-function given by a holomorphic monomial $\phi^k$ and the Higgs branch wave-function given by a Euler character containing $\langle\phi\rangle^k$ will have the same $\CJ_+$ eigenvalues.

Assume that under a permutation a two-atom wave-function, which denote by $\pi$, acquires an anyonic factor:
\be\label{anyon}
\pi(\Box_1,\Box_2)=\varphi_{12}\times \pi(\Box_2,\Box_1) \;.
\ee

It is easy to guess what this factor should look like from the form of expression \eqref{COHA_mul}. The extra factor in the product appearing as a summand is due to effective electro-magnetic degrees of freedom for dyonic quasi-particles on the Coulomb branch, the same factor defines anyonic shift in \eqref{anyon}. Therefore we derive:
$$
\varphi_{12}\sim (-1)^{\delta_{\hat \Box_1,\hat \Box_2}}\frac{(\phi_2-\phi_1)^{|a:\;\hat\Box_1\to \hat \Box_2|}}{(\phi_1-\phi_2)^{|a:\;\hat\Box_2\to \hat \Box_1|}} \;.
$$
The factors in the right hand side are one-loop contributions of chiral field equivariant weights in SQM, therefore shifts by the $\Omega$-background parameters should be added. One should also substitute free parameters $\phi_{1,2}$ for fixed positions in the crystal lattice $\bphi_{\Box}$. Hence we arrive to the following statistical relations:
\bea\label{bf}
\begin{split}
&\pi(\Box_1,\Box_2)=\varphi_{\hat\Box_1,\hat\Box_2}\left(\bphi_{\Box_1}-\bphi_{\Box_2}\right)\times\pi(\Box_2,\Box_1) \;,\\
&\varphi_{v,w}(z)=(-1)^{\delta_{v,w}+|a:\; v\to w|}\frac{\prod\lm_{(a\colon v\to w)\in\CA}\left(z+\epsilon_a\right)}{\prod\lm_{(b:\;w\to v)\in\CA}\left(z-\epsilon_b\right)} \;.
\end{split}
\eea

Now let us consider a crystal $\Lambda$ projected to the $\Phi$-plane and a new atom brought from infinity or taken away to infinity: 
$$
\begin{array}{ccc }
\begin{array}{c}
\begin{tikzpicture}
\begin{scope}[scale=0.5]
\draw[->,dashed] (-7,0) to (-3,0);
\draw[fill=black] (-7,0) circle (0.1);
\begin{scope}
\draw[fill=black] (1,0) circle (0.1) (0.5, 0.866025) circle (0.1) (-0.5, 0.866025) circle (0.1) (-1,0) circle (0.1) (-0.5, -0.866025) circle (0.1) (0.5, -0.866025) circle (0.1);
\draw[thick]  (1,0) -- (0.5, 0.866025) -- (-0.5, 0.866025) -- (-1,0) -- (-0.5, -0.866025) -- (0.5, -0.866025) -- cycle;
\end{scope}
\begin{scope}[shift={(1.5, 0.866025)}]
\draw[fill=black] (1,0) circle (0.1) (0.5, 0.866025) circle (0.1) (-0.5, 0.866025) circle (0.1) (-1,0) circle (0.1) (-0.5, -0.866025) circle (0.1) (0.5, -0.866025) circle (0.1);
\draw[thick]  (1,0) -- (0.5, 0.866025) -- (-0.5, 0.866025) -- (-1,0) -- (-0.5, -0.866025) -- (0.5, -0.866025) -- cycle;
\end{scope}
\begin{scope}[shift={(1.5, -0.866025)}]
\draw[fill=black] (1,0) circle (0.1) (0.5, 0.866025) circle (0.1) (-0.5, 0.866025) circle (0.1) (-1,0) circle (0.1) (-0.5, -0.866025) circle (0.1) (0.5, -0.866025) circle (0.1);
\draw[thick]  (1,0) -- (0.5, 0.866025) -- (-0.5, 0.866025) -- (-1,0) -- (-0.5, -0.866025) -- (0.5, -0.866025) -- cycle;
\end{scope}
\begin{scope}[shift={(-1.5, 0.866025)}]
\draw[fill=black] (1,0) circle (0.1) (0.5, 0.866025) circle (0.1) (-0.5, 0.866025) circle (0.1) (-1,0) circle (0.1) (-0.5, -0.866025) circle (0.1) (0.5, -0.866025) circle (0.1);
\draw[thick]  (1,0) -- (0.5, 0.866025) -- (-0.5, 0.866025) -- (-1,0) -- (-0.5, -0.866025) -- (0.5, -0.866025) -- cycle;
\end{scope}
\begin{scope}[shift={(-1.5, -0.866025)}]
\draw[fill=black] (1,0) circle (0.1) (0.5, 0.866025) circle (0.1) (-0.5, 0.866025) circle (0.1) (-1,0) circle (0.1) (-0.5, -0.866025) circle (0.1) (0.5, -0.866025) circle (0.1);
\draw[thick]  (1,0) -- (0.5, 0.866025) -- (-0.5, 0.866025) -- (-1,0) -- (-0.5, -0.866025) -- (0.5, -0.866025) -- cycle;
\end{scope}
\end{scope}
\end{tikzpicture} 
\end{array}
& 
\quad \textrm{or} \quad
&
\begin{array}{c}
\begin{tikzpicture}
\begin{scope}[scale=0.5]
\draw[->,dashed] (3,0) to (7,0);
\draw[fill=black] (3,0) circle (0.1);
\begin{scope}
\draw[fill=black] (1,0) circle (0.1) (0.5, 0.866025) circle (0.1) (-0.5, 0.866025) circle (0.1) (-1,0) circle (0.1) (-0.5, -0.866025) circle (0.1) (0.5, -0.866025) circle (0.1);
\draw[thick]  (1,0) -- (0.5, 0.866025) -- (-0.5, 0.866025) -- (-1,0) -- (-0.5, -0.866025) -- (0.5, -0.866025) -- cycle;
\end{scope}
\begin{scope}[shift={(1.5, 0.866025)}]
\draw[fill=black] (1,0) circle (0.1) (0.5, 0.866025) circle (0.1) (-0.5, 0.866025) circle (0.1) (-1,0) circle (0.1) (-0.5, -0.866025) circle (0.1) (0.5, -0.866025) circle (0.1);
\draw[thick]  (1,0) -- (0.5, 0.866025) -- (-0.5, 0.866025) -- (-1,0) -- (-0.5, -0.866025) -- (0.5, -0.866025) -- cycle;
\end{scope}
\begin{scope}[shift={(1.5, -0.866025)}]
\draw[fill=black] (1,0) circle (0.1) (0.5, 0.866025) circle (0.1) (-0.5, 0.866025) circle (0.1) (-1,0) circle (0.1) (-0.5, -0.866025) circle (0.1) (0.5, -0.866025) circle (0.1);
\draw[thick]  (1,0) -- (0.5, 0.866025) -- (-0.5, 0.866025) -- (-1,0) -- (-0.5, -0.866025) -- (0.5, -0.866025) -- cycle;
\end{scope}
\begin{scope}[shift={(-1.5, 0.866025)}]
\draw[fill=black] (1,0) circle (0.1) (0.5, 0.866025) circle (0.1) (-0.5, 0.866025) circle (0.1) (-1,0) circle (0.1) (-0.5, -0.866025) circle (0.1) (0.5, -0.866025) circle (0.1);
\draw[thick]  (1,0) -- (0.5, 0.866025) -- (-0.5, 0.866025) -- (-1,0) -- (-0.5, -0.866025) -- (0.5, -0.866025) -- cycle;
\end{scope}
\begin{scope}[shift={(-1.5, -0.866025)}]
\draw[fill=black] (1,0) circle (0.1) (0.5, 0.866025) circle (0.1) (-0.5, 0.866025) circle (0.1) (-1,0) circle (0.1) (-0.5, -0.866025) circle (0.1) (0.5, -0.866025) circle (0.1);
\draw[thick]  (1,0) -- (0.5, 0.866025) -- (-0.5, 0.866025) -- (-1,0) -- (-0.5, -0.866025) -- (0.5, -0.866025) -- cycle;
\end{scope}
\end{scope}
\end{tikzpicture}
\end{array}
\end{array}
\;.
$$
These operations act on the crystal wave-function adding corresponding statistical factors, so we could search for the expression for $e$ and $f$ matrix elements as products of such factors from a moved atom and the rest of the crystal:
\be\label{EF_suggestion}
E(\Lambda\to\Lambda+\Box)=\prod\lm_{\Box'\in\Lambda}\pi_+(\Box,\Box') \;,
\quad 
F(\Lambda+\Box\to\Lambda)=\prod\lm_{\Box'\in\Lambda}\pi_-(\Box,\Box') \;,
\ee
where $\pi_{\pm}$ satisfy relations \eqref{anyon}.
Then for a ratio of matrix elements we have:
\bea
\begin{split}
\frac{E(\Lambda\to\Lambda+\Box_1)E(\Lambda+\Box_1\to\Lambda+\Box_1+\Box_2)}{E(\Lambda\to\Lambda+\Box_2)E(\Lambda+\Box_2\to\Lambda+\Box_1+\Box_2)}=\\
=\frac{\prod\lm_{\Box'\in \Lambda}\pi_+(\Box_1,\Box')\times \prod\lm_{\Box'\in \Lambda+\Box_1}\pi_+(\Box_2,\Box')}{\prod\lm_{\Box'\in \Lambda}\pi_+(\Box_2,\Box')\times \prod\lm_{\Box'\in \Lambda+\Box_2}\pi_+(\Box_1,\Box')}=\\
=\frac{\pi_+(\Box_2,\Box_1)}{\pi_+(\Box_1,\Box_2)}=\left[\varphi_{\hat\Box_1,\hat\Box_2}\left(\bphi_{\Box_1}-\bphi_{\Box_2}\right)\right]^{-1} \;.
\end{split}
\eea

This relation coincides with one of the relations from \eqref{EF0}. We  can derive a similar expression for $f$-generators.

Consider further the following situation. Suppose we have two atoms 1 and 2. First we consider atom $2$ presented in the crystal, then we bring in atom 1 from infinity to position $1'$. Afterwords we take atom $2$ and bring it to position $2'$ at infinity:
$$
\begin{array}{c}
\begin{tikzpicture}
\draw[fill=black] (-2,0) circle (0.05) (0,0) circle (0.05) (0,-0.5) circle (0.05) (2,-0.5) circle (0.05);
\draw[->,dashed] (-2,0) -- (0,0);
\draw[->,dashed] (0,-0.5) -- (2,-0.5);
\node[left] at (-2,0) {$1$};
\node[right] at (2,-0.5) {$2'$};
\node[above] at (0,0) {$1'$};
\node[below] at (0,-0.5) {$2$};
\end{tikzpicture}
\end{array}
$$
The resulting position is the same therefore we conclude:
\be
\pi_+(\Box_1,\Box_2)\pi_-(\Box_2,\Box_1)=1 \;.
\ee
Using this relation we could calculate other relations between matrix elements. For example,
\bea
\begin{split}
	&\frac{E(\Lambda+\Box_1\to \Lambda+\Box_1+\Box_2)F(\Lambda+\Box_1+\Box_2\to\Lambda+\Box_2)}{F(\Lambda+\Box_1\to\Lambda)E(\Lambda\to \Lambda+\Box_2)}=\\
	&\qquad =\frac{\prod\lm_{\Box'\in\Lambda+\Box_1}\pi_+(\Box_2,\Box')\prod\lm_{\Box'\in\Lambda+\Box_2}\pi_-(\Box_1,\Box')}{\prod\lm_{\Box'\in\Lambda}\pi_-(\Box_1,\Box')\prod\lm_{\Box'\in\Lambda}\pi_+(\Box_2,\Box')}=\pi_+(\Box_2,\Box_1)\pi_-(\Box_1,\Box_2)=1 \;,
\end{split}
\eea
and for coincident atoms we have:
\bea
\begin{split}
E(\Lambda\to\Lambda+\Box)F(\Lambda+\Box\to\Lambda)=\prod\lm_{\Box'\in\Lambda}(\pi_+(\Box',\Box)\pi_-(\Box',\Box))=\\
=\prod\lm_{\Box'\in\Lambda}\varphi_{\hat\Box,\hat\Box'}\left(\bphi_{\Box}-\bphi_{\Box'}\right) \;.
\end{split}
\eea
The last expression is somewhat problematic. Indeed the function $\varphi$ \eqref{bf} will contribute to it with poles for some near corner atoms. In addition it does not take into account the fact that a new atom can contribute with an extra potential due to interaction with an empty crystal. The correct resolution of this expression is 
\be
E(\Lambda\to\Lambda+\Box)F(\Lambda+\Box\to\Lambda)=\mathop{\rm res}\lm_{z=\bphi_{\Box}}\psi_{\Lambda}(z) \;,
\ee
where $\psi_{\Lambda}(z)$ is given by \eqref{psihat}. One can lift these matrix elements to the definition of generator series $e$, $f$ and $\psi$. The resulting OPE relations coincide with \eqref{true_OPE}.

Let us push this conjectural ansatz a bit further. Consider a \emph{tensor product} of BPS algebra moduli. For this construction it is enough to consider a framing node of dimension two. To the frozen scalar field associated with the framing node one assigns a diagonal value:
$$
\left(\begin{array}{cc}
m_{\IC,1}&0\\
0& m_{\IC,2}\\
\end{array} \right) \;.
$$
Parameters $m_{\IC,i}$ play the role of complex flavor masses associated to chiral fields of multiplet $I$. Critical points as in the case of 4d instantons (see e.g. \cite{Nekrasov:2003rj}) are labeled by pairs of crystals whose projections are growing not from $0\in\IC$, rather from two points $m_{\IC,1}$ and $m_{\IC,2}$:
$$
\begin{array}{c}
\begin{tikzpicture}
\begin{scope}[scale=0.5]
\begin{scope}
\draw[fill=black] (1,0) circle (0.1) (0.5, 0.866025) circle (0.1) (-0.5, 0.866025) circle (0.1) (-1,0) circle (0.1) (-0.5, -0.866025) circle (0.1) (0.5, -0.866025) circle (0.1);
\draw[thick]  (1,0) -- (0.5, 0.866025) -- (-0.5, 0.866025) -- (-1,0) -- (-0.5, -0.866025) -- (0.5, -0.866025) -- cycle;
\end{scope}
\begin{scope}[shift={(1.5, 0.866025)}]
\draw[fill=black] (1,0) circle (0.1) (0.5, 0.866025) circle (0.1) (-0.5, 0.866025) circle (0.1) (-1,0) circle (0.1) (-0.5, -0.866025) circle (0.1) (0.5, -0.866025) circle (0.1);
\draw[thick]  (1,0) -- (0.5, 0.866025) -- (-0.5, 0.866025) -- (-1,0) -- (-0.5, -0.866025) -- (0.5, -0.866025) -- cycle;
\end{scope}
\begin{scope}[shift={(1.5, -0.866025)}]
\draw[fill=black] (1,0) circle (0.1) (0.5, 0.866025) circle (0.1) (-0.5, 0.866025) circle (0.1) (-1,0) circle (0.1) (-0.5, -0.866025) circle (0.1) (0.5, -0.866025) circle (0.1);
\draw[thick]  (1,0) -- (0.5, 0.866025) -- (-0.5, 0.866025) -- (-1,0) -- (-0.5, -0.866025) -- (0.5, -0.866025) -- cycle;
\end{scope}
\begin{scope}[shift={(-1.5, 0.866025)}]
\draw[fill=black] (1,0) circle (0.1) (0.5, 0.866025) circle (0.1) (-0.5, 0.866025) circle (0.1) (-1,0) circle (0.1) (-0.5, -0.866025) circle (0.1) (0.5, -0.866025) circle (0.1);
\draw[thick]  (1,0) -- (0.5, 0.866025) -- (-0.5, 0.866025) -- (-1,0) -- (-0.5, -0.866025) -- (0.5, -0.866025) -- cycle;
\end{scope}
\begin{scope}[shift={(-1.5, -0.866025)}]
\draw[fill=black] (1,0) circle (0.1) (0.5, 0.866025) circle (0.1) (-0.5, 0.866025) circle (0.1) (-1,0) circle (0.1) (-0.5, -0.866025) circle (0.1) (0.5, -0.866025) circle (0.1);
\draw[thick]  (1,0) -- (0.5, 0.866025) -- (-0.5, 0.866025) -- (-1,0) -- (-0.5, -0.866025) -- (0.5, -0.866025) -- cycle;
\end{scope}
\node {$m_{\IC,1}$};
\begin{scope}[shift={(8,0)}]
\begin{scope}
\draw[fill=black] (1,0) circle (0.1) (0.5, 0.866025) circle (0.1) (-0.5, 0.866025) circle (0.1) (-1,0) circle (0.1) (-0.5, -0.866025) circle (0.1) (0.5, -0.866025) circle (0.1);
\draw[thick]  (1,0) -- (0.5, 0.866025) -- (-0.5, 0.866025) -- (-1,0) -- (-0.5, -0.866025) -- (0.5, -0.866025) -- cycle;
\end{scope}
\begin{scope}[shift={(1.5, 0.866025)}]
\draw[fill=black] (1,0) circle (0.1) (0.5, 0.866025) circle (0.1) (-0.5, 0.866025) circle (0.1) (-1,0) circle (0.1) (-0.5, -0.866025) circle (0.1) (0.5, -0.866025) circle (0.1);
\draw[thick]  (1,0) -- (0.5, 0.866025) -- (-0.5, 0.866025) -- (-1,0) -- (-0.5, -0.866025) -- (0.5, -0.866025) -- cycle;
\end{scope}
\begin{scope}[shift={(1.5, -0.866025)}]
\draw[fill=black] (1,0) circle (0.1) (0.5, 0.866025) circle (0.1) (-0.5, 0.866025) circle (0.1) (-1,0) circle (0.1) (-0.5, -0.866025) circle (0.1) (0.5, -0.866025) circle (0.1);
\draw[thick]  (1,0) -- (0.5, 0.866025) -- (-0.5, 0.866025) -- (-1,0) -- (-0.5, -0.866025) -- (0.5, -0.866025) -- cycle;
\end{scope}
\begin{scope}[shift={(-1.5, 0.866025)}]
\draw[fill=black] (1,0) circle (0.1) (0.5, 0.866025) circle (0.1) (-0.5, 0.866025) circle (0.1) (-1,0) circle (0.1) (-0.5, -0.866025) circle (0.1) (0.5, -0.866025) circle (0.1);
\draw[thick]  (1,0) -- (0.5, 0.866025) -- (-0.5, 0.866025) -- (-1,0) -- (-0.5, -0.866025) -- (0.5, -0.866025) -- cycle;
\end{scope}
\begin{scope}[shift={(-1.5, -0.866025)}]
\draw[fill=black] (1,0) circle (0.1) (0.5, 0.866025) circle (0.1) (-0.5, 0.866025) circle (0.1) (-1,0) circle (0.1) (-0.5, -0.866025) circle (0.1) (0.5, -0.866025) circle (0.1);
\draw[thick]  (1,0) -- (0.5, 0.866025) -- (-0.5, 0.866025) -- (-1,0) -- (-0.5, -0.866025) -- (0.5, -0.866025) -- cycle;
\end{scope}
\node {$m_{\IC,2}$};
\end{scope}
\draw[->,dashed] (-6,0) to (-3,0);
\draw[->, dashed] (-6,0) to[out=0,in=180] (-2,2) -- (2,2) to[out=0,in=180] (5,0);
\draw[->,dashed] (11,0) to (14,0);
\draw[->, dashed] (3,0) to[out=0,in=180] (6,-2) -- (10,-2) to[out=0,in=180] (14,0);
\draw[fill=black] (-6,0) circle (0.1) (14,0) circle (0.1);
\end{scope}
\end{tikzpicture}
\end{array}
$$

A suggested form for the matrix coefficients \eqref{EF_suggestion} implies that in this contribution atoms belonging to a crystal do not interfere with each other, therefore the expression is just a product of single interactions of a new atom with each atom in the crystal. Now it is natural to guess that in a situation when we have two grown crystals $\Lambda_1$ and $\Lambda_2$ in the $\Phi$-plane (as in the case depicted above) there will be no mutual interference between atoms of $\Lambda_1$ and $\Lambda_2$. Therefore we can expect  that the matrix coefficient in this case is just a product of statistical factors from interaction of a new atom with $\Lambda_1$ and $\Lambda_2$:
\bea\label{ef_tensor}
\begin{split}
	\hatE|\Lambda_1,\Lambda_2\rangle=\sum\lm_{\Box\in \Lambda_1^+}\left(\prod\lm_{\Box'\in\Lambda_1}\pi_+(\Box,\Box')\right)\left(\prod\lm_{\Box'\in\Lambda_2}\pi_+(\Box,\Box')\right)|\Lambda_1+\Box,\Lambda_2\rangle+\\
	+\sum\lm_{\Box\in \Lambda_2^+}\left(\prod\lm_{\Box'\in\Lambda_1}\pi_+(\Box,\Box')\right)\left(\prod\lm_{\Box'\in\Lambda_2}\pi_+(\Box,\Box')\right)|\Lambda_1,\Lambda_2+\Box\rangle\;,\\
	\hatF|\Lambda_1,\Lambda_2\rangle=\sum\lm_{\Box\in \Lambda_1^-}\left(\prod\lm_{\Box'\in\Lambda_1}\pi_-(\Box,\Box')\right)\left(\prod\lm_{\Box'\in\Lambda_2}\pi_-(\Box,\Box')\right)|\Lambda_1-\Box,\Lambda_2\rangle+\\
	+\sum\lm_{\Box\in \Lambda_2^-}\left(\prod\lm_{\Box'\in\Lambda_1}\pi_-(\Box,\Box')\right)\left(\prod\lm_{\Box'\in\Lambda_2}\pi_-(\Box,\Box')\right)|\Lambda_1,\Lambda_2-\Box\rangle\;.
\end{split}
\eea
Here by $|\Lambda_1,\Lambda_2\rangle$ we mean a wave-function obtained by multiplying wave-functions of crystals $\Lambda_1$ and $\Lambda_2$. The norm of this state is thus simply a product of norms of $\Lambda_1$ and $\Lambda_2$. It would be natural to incorporate a mutual statistical factor between $\Lambda_1$ and $\Lambda_2$ into this norm. Let us assume that we grow the crystal $\Lambda_1$ first this will produce a state $|\Lambda_1,\emptyset\rangle$. We next bring in a new crystal $\Lambda_2$ in $\Phi$-plane, which will produce a state $|\Lambda_1,\Lambda_2\rangle$ with a multiplier given by a mutual statistical factor. We call the resulting state a tensor product of states:
\be
|\Lambda_1\rangle\otimes |\Lambda_2\rangle:=\left(\prod\lm_{\substack{\Box_1\in\Lambda_1,\\\Box_2\in\Lambda_2}}\pi_+(\Box_2,\Box_1)\right)|\Lambda_1,\Lambda_2\rangle\;.
\ee
Using this definition it is simple to rewrite \eqref{ef_tensor} in the new basis. In this way we define a co-product structure on the BPS algebra:
\bea\label{pre_co-mul}
\begin{split}
\Delta(\hatE)\;|\Lambda_1\rangle\otimes |\Lambda_2\rangle=\sum\lm_{\Box\in\Lambda_2^+}E(\Lambda_2\to \Lambda_2+\Box)\;|\Lambda_1\rangle\otimes |\Lambda_2+\Box\rangle+\quad\quad\quad\quad\\
+\sum\lm_{\Box\in\Lambda_1^+} E(\Lambda_1\to \Lambda_1+\Box)\, \psi^{(\hat\Box)}_{\Lambda_2}\left(m_{\IC,1}-m_{\IC,2}\right)\;|\Lambda_1+\Box\rangle\otimes |\Lambda_2\rangle\;,\\
\Delta(\hatF)\;|\Lambda_1\rangle\otimes |\Lambda_2\rangle=\sum\lm_{\Box\in\Lambda_1^-}F(\Lambda_1\to \Lambda_1-\Box)\;|\Lambda_1-\Box\rangle\otimes |\Lambda_2\rangle+\quad\quad\quad\quad\\
+\sum\lm_{\Box\in\Lambda_2^-} E(\Lambda_2\to \Lambda_2-\Box)\, \psi^{(\hat\Box)}_{\Lambda_1}\left(m_{\IC,2}-m_{\IC,1}\right)\;|\Lambda_1\rangle\otimes |\Lambda_2-\Box\rangle\;.
\end{split}
\eea

Obviously, we could have chosen crystal $\Lambda_2$ as a primary crystal, and add $\Lambda_1$ as a secondary one. If this alternative choice is made resulting co-product $\tilde \Delta$ will be \emph{different} from $\Delta$ with roles of tensor factors interchanged.

The structure \eqref{pre_co-mul} on module tensor products can be lifted to the co-product structure on the generator series in a \emph{naive} way:
\bea\label{co-mul_0}
\begin{split}
\Delta_0[e^{(v)}(z)]&=e^{(v)}(z)\otimes \psi^{(v)}(z)+1\otimes e^{(v)}(z) \;,\\
\Delta_0[f^{(v)}(z)]&=f^{(v)}(z)\otimes 1+\psi^{(v)}(z)\otimes f^{(v)}(z) \;, \\
\Delta_0[\psi^{(v)}(z)]&=\psi^{(v)}(z)\otimes \psi^{(v)}(z) \;.
\end{split}
\eea
This co-product structure seems to coincide with a reduction of the co-multiplication from the quantum toroidal superalgebras $\mathfrak{gl}_{m|n}$ \cite{Bezerra:2019dmp}. However, a more thorough analysis of \cite[Section 2.5]{Prochazka:2015deb} shows that co-multiplication $\Delta_0$ is incompatible with a Yangian structure and should be corrected, and a closed form of this correction is unknown. One notices that the right hand side of \eqref{co-mul_0} acquires extra pole contributions compared to the left hand side. We will leave a problem of lifting co-product \eqref{pre_co-mul} to generator generating series and its relation to the co-product in Yangians for further investigation elsewhere.

Having two co-multiplication structures $\Delta$ and $\tilde\Delta$ one defines the $R$-matrix as an intertwining operator mapping between these two co-product structures:
\be
\Delta\circ R=R\circ \tilde \Delta \;.
\ee

It is natural to expect that the $R$-matrix in the case of Calabi-Yau three-fold quivers will represent a generalization of four-dimensional instanton $R$-matrix \cite{MO,Smirnov:2013hh,Awata:2016mxc} to six-dimensional instantons.

%%%%%%%%%%%%%%%%%%%%%%%%%%%%%%%%%%%%%%%%%%%%%%%%%%%%%%%%%%%%%%%%%%%%%%%%%%%%%%%%%%%%
\section{Examples of Affine Yangians as BPS Algebras}\label{s:Examples}
%%%%%%%%%%%%%%%%%%%%%%%%%%%%%%%%%%%%%%%%%%%%%%%%%%%%%%%%%%%%%%%%%%%%%%%%%%%%%%%%%%%%

In this section we explicitly work out many examples of the toric Calabi-Yau geometries.
In each of the examples we find that
the algebra relations of the quiver Yangian \eqref{q_OPE}
are satisfied in our representations.

%%%%%%%---------------------------------------------------------------------------------------------------------------------------------------------------------------------------------------------
\subsection{\texorpdfstring{$\IC^2$ Appetizer---$Y\left(\widehat{\fg\fl}_1\right)$  Fock Modules}{A C2 appetizer ---Y(gl1)  Fock Modules}} 
\label{s:gl2}

As a basic example of the proposed construction we start with the canonical example of the ADHM quiver for Hilbert schemes on  $\IC^2$ \cite{Nakajima_book}. In this case ADHM quivers describe moduli spaces of instantons in four-dimensional $\U(1)$ Yang-Mills theories on a non-commutative space-time. To describe these theories in the context of Calabi-Yau three-folds we have to consider specific deformation of branes on $\IC^3$ that are allowed to move only inside a $\IC^2$-hyperplane.

The geometry $\IC^3$ is described by a trefoil quiver with a superpotential \eqref{trefoil_early}. 
To confine branes in $\IC^2$-hyperplane spanned by $B_1$ and $B_2$ we modify the quiver adding a framing node with dimension 1 and the superpotential:
\be
\begin{array}{c}
	\begin{tikzpicture}
	\draw[->] (0.173,-0.1) to[out=330,in=270] (1,0) to[out=90,in=30] (0.173,0.1);
	\node[right] at (1,0) {$B_1$};
	\begin{scope}[rotate=90]
	\draw[->] (0.173,-0.1) to[out=330,in=270] (1,0) to[out=90,in=30] (0.173,0.1);
	\node[above] at (1,0) {$B_2$};
	\end{scope}
	\begin{scope}[rotate=180]
	\draw[->] (0.173,-0.1) to[out=330,in=270] (1,0) to[out=90,in=30] (0.173,0.1);
	\node[left] at (1,0) {$B_3$};
	\end{scope}
	\draw (0,0) circle (0.2);
	\draw[->] (-0.05,-0.194) -- (-0.05,-1);
	\draw[->] (0.05,-1) --  (0.05,-0.194);
	\draw (-0.1,-1) -- (0.1,-1) -- (0.1,-1.2) -- (-0.1,-1.2) -- (-0.1,-1);
	\node[right] at (0.05,-0.6) {$I$};
	\node[left] at (-0.05,-0.6) {$J$};
	\end{tikzpicture} 
\end{array},\quad W=\Tr\left\{\left(\left[B_1,B_2\right]+IJ\right)B_3\right\} \;.
\ee

In this setup the field $B_3$ plays the role of the Lagrange multiplier. We expect it will have a vacuum expectation value $\langle B_3\rangle=0$. The weights of $\Omega$-background we assign to fields $B_i$ are $\epsilon_i$. To the fields $I$ and $J$ we assign $\Omega$-background weights $a$ and $b$ correspondingly. The gauge invariance of the superpotential implies the following relation between weights:
$$
\epsilon_3=-\epsilon_1-\epsilon_2 \;,\quad b=\epsilon_1+\epsilon_2-a \;.
$$
The vacuum expectation values of remaining fields are defined by the following set of equations:
\be\label{ADHM}
\left\{
\begin{array}{c}
	\mu_{\IR}=\left[B_1,B_1^{\dagger}\right]+\left[B_2,B_2^{\dagger}\right]+II^{\dagger}-J^{\dagger}J=\theta \;,\\
	\p_{B_3}W=\mu_{\IC}=\left[B_1,B_2\right]+IJ=0 \;,\\
	V(B_1)\sim\left[\Phi,B_1\right]-\epsilon_1B_1=0 \;,\quad V(B_2)\sim\left[\Phi,B_2\right]-\epsilon_2B_2=0 \;,\\
	V(I)\sim \Phi I-a I=0 \;,\quad V(J)\sim J\Phi-(\epsilon_1+\epsilon_2-a)J=0 \;.
\end{array}
\right\}
\ee
This is the canonical set of ADHM equations \cite{Nekrasov:2003rj}.

Currently rather than 3d crystals we have 2d crystals labeling the SQM vacua. These crystals can be enumerated by partitions, where a partition of $n$ is defined as a sequence:
$$
\lambda=\lambda_1\geq\lambda_2\geq\lambda_3\geq\ldots\geq \lambda_k\geq 0  \;,\quad \sum\lm_{i=1}^k\lambda_i=n \;.
$$

A crystal $\Lambda$ corresponds to a Young diagram of a given partition. For example, a particular partition of $8$ reads:
$$
\{4,3,1\}\;\leftrightarrow\;\begin{array}{c}
\begin{tikzpicture}
\node at (0,0) {$(0,0)$};
\draw (-0.5,-0.5) -- (0.5,-0.5) -- (0.5,0.5) -- (-0.5,0.5) -- cycle;
\begin{scope}[shift={(1,0)}]
\node at (0,0) {$(1,0)$};
\draw (-0.5,-0.5) -- (0.5,-0.5) -- (0.5,0.5) -- (-0.5,0.5) -- cycle;
\end{scope}
\begin{scope}[shift={(2,0)}]
\node at (0,0) {$(2,0)$};
\draw (-0.5,-0.5) -- (0.5,-0.5) -- (0.5,0.5) -- (-0.5,0.5) -- cycle;
\end{scope}
\begin{scope}[shift={(3,0)}]
\node at (0,0) {$(3,0)$};
\draw (-0.5,-0.5) -- (0.5,-0.5) -- (0.5,0.5) -- (-0.5,0.5) -- cycle;
\end{scope}
\begin{scope}[shift={(0,1)}]
\node at (0,0) {$(0,1)$};
\draw (-0.5,-0.5) -- (0.5,-0.5) -- (0.5,0.5) -- (-0.5,0.5) -- cycle;
\end{scope}
\begin{scope}[shift={(1,1)}]
\node at (0,0) {$(1,1)$};
\draw (-0.5,-0.5) -- (0.5,-0.5) -- (0.5,0.5) -- (-0.5,0.5) -- cycle;
\end{scope}
\begin{scope}[shift={(2,1)}]
\node at (0,0) {$(2,1)$};
\draw (-0.5,-0.5) -- (0.5,-0.5) -- (0.5,0.5) -- (-0.5,0.5) -- cycle;
\end{scope}
\begin{scope}[shift={(0,2)}]
\node at (0,0) {$(0,2)$};
\draw (-0.5,-0.5) -- (0.5,-0.5) -- (0.5,0.5) -- (-0.5,0.5) -- cycle;
\end{scope}
\end{tikzpicture}
\end{array}
$$

One assigns to Young diagram boxes integer coordinates $({\bf x},{\bf y})$ starting with $(0,0)$ from the left bottom corner. In this case a solution to ADHM equations \eqref{ADHM} for fields $\Phi$ has the following form. For weight function we have
$$
\bphi_{\Box}=a+\epsilon_1{\bf x}(\Box) +\epsilon_2 {\bf y}(\Box) \;.
$$

The complement $\Lambda^+$ defines an ideal in a ring of polynomials in two variables $t_1$ and $t_2$:
\be
\CI_{\Lambda}=\bigoplus\lm_{\Box\in\Lambda^+}\IC[t_1,t_2]\cdot t_1^{{\bf x}(\Box)}t_2^{{\bf y}(\Box)}  \;.
\ee
The quotient vector space
\be
V_{\Lambda}\coloneqq \IC[t_1,t_2]/\CI_{\Lambda}
\ee
has dimension $|\Lambda|$.

To construct a BPS wave-function following Section \ref{s:Higgs} one has to consider tangent directions at a fixed point in the configuration space. We will do this in a flat Euclidean space spanned by matrices: $X^3$, $\Phi$, $B_1$, $B_2$, $B_3$, $I$, $J$.

We can split tangent degrees of freedom in the following groups:
\begin{enumerate}
	\item $X^3$, $B_{1,2}$ and $I$ tangent to expectation values.
	\item $\Phi$ and $\GL(n,\IC)$ gauge freedom in $B_{1,2}$ and $I$.
	\item Those degrees of freedom that deviate from $dW=0$. This includes $B_3$ and $J$.
	\item The remaining degrees of freedom.
\end{enumerate}

We illustrate appearance of all these groups in a concrete example in Appendix \ref{s:exapmle}. We notice that groups 1 -- 3 have frequencies $\omega\sim|\theta|^{\frac{1}{2}}$ and contribute to $\Psi_{|\omega|>\Lambda}$. Therefore the only degrees of freedom contributing to the effective IR wave-function $\Psi_{|\omega|<\Lambda}$ are in group 4.

Summarizing, we find that the degrees of freedom contributing to $\Psi_{|\omega|<\Lambda}$ are those tangent $B_1$, $B_2$ and $I$ modulo the complexified gauge group $\GL(n,\IC)$ and such that $dW=0$ on these tangent directions. These degrees of freedom parameterize a gauge invariant manifold in IR limit, and in Section \ref{s:Higgs} we called them mesons $m_i$.

An example of detailed calculations for this setup is presented in Appendix \ref{s:ideals}.

The expression for the wave-function is nothing more than a Thom representative of the corresponding Euler class:
\be
{\rm Eul}_{\Lambda}=\prod\lm_iw_i \;,
\ee
where $w_i$ are effective equivariant weights of $m_i$.

Now we would like to construct an operator bringing in a new instanton impurity. 

As we have discussed in Section \ref{s:Hecke} a process of bringing in a new D0-brane induces a Hecke modification, and we switch from the quiver representation $\CR_n$ with $n$ instantons to a  new representation $\CR_{n+1}$. The information that the two D0 configurations differ by a single instanton is reflected in a fact that $\CR_n$ is a subrepresentation of $\CR_{n+1}$. Compare two representations corresponding to fixed points $\Lambda$ and $\Lambda+\Box$. A subrepresentation constraint becomes a constraint for ideals:
\be
\CI_{\Lambda+\Box}\subset \CI_{\Lambda} \;.
\ee

This constraint can be reduced to a purely algebraic constraint defining the so-called incidence locus \cite{Nakajima_book}:
\be\label{incidence}
\CI_{\Lambda+\Box}\; {\rm mod}\; \CI_{\Lambda}=0 \;.
\ee
Obviously, ideals $\CI_{\Lambda}$ and $\CI_{\Lambda+\Box}$ corresponding to fixed points when mesonic fields are turned off $m_i=0$ lie on this locus. Suppose that  tangent directions to $\CI_{\Lambda}$ are spanned by $m_i$, and tangent directions to $\CI_{\Lambda+\Box}$ are spanned by $m_i'$. Expanding the incidence locus constraint \eqref{incidence} up to the linear order  we derive a  hyperplane constraint for a tangent space to the locus \eqref{incidence} at the fixed point. The tangent vectors to this hyperplane are vectors in the union of the two spaces spanned by $\{m_i \}$ and $\{m_i' \}$, therefore they are also weighted by equivariant parameters. Let us denote these weights by $s_j$. Thus we are able to present a definition for the Euler character of the tangent space to the incidence locus in a fixed point:
\be
{\rm Eul}_{\Lambda,\Lambda+\Box}=\prod\lm_{j}s_j \;.
\ee

We give an example of explicit calculation in Appendix \ref{s:ideals}. 

Generic expressions for Euler characters and generator matrix elements in the fixed point basis are known in the literature (see \cite{Nakajima, FT}), those expressions we can also validate numerically.

An Euler character associated with a partition $\Lambda$ reads:
\be\label{Eul}
{\rm Eul}_{\Lambda}=\prod\lm_{\Box\in \lambda}\left(\epsilon_1\left({\rm leg}(\Box)+1\right)-\epsilon_2\;{\rm arm}(\Box)\right)\left(\epsilon_2\left({\rm arm}(\Box)+1\right)-\epsilon_1\;{\rm leg}(\Box)\right) \;,
\ee
where arm- and leg-functions for a box $\Box$ inside a partition are defined according to the following diagram:
$$
\begin{array}{c}
\begin{tikzpicture}
\begin{scope}
\draw[<->] (0,4) -- (0,0) -- (4,0);
\node[right] at (4,0) {$x$};
\node[above] at (0,4) {$y$};
\draw[thick] (0,3) -- (1,3) -- (1,2) -- (2,2) -- (2,1) -- (3,1) -- (3,0);
\draw[thick, fill=red] (0.4,0.4) -- (0.4,0.6) -- (0.6,0.6) -- (0.6,0.4) -- cycle;
\draw[thick, fill=black!20!green] (0.4,0.6) -- (0.4,3) -- (0.6,3) -- (0.6,0.6) -- cycle;
\draw[thick, fill=black!20!green] (0.6,0.4) -- (3,0.4) -- (3,0.6) -- (0.6,0.6) -- cycle;
\draw (0.4,0.6) to[out=180,in=0] (-0.2,1.8) to[out=0,in=180] (0.4,3);
\node[left] at (-0.2,1.8) {arm};
\draw (0.6,0.4) to[out=270,in=90] (1.8,-0.2) to[out=90,in=270] (3,0.4);
\node[below] at (1.8,-0.2) {leg};
\end{scope}
\end{tikzpicture}
\end{array}
\;.
$$

One defines the norm of vectors by
\be \label{2d_norm}
\langle\Lambda,\Lambda'\rangle=\delta_{\Lambda,\Lambda'}{\rm Eul}_{\Lambda} \;.
\ee

Let us denote by $\lambda+k$ a new partition acquired from partition $\lambda$ by adding a box to $(k,\lambda_k+1)$ position if it makes a partition again. Denote by $\CH_{(i,j)}(\lambda)$ all boxes with coordinates $(k<i,j)$ and by $\CV_{(i,j)}(\lambda)$ all boxes with $(i,k<j)$. In these terms the action of the raising generator \eqref{gen_e_f} reads (see \cite{FT})
\bea\label{e_z_Fock}
\begin{split}
	e(z)|\lambda\rangle&=\sum\lm_{k}\frac{{\rm Eul}_{\lambda+k}}{z-\epsilon_1\lambda_k-\epsilon_2(k-1)}\times\\
	&\times\prod\lm_{s\in \CV_{(k,\lambda_k+1)}(\lambda+k)}\frac{\epsilon_1({\rm leg}(s)-1)-\epsilon_2({\rm arm}(s)+1)}{\epsilon_1({\rm leg}(s))-\epsilon_2({\rm arm}(s)+1)}\times \\ 
	&\times\prod\lm_{s\in \CH_{(k,\lambda_k+1)}(\lambda+k)}\frac{\epsilon_1({\rm leg}(s)+1)-\epsilon_2({\rm arm}(s)-1)}{\epsilon_1({\rm leg}(s)+1)-\epsilon_2({\rm arm}(s))}|\lambda+k\rangle \;.
\end{split}
\eea
The action of the lowering generators $f(z)$ can be determined by its property of being conjugated to $e(z)$ by the vector norm \eqref{2d_norm}.

This construction allows one to establish an isomorphism between a basis of Fock module and Jack polynomials $J_{\lambda}^{(\beta)}$. In particular, the action of the generator $e_0 \coloneqq \lim\lm_{z\to\infty}z\; e(z)$ on the vectors of the Fock module is mimicked by multiplication by $p_1$ on the Jack polynomial side.

Jack polynomials give the canonical Fock representation of affine Yangian algebra $Y\left(\widehat{\fg\fl}_1\right)$. This is an associative algebra with three infinite sets of generators $e_i$, $f_i$ and $\psi_i$, for $i=0,1,2,\ldots$, satisfying the following set of relations:
\bea
\begin{split}
	0&=\left[\psi_i,\psi_k\right] \;,\\
	0&=\left[e_{j+3},e_k\right]-3\left[e_{j+2},e_{k+1}\right]+\left[e_{j+1},e_{k+2}\right]-\left[e_{j},e_{k+3}\right]+\\
	&\qquad\quad+\sigma_2\left[e_{j+1},e_{k}\right]
	-\sigma_2\left[e_{j},e_{k+1}\right]-\sigma_3\left\{e_j,e_k\right\} \;,\\
	0&=\left[f_{j+3},f_k\right]-3\left[f_{j+2},f_{k+1}\right]+\left[f_{j+1},f_{k+2}\right]-\left[f_{j},f_{k+3}\right]+\\
	&\qquad\quad+\sigma_2\left[f_{j+1},f_{k}\right]-\sigma_2\left[f_{j},f_{k+1}\right]+\sigma_3\left\{f_j,f_k\right\} \;,\\
	0&=\left[e_j,f_k\right]-\psi_{j+k} \;,\\
	0&=\left[\psi_{j+3},e_k\right]-3\left[\psi_{j+2},e_{k+1}\right]+\left[\psi_{j+1},e_{k+2}\right]-\left[\psi_{j},e_{k+3}\right]+\\
	&\qquad\quad+\sigma_2\left[\psi_{j+1},e_{k}\right]-\sigma_2\left[\psi_{j},e_{k+1}\right]-\sigma_3\left\{\psi_j,e_k\right\} \;,\\
	0&=\left[\psi_{j+3},f_k\right]-3\left[\psi_{j+2},f_{k+1}\right]+\left[\psi_{j+1},f_{k+2}\right]-\left[\psi_{j},f_{k+3}\right]+\\
	&\qquad\quad+\sigma_2\left[\psi_{j+1},f_{k}\right]-\sigma_2\left[\psi_{j},f_{k+1}\right]-\sigma_3\left\{\psi_j,f_k\right\}  \;.
\end{split}
\eea
Relations for $\psi$-generators with a small order number read:
\be
\begin{array}{lllll}
	\left[\psi_0,e_j\right]=0 \;,&\quad& \left[\psi_1,e_j\right]=0 \;,&\quad & \left[\psi_2,e_j\right]=2e_j \;.\\
	\left[\psi_0,f_j\right]=0  \;,&& \left[\psi_1,f_j\right]=0& \;,&\left[\psi_2,f_j\right]=-2f_j\;.\\
\end{array}
\ee
Generators furthermore satisfy cubic Serre relations:
\be\label{Y_gl_1_Serre}
{\rm Sym}_{(j_1,j_2,j_3)}\left[e_{j_1},\left[e_{j_2},e_{j_3}\right]\right]={\rm Sym}_{(j_1,j_2,j_3)}\left[f_{j_1},\left[f_{j_2},f_{j_3}\right]\right]=0 \;.
\ee

Three parameters $\epsilon_1$, $\epsilon_2$ and $\epsilon_3$ are gathered in the following combinations:
\be
\sigma_1=\epsilon_1+\epsilon_2+\epsilon_3=0 \;,\quad
\sigma_2=\epsilon_1\epsilon_2+\epsilon_2\epsilon_3+\epsilon_3\epsilon_1 \;,\quad \sigma_3=\epsilon_1\epsilon_2\epsilon_3 \;.
\ee

These generators are combined in the following series:
\be
e(z)=\sum\lm_{j=0}^{\infty} \frac{e_j}{z^{j+1}} \;,\quad f(z)=\sum\lm_{j=0}^{\infty} \frac{f_j}{z^{j+1}}
\;, \quad \psi(z)=1+\sigma_3\sum\lm_{j=0}^{\infty} \frac{\psi_j}{z^{j+1}} \;.
\ee

In terms of generating functions algebraic relations take a simpler form:
\begin{equation}\label{Y_gl_1}
\begin{split}
	e(z)e(w)&\sim\varphi(z-w)e(w)e(z)\;,\\
	f(z)f(w)&\sim\varphi(u-v)f(w)f(z) \;,\\
	\psi(z)e(w)&\simeq\varphi(z-w)e(w)\psi(z) \;,\\
	\psi(z)f(w)&\simeq\varphi(z-w)f(w)\psi(z) \;,\\
	\sigma_3\left[e(z),f(w)\right]&\sim-\frac{\psi(z)-\psi(w)}{z-w} \;,
\end{split}
\end{equation}
where
$$
\varphi(z)\coloneqq \frac{(z+\epsilon_1)(z+\epsilon_2)(z+\epsilon_3)}{(z-\epsilon_1)(z-\epsilon_2)(z-\epsilon_3)} \;.
$$
%and by ``$\sim$" one implies equivalence up to positive powers of $z$ and $w$, or, alternatively, we can claim that the LHS and RHS have the same poles and residues.

This algebra is a natural $q\to 1$ limit of the quantum version of $Y\left(\widehat{\fg\fl_1}\right)$, which is known in the literature under different names: quantum toroidal $\fg\fl(1)$ \cite{MR1324698}, Ding-Iohara-Miki algebra \cite{Ding:1996mq,Miki2007, Mironov:2016yue, Awata:2016riz} (defined through quadratic relations between currents); it is also contained \cite{di2019mathbf} in the spherical DAHA algebra \cite{Cherednick}. The quantum version is also parametrized by three parameters
$$
t_1=q^{\epsilon_1},\quad t_2=q^{\epsilon_2},\quad t_3=q^{\epsilon_3} \;,
$$
together with a constraint $t_1t_2t_3=1$.

%%%%%%%---------------------------------------------------------------------------------------------------------------------------------------------------------------------------------------------
\subsection{\texorpdfstring{$\IC^3$ Example---$Y\left(\widehat{\fg\fl}_1\right)$ MacMahon Modules}{C3 Example---Y(gl1) MacMahon Modules}}
\label{s:C3}

Our next model is a Hilbert scheme on $\IC^3$, and to describe it we use the standard trefoil quiver with a canonical superpotential \eqref{trefoil_early}.

The moduli space of a trefoil quiver representations is given by 
$$
{\rm Spec}\;\IC[B_1,B_2,B_3]/\left\langle \left[B_i,B_j\right]=0\right\rangle \;.
$$
However this space is rather singular. In particular, one will not be able to define zeroes to the real moment map on this manifold. Therefore, in practice, one usually introduces a resolution by chiral framing matter $I$:
\be\label{trefoil}
\begin{array}{c}
	\begin{tikzpicture}
	\draw[->] (0.173,-0.1) to[out=330,in=270] (1,0) to[out=90,in=30] (0.173,0.1);
	\node[right] at (1,0) {$B_1$};
	\begin{scope}[rotate=90]
	\draw[->] (0.173,-0.1) to[out=330,in=270] (1,0) to[out=90,in=30] (0.173,0.1);
	\node[above] at (1,0) {$B_2$};
	\end{scope}
	\begin{scope}[rotate=180]
	\draw[->] (0.173,-0.1) to[out=330,in=270] (1,0) to[out=90,in=30] (0.173,0.1);
	\node[left] at (1,0) {$B_3$};
	\end{scope}
	\draw (0,0) circle (0.2);
	\draw[->] (0,-1) --  (0,-0.2);
	\draw (-0.1,-1) -- (0.1,-1) -- (0.1,-1.2) -- (-0.1,-1.2) -- (-0.1,-1);
	\node[right] at (0,-0.6) {$I$};
	\end{tikzpicture}
\end{array},\quad W=\Tr\left(\left[B_1,B_2\right]B_3\right) \;.
\ee 

For such situation we can apply an analog of Hitchin-Kobayashi correspondence and state that torus fixed points on a locus $\mu_{\IR}=\mu_{\IC}=0$ are equivalent to the torus fixed points on the following manifold:
\be\label{moduli}
\left\{ \begin{array}{c}
	\left[B_i,B_j\right]=0\\
	I\mbox{ -- cyclic}
\end{array}\right\}/\{\mbox{complex gauge}\} \;.
\ee

\begin{figure}[ht!]
	\begin{center}
		\begin{tikzpicture}
		\node{\includegraphics[trim=0 100 50 0,clip,scale=0.6]{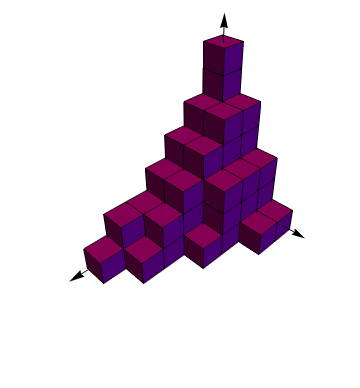}};
		\node at (-2.2,-3) {\bf x};
		\node at (3.5,-2.2) {\bf y};
		\node at (1.7,3) {\bf z};
		\end{tikzpicture}
	\end{center}
	\caption{Plane partition $\{\{7, 6, 4, 2\}, \{6, 5, 3\}, \{4, 4, 3\}, \{3, 3\}, \{2, 2\}, \{1\}, \{1\}\}$}\label{fig:plane_partition}
\end{figure}
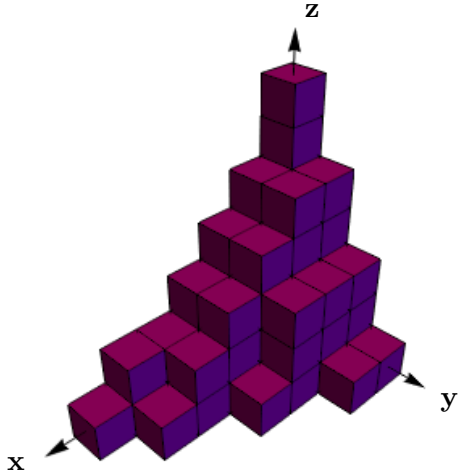

The lift of \eqref{trefoil} to the torus and afterwards to $\IR^2$ covering the torus reveals a triangular lattice with triangle edges corresponding to multiplications by $B_i$. Since in the fixed point \eqref{moduli} $B_i$ are commuting we associate any word of the form $B_1^aB_2^bB_3^cI$ to a monomial $u_1^au_2^bu_3^c$ in commutative variables $u_i$. Possible crystals in this case span vector spaces identified with
$$
\CV=\IC[u_1,u_2,u_3]\;{\rm mod}\; \CI  \;,
$$
where $\CI$ is an ideal of length $n$. The torus action on variables $u_1$, $u_2$, $u_3$ can be read from the equivariant action on $B_i$ having weights $\epsilon_i$. To visualize the crystals let us treat a monomial $u_1^au_2^bu_3^c$ as an atom with coordinates $(a,b,c)$ in the 3d octant $\bf O$. One associates the very octant with the monomial generators of the ring $\IC[u_1,u_2,u_3]$. Each ideal generator $u_1^au_2^bu_3^c$ cuts from the octant all atoms with coordinates $(a,b,c)+{\bf O}$. The remaining atoms correspond to basis vectors of $\CV$, and one should have $n=|\CV|$ of them. Therefore we conclude that the torus fixed points on \eqref{moduli} are in one-to-one correspondence with the plane partitions that enumerate all possible crystals (see Fig.~\ref{fig:plane_partition}).

The equivariant weight parameters are subjected to the superpotential nilpotency constraint \eqref{W-inv}:
$$
\epsilon_1+\epsilon_2+\epsilon_3=0  \;,
$$
and the weight function for an atom $\Box$ with 3d coordinates $({\bf x},{\bf y},{\bf z})$ reads:
\be
\bphi_{\Box}=\epsilon_1{\bf x}+\epsilon_2{\bf y}+\epsilon_3{\bf z} \;.
\ee

If one follows the same steps taken in Section \ref{s:Hecke} the matrix elements should take the from of equivariant integrals leading to ratios of Euler characters of the tangent space to the fixed point and of the tangent space to the incidence locus:
\be\label{struc_consts}
E(\Lambda\to\Lambda+\Box)=\frac{\widetilde{\rm Eul}_{\Lambda}}{\widetilde{\rm Eul}_{\Lambda,\Lambda+\Box}} \;,\quad 
F(\Lambda\to\Lambda-\Box)=\frac{\widetilde{\rm Eul}_{\Lambda}}{\widetilde{\rm Eul}_{\Lambda-\Box,\Lambda}}  \;.
\ee
The only problem in this construction is that the Hilbert scheme on $\IC^3$ is singular, therefore Euler characters of the tangent spaces are not well-defined and require some regularization. We denote the regularization we will introduce in what follows by the tilde.

In the math literature an issue with singularity of the space of Hilbert schemes on $\IC^3$ is resolved by consideration of a cohomology of the moduli space with values in the vanishing cycle of $dW$, see \cite{Rapcak:2018nsl}.

We will take the following strategy to deal with this situation. We can consider different applications of the RG flow. In \eqref{family} we used as inflation coefficients a scale $\bf s$ for all parts of the supercharge. For example, conjugation with a single operator $\bar\psi \psi$ scales conjugate fermion operators $\psi$ and $\bar \psi$ in opposite ways:
\be
e^{\lambda\bar\psi\psi}\;\psi\; e^{-\lambda\bar\psi\psi}=e^{-\lambda}\psi \;,
\quad 
e^{\lambda\bar\psi\psi}\;\bar\psi \;e^{-\lambda\bar\psi\psi}=e^{\lambda}\bar\psi
 \;.
\ee

One has exactly conjugate fields in terms of \eqref{family} corresponding to the vector field and external multiplication by $dW$. Therefore, we can introduce a modified two-parametric family defined by%n a modified two-parametric way:
\be
\bar Q_{\dot 1}^{({\bf s}_1,{\bf s}_2)}=e^{-{\bf s}_1 \fH}\left(d_{X^3}+\bar\p_{\Phi,q}+\iota_{{\bf s}_1V}+{\bf s}_2dW\wedge\right)e^{{\bf s}_1\fH}  \;.
\ee
Finally, we can rescale ${\bf s}_1$ first, and then ${\bf s}_2$ afterwards. This action does not change the torus action fixed points we determined before and the BPS Hilbert space, still it is spanned by wave-functions approximated by Gaussian fluctuations around fixed points identified with plane partitions. However, the introduced re-scaling procedure modifies frequencies, or weights of tangent directions in the moduli space, and, eventually, it modifies what we called the effective IR wave-function in Section \ref{s:RG}.

As before we decompose all the chiral fields around their expectation values defined by the crystal fixed point  up to a linear order:
\be
B_1=\langle B_1\rangle+\delta B_1 \;, \; B_2=\langle B_2\rangle+\delta B_2 \;,\; B_3=\langle B_3\rangle+\delta B_3 \;,\; I=\langle I\rangle+\delta I  \;.
\ee

The linearized complexified gauge action parameterized by matrix $G$ acts on the tangent fields according to the following pattern:
\be
\delta B_i\to \delta B_i+\left[G,\langle B_i\rangle\right],\quad \delta I\to\delta I+G\langle I\rangle  \;.
\ee

Weights of tangent directions under the equivariant torus action are defined by the eigenvalues of the following operators:
\be
\delta B_i:\; \left[\Phi,* \right]-\epsilon_i\cdot *,\quad \delta I:\;\Phi\cdot *  \;.
\ee

We expect that re-scaling first ${\bf s}_1$ will lead to the fact that tangent degrees of freedom corresponding to the gauge directions will acquire large frequencies and will not contribute to the effective wave-function as it was in the previous example of ${\rm Hilb}^n(\IC^2)$. The superpotential does not contribute at the first order, therefore we do not impose this extra constraint on the tangent degrees of freedom, and non-zero weights are not affected by it and remain intact. Let us call this graded vector space $\CU_{\Lambda}$. One will however encounter certain direction in the tangent space with 
zero values for the weights of the equivariant torus action. For those directions we have to consider higher corrections including the superpotential. 

First of all we claim that the tangent space to the fixed point  after factoring out gauge degrees of freedom, which we called $\CU_{\Lambda}$, has an even number of zero-weighted basis vectors. Moreover one can combine these vectors in pairs $(r_a,s_a)$, so that the superpotential in this local coordinates will have the following form:
\be\label{eff_W}
W=\sum\lm_{a} r_as_a  \;.
\ee
The reason for this decomposition is the following. The superpotential is invariant with respect to the gauge and flavor symmetries, and in terms of the tangent local coordinates it has a quadratic form. In this quadratic form only degrees of freedom corresponding to mutually opposite weights can contribute as monomials. The superpotential contribution corresponding to the zero-weighted degrees of freedom can not have quadratic terms like $r_a^2$ since the initial superpotenial does not have quadratic terms in $B_i$ and has to be a non-degenerate quadratic form; otherwise fixed points we identified with the plane partitions will not be fixed along directions where $W$ degenerates.

The supercharge will reduce effectively to the twisted differential:
$$
\bar \p+dW\wedge  \;.
$$

If one defines a wave-function corresponding to the cohomology of such supercharge it will not have the form  of a Thom representative of the corresponding Euler class. To put the situation back on track let us consider a regularization of zero equivariant weights of field pairs $(r_a,s_a)$. Assume to $r_a$ we assign some weight $\epsilon_a$, then to $s_a$ one has to assign weight $-\epsilon_a$ to preserve the gauge invariance of \eqref{eff_W}. Therefore we conclude that a pair of fields $(r_a,s_a)$ contribute as $-\epsilon_a^2$ to the Euler character. Let us omit the normalization and state that the pair of fields with the zero weight contributes as sign multiplier $(-1)$ to the Euler character. 

In the case of tangent space to the fixed point the number of fields with zero equivariant weight is always even, as we have argued already. 
In that case the number of pairs is just half the number of those fields. In general, for example when we will discuss incidence loci, we will encounter situations when the number of zero-weighted fields is odd, in that case we take as the number of pairs the integer part of half the number of zero-weighted fields.

Summarizing, assume there is a vector space $\CN$ with a basis where vectors are graded by the equivariant torus action weights:
\be\label{CN}
\CN:\;\left\{(m_i,w_i)\right\}_{i=1}^{{\rm dim}\;\CN}  \;.
\ee
For such a space we define the regularized Euler character as:
\be\label{res_Eul}
\widetilde{\rm Eul}(\CN) \coloneqq (-1)^{\left[\frac{1}{2}\#(i:\;w_i=0)\right]}\prod\lm_{i:\;w_i\neq 0}w_i  \;.
\ee
Here by $\#(i \colon w_i=0)$ we denote the cardinality of the subset of the zero-weighted vectors.

Therefore we identify naturally:
\be
\widetilde{\rm Eul}_{\Lambda} \coloneqq \widetilde{\rm Eul}(\CU_{\Lambda})  \;.
\ee

The only remaining ingredient in this construction is a description of the incidence locus. We perform this construction by incorporating a homomorphism $\tau$ as we stated in Section \ref{s:Hecke}:
\be
\begin{array}{c}
	\begin{tikzpicture}
	\node(A) at (0,0) {$\CV_{\Lambda+\Box}$};
	\node(B) at (2,0) {$\CV_{\Lambda}$};
	\path (A) edge[->] node[above] {$\tau$} (B);
	\begin{scope}[shift={(-0.6,0)}]
	\draw[->] (0,0.2) to[out=150,in=90] (-0.7,0) to[out=270,in=210] (0,-0.2); 
	\node[left] at (-0.7,0) {$B_i^{(2)}$};
	\end{scope}
	\begin{scope}[xscale=-1,shift={(-2.3,0)}]
	\draw[->] (0,0.2) to[out=150,in=90] (-0.7,0) to[out=270,in=210] (0,-0.2); 
	\node[right] at (-0.7,0) {$B_i^{(1)}$};
	\end{scope}
	\end{tikzpicture}
\end{array}\;.
\ee

We will denote matrices $B_i$ acting on the representations $\CV_{\Lambda}$ and $\CV_{\Lambda+\Box}$ as $B_i^{(1)}$ and $B_i^{(2)}$. Then morphism $\tau$ satisfies the following set of relations:
\be\label{incidence_2}
\tau \cdot B_i^{(2)}=B_i^{(1)}\cdot \tau,\quad i=1,2,3  \;.
\ee
In the case of ${\rm Hilb}^n(\IC^3)$ this constraint substitutes an analogous constraint \eqref{incidence} in the case of ${\rm Hilb}^n(\IC^2)$.

In the fixed bases in spaces $\CV$ associated to partitions it is quite simple to describe the vacuum expectation value of morphism $\tau$. Since $\CV_{\Lambda+\Box}$ has dimension higher than $\CV_{\Lambda}$, $\langle\tau\rangle$ is a projection:
\be
\langle\tau\rangle|\Box\rangle=\left\{\begin{array}{ll}
	|\Box\rangle, & {\rm if}\; \Box\in\Lambda \;,\\
	0,& {\rm otherwise} \;.
\end{array} \right.
\ee
By linearizing \eqref{incidence_2} around fixed points:
$$
\langle\tau\rangle\cdot \delta B_i^{(2)}+\delta \tau\cdot\langle B_i^{(2)}\rangle=\delta B_i^{(1)}\cdot\langle\tau\rangle +\langle B_i^{(1)}\rangle\cdot\delta \tau,\quad i=1,2,3 \;,
$$
one is able to derive a hyperplane $\Sigma$ in the space of all linear deformations $\delta B_i$ and $\delta\tau$. We define the incidence locus in this case as:
\be
\CU_{\Lambda,\Lambda+\Box}\coloneqq \left(\CU_{\Lambda}\cup\CU_{\Lambda+\Box}\right)\cap\Sigma  \;.
\ee

And the corresponding Euler character reads:
\be
\widetilde{\rm Eul}_{\Lambda,\Lambda+\Box}\coloneqq \widetilde{\rm Eul}(\CU_{\Lambda,\Lambda+\Box})  \;.
\ee

Unfortunately, in this setup we are unable to present a combinatorial expression for matrix elements \eqref{struc_consts} in the manner of \eqref{e_z_Fock}. On the other hand, we are able to perform numerical checks for plane partitions with a rather high number of atoms. We will present an example of a matrix element calculation in Appendix \ref{s:ex_gl_1}. 

We find numerically that the resulting matrix elements satisfy the following set of equations (compare to (4.45) -- (4.49) in \cite{Prochazka:2015deb}):
\bea
\begin{split}
	&E(\Lambda\to\Lambda+\Box)F(\Lambda+\Box\to\Lambda)=-\frac{1}{\epsilon_1\epsilon_2\epsilon_3}\;\mathop{\rm Res}\lm_{z=\bphi_{\Box}}\psi_{\Lambda}(z)\\
	&\frac{E(\Lambda\to\Lambda+\Box_1)E(\Lambda+\Box_1\to\Lambda+\Box_1+\Box_2)}{E(\Lambda\to\Lambda+\Box_2)E(\Lambda+\Box_2\to\Lambda+\Box_1+\Box_2)}=\varphi\left(\bphi_{\Box_2}-\bphi_{\Box_1}\right)\\
	&\frac{F(\Lambda+\Box_1+\Box_2\to\Lambda+\Box_1)F(\Lambda+\Box_1\to\Lambda)}{F(\Lambda+\Box_1+\Box_2\to\Lambda+\Box_2)F(\Lambda+\Box_2\to\Lambda)}=\varphi\left(\bphi_{\Box_2}-\bphi_{\Box_1}\right)\\
	&\sum\lm_{\pi\in S_3}\left(\bphi_{\Box_{\pi(1)}}-2\bphi_{\Box_{\pi(2)}}+\bphi_{\Box_{\pi(3)}}\right)E(\Lambda\to \Lambda+\Box_{\pi(1)})\times\\
	&\qquad\times E(\Lambda+\Box_{\pi(1)}\to \Lambda+\Box_{\pi(1)}+\Box_{\pi(2)})\times\\
	&\qquad \times E(\Lambda+\Box_{\pi(1)}+\Box_{\pi(2)}\to \Lambda+\Box_{\pi(1)}+\Box_{\pi(2)}+\Box_{\pi(3)})=0  \;.
\end{split}
\eea

Lifting these relations to generators $e(z)$, $f(z)$ and $\psi(z)$ we find that BPS algebra in this case is again $Y(\widehat{\fg\fl}_{1})$ defined by OPE \eqref{Y_gl_1} and qubic Serre relations \eqref{Y_gl_1_Serre}. However together with the BPS algebra we receive as a byproduct a space of BPS particles---a representation module. In this case the representation is the MacMahon module where vectors are labeled by plane partitions \cite{FJMM}.

%%%%%%%---------------------------------------------------------------------------------------------------------------------------------------------------------------------------------------------
\subsection{\texorpdfstring{Orbifold $(\IC^2/\IZ_2)\times \IC$---$Y(\widehat{\fg\fl}_{2})$ Colored MacMahon Modules}{Orbifold (C2/Z2)x C--Y(gl2) Colored MacMahon Modules}}

Let us next discuss a more complicated geometry $(\IC^2/\IZ_2)\times \IC$.
We can derive the $(\IC^2/\IZ_2)\times \IC$-orbifold quiver from $\IC^3$ quiver using the standard truncation procedure \cite{Ito_Nakajima}. The result resembles affine Dynkin diagram $\hat A_1$ by the McKay correspondence:
\be\label{C3_Z2_quiver}
\begin{array}{c}
	\begin{tikzpicture}
	\draw (0,0) circle (0.2); 
	\draw[fill=black] (4,0) circle (0.2);
	\draw[->] (0.173205, 0.1) to[out=30,in=150] (3.82679, 0.1);
	\draw[->] (0.1, 0.173205) to[out=60,in=120] (3.9, 0.173205);
	\node at (2,1.5) {$A_2$};
	\node at (2,0.9) {$A_3$};
	\draw[<-] (0.173205, -0.1) to[out=330,in=210] (3.82679, -0.1);
	\draw[<-] (0.1, -0.173205) to[out=300,in=240] (3.9, -0.173205);
	\node at (2,-0.4) {$B_2$};
	\node at (2,-0.9) {$B_3$};
	\draw[->] (-0.173205, -0.1) to[out=225,in=270] (-1.5,0) to[out=90,in=135] (-0.173205, 0.1);
	\draw[->] (4.173205, -0.1) to[out=315,in=270] (5.5,0) to[out=90,in=45] (4.173205, 0.1);
	\node[left] at (-1.5,0) {$A_1$};
	\node[right] at (5.5,0) {$B_1$};
	\draw[->] (0,1.5) -- (0,0.2);
	\begin{scope}[shift={(0,1.6)}]
	\draw (-0.1,-0.1) -- (-0.1,0.1) -- (0.1,0.1) -- (0.1,-0.1) -- (-0.1,-0.1);
	\end{scope}
	\node[left] at (0,1) {$I$};
	\end{tikzpicture}
\end{array} \;,
\ee
\be
W=\Tr\left[B_3A_2A_1+A_3B_2B_1-B_2A_3A_1-A_2B_3B_1\right] \;.
\ee
To stabilize the quiver we introduce a simple framing to one node; we will call that node \emph{white}, or node $(1)$, the other node is \emph{black}, or node $(2)$.

We assign equivariant torus weights $\alpha_i$ and $\beta_i$ to maps $A_i$, and $B_i$ correspondingly.

The nilpotency requirement for supercharge \eqref{W-inv} constraints the weights of the chiral fields:
\be
\alpha_1=\beta_1=-(\alpha_2+\beta_3)=-(\alpha_3+\beta_2)=:\gamma  \;.
\ee
At this stage we clearly observe an appearance of a new relevant torus parameter combination we denote as $\gamma$.

\begin{figure}[ht!]
	\begin{center}
		\begin{tikzpicture}
		\node{\includegraphics[trim=0 100 50 0,clip,scale=0.6]{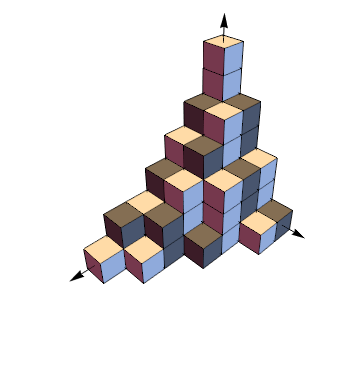}};
		\node at (-2.2,-3) {\bf x};
		\node at (3.5,-2.2) {\bf y};
		\node at (1.7,3) {\bf z};
		\end{tikzpicture}
	\end{center}
	\caption{Colored plane partition $\{\{7, 6, 4, 2\}, \{6, 5, 3\}, \{4, 4, 3\}, \{3, 3\}, \{2, 2\}, \{1\}, \{1\}\}$}\label{fig:col_plane_partition}
\end{figure}
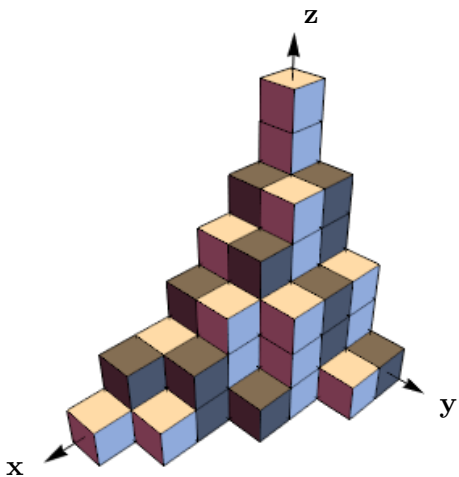

The fixed points are labelled by colored plane partitions. To a box with coordinates $({\bf x},{\bf y},{\bf z})$ one assigns a white color or black color depending on whether the combination ${\bf x}+{\bf y}$ is an even or odd integer. See an example in \fig\ref{fig:col_plane_partition}.

As before, in the fixed point we choose a specific basis in the quiver representation where the basis vectors are labeled by boxes in the partition. White boxes are associated to the vectors of the white node vector space $V^{(1)}$, while black boxes are associated to the vectors of the black node vector space $V^{(2)}$ in quiver \eqref{Conif_quiver}:
\be
\dim \; V^{(1)}=\left|\Lambda^{(1)}\right|,\quad \dim \; V^{(2)}=\left|\Lambda^{(2)}\right|  \;.
\ee

In the chosen basis the expectation values of various maps are given by
\bea
\begin{split}
	\langle I\rangle &=|0,0,0,0\rangle\;,\\
	\langle A_1\rangle |{\bf x},{\bf y},{\bf z}\rangle &=|{\bf x},{\bf y},{\bf z}+1\rangle,\quad \langle B_1\rangle |{\bf x},{\bf y},{\bf z}\rangle=|{\bf x},{\bf y},{\bf z}+1\rangle\;,\\
	\langle A_2\rangle |{\bf x},{\bf y},{\bf z}\rangle&=|{\bf x}+1,{\bf y},{\bf z}\rangle,\quad \langle B_2\rangle |{\bf x},{\bf y},{\bf z}\rangle=|{\bf x}+1,{\bf y},{\bf z}\rangle\;,\\
	\langle A_3\rangle |{\bf x},{\bf y},{\bf z}\rangle&=|{\bf x},{\bf y}+1,{\bf z}\rangle,\quad \langle B_3\rangle |{\bf x},{\bf y},{\bf z}\rangle=|{\bf x},{\bf y}+1,{\bf z}\rangle\;.\\
\end{split}
\eea
Despite a similarity in actions of $\langle A_i\rangle$ and $\langle B_i\rangle$, $\langle A_i\rangle$ act only upon the white box basis vectors, and $\langle B_i\rangle$ act only upon the black box basis vectors: the resulting color under such action may be either.
If the action of either generator shifts to a box outside the partition it is implied that the corresponding matrix element is zero.

The weight function on an atom $\Box$ with coordinates $({\bf x},{\bf y},{\bf z})$ is defined by a box weight function:
\bea
\begin{split}
	\bphi_{\Box}&={\bf z}\;\gamma+\frac{{\bf x}+({\bf x}\;{\rm mod}\; 2) }{2}\alpha_2+\frac{{\bf x}-({\bf x}\;{\rm mod}\; 2)}{2}\beta_2 + \\
	&+\frac{{\bf y}+({\bf y}\;{\rm mod}\;2)}{2}{\bf if}({\bf x}\;{\rm mod}\;2;\alpha_3,\beta_3)+\frac{{\bf y}-({\bf y}\;{\rm mod}\;2)}{2}{\bf if}({\bf x}\;{\rm mod}\;2;\beta_3,\alpha_3) \;,
\end{split}
\eea
where function ${\bf if}$ is defined as:
\be
{\bf if}(a;x,y)\coloneqq \left\{\begin{array}{ll}
	x, & {\rm if}\; a=0 \;,\\
	y, & {\rm if}\; a=1\;.\\
\end{array} \right.
\ee

Denote by $\Lambda^+$ and $\Lambda^-$ a collection of boxes that can be added to or subtracted from the partition so that it becomes a partition again. We split them in subsets of white boxes $\Lambda^{\pm(1)}$ and black boxes $\Lambda^{\pm(2)}$ correspondingly.

As explained in Section \ref{s:Hecke} we construct BPS operators adding and subtracting D0-branes using regularized Euler characters \eqref{res_Eul}:
\bea
\begin{split}
	e^{(v)}(z)|\Lambda\rangle&=\sum\lm_{\Box\in \Lambda^{+(v)}}\frac{1}{z-\bphi_{\Box}}\times\frac{\widetilde{\rm Eul}(\Lambda)}{\widetilde{\rm Eul}(\Lambda,\Lambda+\Box)}|\Lambda+\Box\rangle \;,\\
	f^{(v)}(z)|\Lambda\rangle&=\sum\lm_{\Box\in \Lambda^{-(v)}}\frac{1}{z-\bphi_{\Box}}\times\frac{\widetilde{\rm Eul}(\Lambda)}{\widetilde{\rm Eul}(\Lambda-\Box,\Lambda)}|\Lambda-\Box\rangle  \;.
\end{split}
\eea

We derive these generators to satisfy the following OPE \cite{Li:2020rij}:
\bea
\begin{split}
	\left[e^{(v)}(x),f^{(w)}(y)\right]\; &\sim\;-\delta_{vw}\frac{1}{\gamma}\frac{\psi^{(v)}(x)-\psi^{(v)}(y)}{x-y}\;,\\
	e^{(v)}(x)e^{(v)}(y)\;&\sim\; \varphi^{(0)}(x-y)\; e^{(v)}(y)e^{(v)}(x)\;,\\
	e^{(1)}(x)e^{(2)}(y)\;&\sim\; \varphi^{(1)}(x-y)\; e^{(2)}(y)e^{(1)}(x)\;,\\
	e^{(2)}(x)e^{(1)}(y)\;&\sim\; \varphi^{(2)}(x-y)\; e^{(1)}(y)e^{(2)}(x)\;,\\
	\psi^{(v)}(x)e^{(v)}(y)\;&\sim\; \varphi^{(0)}(x-y)\; e^{(v)}(y)\psi^{(v)}(x)\;,\\
	\psi^{(1)}(x)e^{(2)}(y)\;&\sim\; \varphi^{(2)}(x-y)\; e^{(2)}(y)\psi^{(1)}(x)\;,\\
	\psi^{(2)}(x)e^{(1)}(y)\;&\sim\; \varphi^{(1)}(x-y)\; e^{(1)}(y)\psi^{(2)}(x)\;,\\
	f^{(v)}(x)f^{(v)}(y)\;&\sim\; \left[\varphi^{(0)}(x-y)\right]^{-1}\; f^{(v)}(y)f^{(v)}(x)\;,\\
	f^{(1)}(x)f^{(2)}(y)\;&\sim\; \left[\varphi^{(1)}(x-y)\right]^{-1}\; f^{(2)}(y)f^{(1)}(x)\;,\\
	f^{(2)}(x)f^{(1)}(y)\;&\sim\; \left[\varphi^{(2)}(x-y)\right]^{-1}\; f^{(1)}(y)f^{(2)}(x)\;,\\
	\psi^{(a)}(x)f^{(a)}(y)\;&\sim\; \left[\varphi^{(0)}(x-y)\right]^{-1}\; f^{(a)}(y)\psi^{(a)}(x)\;,\\
	\psi^{(1)}(x)e^{(2)}(y)\;&\sim\; \left[\varphi^{(2)}(x-y)\right]^{-1}\; f^{(2)}(y)\psi^{(1)}(x)\;,\\
	\psi^{(2)}(x)f^{(1)}(y)\;&\sim\; \left[\varphi^{(1)}(x-y)\right]^{-1}\; f^{(1)}(y)\psi^{(2)}(x)\;,
\end{split}
\eea
where the functions read \cite{Li:2020rij}:
\be
\varphi^{(0)}(z)=\frac{z+\gamma}{z-\gamma} \;,\quad 
\varphi^{(1)}(z)=\frac{(z+\alpha_2)(z+\alpha_3)}{(z-\beta_2)(z-\beta_3)} \;,\quad 
\varphi^{(2)}(z)=\frac{(z+\beta_2)(z+\beta_3)}{(z-\alpha_2)(z-\alpha_3)}  \;.
\ee

In the the fixed point basis Cartan elements $\psi^{(a)}(z)$ acquire eigenvalues:
\bea
\begin{split}
	\psi^{(1)}(z)|\Lambda\rangle&=\frac{1}{z}\prod\lm_{\Box\in\Lambda^{(1)}}\varphi^{(0)}(z-\bphi_{\Box})\prod\lm_{\Box\in\Lambda^{(2)}}\varphi^{(2)}(z-\bphi_{\Box})|\Lambda\rangle \;,\\
	\psi^{(2)}(z)|\Lambda\rangle&=\prod\lm_{\Box\in\Lambda^{(2)}}\varphi^{(0)}(z-\bphi_{\Box})\prod\lm_{\Box\in\Lambda^{(1)}}\varphi^{(1)}(z-\bphi_{\Box})|\Lambda\rangle \;.
\end{split}
\eea

Additionally higher order Serre relations are satisfied for this algebra:
\be
{\rm Sym}_{x_1,x_2,x_3}\left[e^{(v)}(x_1),\left[e^{(v)}(x_2),\left[e^{(v)}(x_3),e^{(v\pm 1)}(y)\right]\right]\right]=0  \;.
\ee

This BPS algebra becomes the affine Yangian $Y_{\hbar_1,\hbar_2}(\widehat{\fg\fl}_{2})$ if we choose the following identification of the equivariant parameters:
\be
\alpha_1=\beta_1=\hbar_1 \;,\quad \alpha_2=\beta_2=\hbar_2 \;,\quad \alpha_3=\beta_3=-\hbar_1-\hbar_2  \;.
\ee

%%%%%%%---------------------------------------------------------------------------------------------------------------------------------------------------------------------------------------------
\subsection{\texorpdfstring{Conifold $z_1z_2-z_3z_4=0$---$Y(\widehat{\fg\fl}_{1|1})$ Pyramid Partitions}{Conifold z1 z2-z3 z4=0---Y(gl(1|1)) Pyramid Partitions}}

A classical example of a Calabi-Yau orbifold is a hypersurface in $\IC^4$ subjected to the following constraint:
\be \label{CYo_surf}
z_1z_2-z_3z_4=0  \;,
\ee
where $z_i$ are complex coordinates in $\IC^4$. %The Calabi-Yau metric on this orbifold is the induced metric on \eqref{CYo_surf}. 
%In the quiver construction for Calabi-Yau three-folds this orbifold corresponds to the following quiver with a superpotential:
This geometry corresponds to the following quiver with a superpotential:
\be\label{Conif_quiver}
\begin{array}{c}
	\begin{tikzpicture}
	\begin{scope}[shift={(-1.5,0)}]
	\draw (-0.1,-0.1) -- (-0.1,0.1) -- (0.1,0.1) -- (0.1,-0.1) -- (-0.1,-0.1);
	\end{scope}
	\draw[->] (-1.4,0) -- (-0.2,0);
	\node[above] at (-0.75,0) {$I$};
	\draw (0,0) circle (0.2); 
	\draw[fill=black] (4,0) circle (0.2);
	\draw[->] (0.173205, 0.1) to[out=30,in=150] (3.82679, 0.1);
	\draw[->] (0.1, 0.173205) to[out=60,in=120] (3.9, 0.173205);
	\node at (2,1.5) {$A_1$};
	\node at (2,0.9) {$A_2$};
	\draw[<-] (0.173205, -0.1) to[out=330,in=210] (3.82679, -0.1);
	\draw[<-] (0.1, -0.173205) to[out=300,in=240] (3.9, -0.173205);
	\node at (2,-0.4) {$B_1$};
	\node at (2,-0.9) {$B_2$};
	\end{tikzpicture}
\end{array},\quad W=\Tr\left[A_1B_1A_2B_2-A_1B_2A_2B_1\right] \;.
\ee

This quiver is simpler than one for $(\IC^2/\IZ_2)\times\IC$, however in this case the superpotential is quartic. And again to stabilize the quiver we introduce a simple framing to one node, we will call that node \emph{white}, or node $(1)$, the other node is \emph{black}, or node $(2)$.

We assign equivariant torus weights $\alpha_1$, $\alpha_2$, $\beta_1$, $\beta_2$ to $A_1$, $A_2$, $B_1$, $B_2$ correspondingly.
The constraint \eqref{W-inv} is translated into the following constraint for the equivariant weights:
\be
\alpha_1+\alpha_2+\beta_1+\beta_2=0  \;.
\ee

The fixed points can be identified with pyramid partitions $\Lambda$.

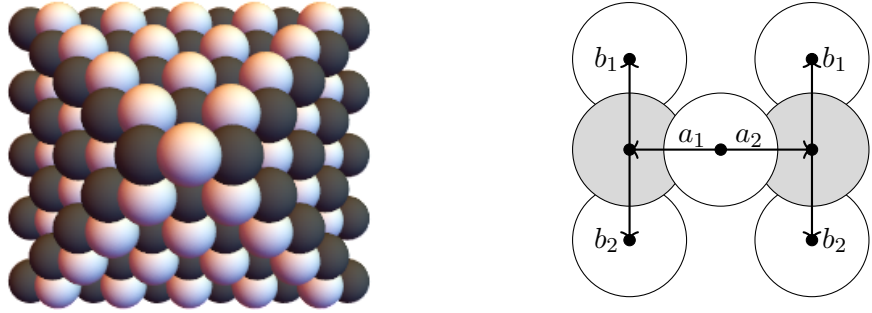
\begin{figure}[ht!]
	\begin{center}
		\begin{tikzpicture}
		\node{\includegraphics[trim=0 50 0 50,clip,scale=0.6]{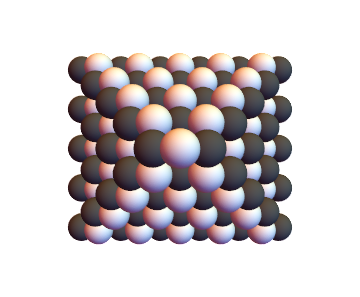}};
		\node at (7,0) {$\begin{array}{c}
			\begin{tikzpicture}
			\begin{scope}[scale=1.5]
			\draw[fill=white] (-0.8,-0.8) circle (0.5) (-0.8,0.8) circle (0.5) (0.8,0.8) circle (0.5) (0.8,-0.8) circle (0.5);
			\draw[fill=gray!30!white] (-0.8,0) circle (0.5) (0.8,0) circle (0.5);
			\draw[fill=white] (0,0) circle (0.5);
			\draw[thick,->] (0,0) -- (-0.8,0);
			\draw[thick,->] (0,0) -- (0.8,0);
			\draw[thick,->] (-0.8,0) -- (-0.8,0.8);
			\draw[thick,->] (-0.8,0) -- (-0.8,-0.8);
			\draw[thick,->] (0.8,0) -- (0.8,0.8);
			\draw[thick,->] (0.8,0) -- (0.8,-0.8);
			\draw[fill=black] (0,0) circle (0.05) (-0.8,0) circle (0.05) (0.8,0) circle (0.05) (-0.8,0.8) circle (0.05) (0.8,0.8) circle (0.05) (-0.8,-0.8) circle (0.05) (0.8,-0.8) circle (0.05);
			\node at (-0.25,0.1) {$a_1$};
			\node at (0.25,0.1) {$a_2$};
			\node[left] at (-0.8,0.8) {$b_1$};
			\node[left] at (-0.8,-0.8) {$b_2$};
			\node[right] at (0.8,0.8) {$b_1$};
			\node[right] at (0.8,-0.8) {$b_2$};
			\end{scope}
			\end{tikzpicture}
			\end{array}$};
		\end{tikzpicture}
	\end{center}
	\caption{Pyramid arrangement of atoms}\label{fig:pyramid}
\end{figure}

A pyramid arrangement of atoms starts with a single white atom on the very zero-level top layer and goes down in layers. Each even layer is filled with atoms of white color, each odd layer is filled with atoms of black color. Each atom on a higher layer is supported by a pair of atoms on a lower layer, one can move from atoms on the higher layer to the atoms on the lower layer along the following vectors:
\be
a_1=(1,0,-1) \;,\quad a_2=(-1,0,-1) \;,\quad b_1=(0,1,-1)\;,\quad b_2=(0,-1,-1)  \;.
\ee
See a detailed diagram of the pyramid arrangement of atoms in \fig \ref{fig:pyramid}.

\begin{figure}[ht!]
	\begin{center}
		\begin{tikzpicture}
		\node{\includegraphics[trim=0 50 0 50,clip,scale=0.4]{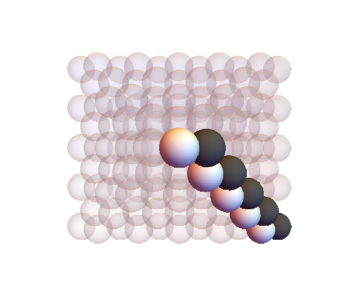}};
		\node at (4.5,0) {\includegraphics[trim=0 50 0 50,clip,scale=0.4]{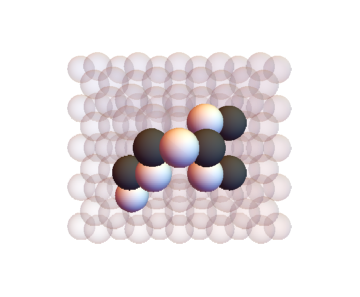}};
		\node at (9,0) {\includegraphics[trim=0 50 0 50,clip,scale=0.4]{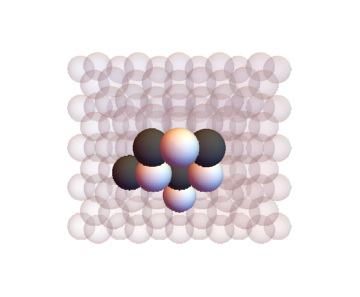}};
		\end{tikzpicture}
	\end{center}
	\caption{Some examples of pyramid partitions} \label{fig:pyr_ex}
\end{figure}

We call a pyramid partition $\Lambda$ a group of atoms that can be eliminated from the top of the pyramid so that the remaining construction remains stable, i.e. each stone on a given layer is supported by a pair of atoms on a lower layer (see \fig\ref{fig:pyr_ex}).

A subset of white atoms in a pyramid partition $\Lambda$ we will denote as $\Lambda^{(1)}$, and a subset of black atoms as $\Lambda^{(2)}$ correspondingly.

The generating functions of pyramid partitions reads:
\be
\sum\lm_{\Lambda}q_1^{|\Lambda^{(1)}|}q_2^{|\Lambda^{(2)}|}=\prod\lm_{k=1}^{\infty}\frac{\left(1+q_1^kq_2^{k-1}\right)^k\left(1+q_1^kq_2^{k+1}\right)^k}{\left(1-\left(q_1q_2\right)^k\right)^{2k}} \;.
\ee

A real space position $\vec r$ of a stone in a pyramid partition can be uniquely decomposed in the following form:
$$
\vec r={\bf x}\;\vec a_1+{\bf y}\;\vec a_2+{\bf z}\;\vec b_1+{\bf w}\;\vec  b_2,\quad {\bf x},{\bf y},{\bf z},{\bf w}\in\IZ_{\geq 0}  \;.
$$

At the fixed point we choose a specific basis in the quiver representation where the basis vectors are labeled by atoms in the partition. White atoms are associated to the vectors of the white node vector space $V^{(1)}$, black atoms are associated to the vectors of the black node vector space $V^{(2)}$ in quiver \eqref{Conif_quiver}:
\be
\dim \; V^{(1)}=\left|\Lambda^{(1)}\right|,\quad \dim \; V^{(2)}=\left|\Lambda^{(2)}\right| \;.
\ee

In the chosen basis the action of various map expectation values are given by the following expressions:
\bea
\begin{split}
	\langle I\rangle&=|0,0,0,0\rangle \;,\\
	\langle A_1\rangle |{\bf x},{\bf y},{\bf z},{\bf w}\rangle&=|{\bf x}+1,{\bf y},{\bf z},{\bf w}\rangle \;,\\
	\langle A_2\rangle |{\bf x},{\bf y},{\bf z},{\bf w}\rangle&=|{\bf x},{\bf y}+1,{\bf z},{\bf w}\rangle \;,\\
	\langle B_1\rangle |{\bf x},{\bf y},{\bf z},{\bf w}\rangle&=|{\bf x},{\bf y},{\bf z}+1,{\bf w}\rangle \;,\\
	\langle B_2\rangle |{\bf x},{\bf y},{\bf z},{\bf w}\rangle&=|{\bf x},{\bf y},{\bf z},{\bf w}+1\rangle \;.
\end{split}
\eea
If the action of either generator shifts to a stone outside the partition, it is then implied that the corresponding matrix element is zero.

The weight function on atom $\Box$ is given by%:
\be
\bphi_{\Box}={\bf x}\;\alpha_1+{\bf y}\;\alpha_2+{\bf z}\;\beta_1+{\bf w}\;\beta_2  \;.
\ee

Denote by $\Lambda^+$ and $\Lambda^-$ a collection of atoms that can be added to or subtracted from the partition so that it becomes a partition again. We split them in subsets of white atoms $\Lambda^{\pm(1)}$ and black atoms $\Lambda^{\pm(2)}$ correspondingly.

BPS Hecke operators are constructed in the same fashion as explained in Section~\ref{s:Hecke} with regularized Euler characters \eqref{res_Eul}:
\bea
\begin{split}
	\hat e^{(v)}(z)|\Lambda\rangle&=\sum\lm_{\Box\in \Lambda^{+(v)}}\frac{1}{z-\bphi_{\Box}}\times\frac{\widetilde{\rm Eul}(\Lambda)}{\widetilde{\rm Eul}(\Lambda,\Lambda+\Box)}|\Lambda+\Box\rangle \;,\\
	\hat f^{(v)}(z)|\Lambda\rangle&=\sum\lm_{\Box\in \Lambda^{-(v)}}\frac{1}{z-\bphi_{\Box}}\times\frac{\widetilde{\rm Eul}(\Lambda)}{\widetilde{\rm Eul}(\Lambda-\Box,\Lambda)}|\Lambda-\Box\rangle  \;.
\end{split}
\eea

We derive these generators to satisfy the following OPE \cite{Li:2020rij}:
\bea\label{e_f_gl_1_1}
\begin{split}
	\left[\hat e^{(v)}(x),\hat f^{(w)}(y)\right]\;&\sim\;\delta_{vw}\frac{\hat \psi^{(v)}(x)-\hat \psi^{(y)}(w)}{x-y}\;,\\
	\hat \psi^{(v)}(x)\hat e^{(v)}(y)\;&\sim\; -\hat  e^{(v)}(y)\hat \psi^{(v)}(x)\;,\\
	\hat e^{(v)}(x)\hat e^{(v)}(y)\;&\sim\; -\hat e^{(v)}(y)\hat e^{(v)}(x)\;,\\
	\hat e^{(1)}(x)\hat e^{(2)}(y)\;&\sim\; \varphi^{(1)}(x-y)\; \hat e^{(2)}(y)\hat e^{(1)}(z)\;,\\
	\hat e^{(2)}(x)\hat e^{(1)}(y)\;&\sim\; \varphi^{(2)}(x-y)\; \hat e^{(1)}(y)\hat e^{(2)}(x)\;,\\
	\hat \psi^{(1)}(x)\hat e^{(2)}(y)\;&\sim\; \varphi^{(2)}(x-y)\; \hat e^{(2)}(y)\hat \psi^{(1)}(x)\;,\\
	\hat \psi^{(2)}(x)\hat e^{(1)}(y)\;&\sim\; \varphi^{(1)}(x-y)\; \hat e^{(1)}(y)\hat \psi^{(2)}(x)\;,\\
	\hat \psi^{(v)}(x)\hat f^{(v)}(y)\;&\sim\; - \hat f^{(v)}(y)\hat \psi^{(v)}(x)\;,\\
	\hat f^{(v)}(x)\hat f^{(v)}(y)\;&\sim\; -\hat f^{(v)}(y)\hat f^{(v)}(x)\;,\\
	\hat f^{(1)}(x)\hat f^{(2)}(y)\;&\sim\;\left[ \varphi^{(1)}(x-y)\right]^{-1}\; \hat f^{(2)}(y)\hat f^{(1)}(x)\;,\\
	\hat f^{(2)}(x)\hat f^{(1)}(y)\;&\sim\; \left[\varphi^{(2)}(x-y)\right]^{-1}\; \hat f^{(1)}(y)\hat f^{(2)}(x)\;,\\
	\hat \psi^{(1)}(x)\hat f^{(2)}(y)\;&\sim\; \left[\varphi^{(2)}(x-y)\right]^{-1}\; \hat f^{(2)}(y)\hat \psi^{(1)}(x)\;,\\
	\hat \psi^{(2)}(x)\hat f^{(1)}(y)\;&\sim\; \left[\varphi^{(1)}(x-y)\right]^{-1}\;\hat  f^{(1)}(y)\hat \psi^{(2)}(x) \;,\\
\end{split}
\eea
where the binding factors are given by the following expressions \cite{Li:2020rij}:
\be \varphi^{(1)}(z)=\frac{(z+\alpha_1)(z+\alpha_2)}{(z-\beta_1)(z-\beta_2)} \;,\quad \varphi^{(2)}(z)=\frac{(z+\beta_1)(z+\beta_2)}{(z-\alpha_1)(z-\alpha_2)} \;.
\ee

In the fixed point basis Cartan elements $\psi^{(a)}(z)$ acquire eigenvalues:
\bea
\begin{split}
	\hat\psi^{(1)}(z)|\Lambda\rangle&=\frac{1}{z}(-1)^{|\Lambda^{(1)}|}\prod\lm_{\Box\in\Lambda^{(2)}}\varphi^{(2)}(z-\bphi_{\Box})|\Lambda\rangle \;,\\
	\hat\psi^{(2)}(z)|\Lambda\rangle&=(-1)^{|\Lambda^{(2)}|}\prod\lm_{\Box\in\Lambda^{(1)}}\varphi^{(1)}(z-\bphi_{\Box})|\Lambda\rangle  \;.
\end{split}
\eea

In addition higher order Serre relations are satisfied for this algebra:
\be
{\rm Sym}_{x_1,x_2}{\rm Sym}_{y_1,y_2}\left\{\hat e^{(v)}(x_1),\left[\hat e^{(v\pm 1)}(y_1),\left\{\hat e^{(v)}(x_2),\hat e^{(v\pm 1)}(y_2)\right\}\right]\right\}=0  \;.
\ee

An attentive reader would notice that these OPE and Serre relations are different from those expected for the affine Yangian $Y(\widehat{\fg\fl}_{1|1})$. We will cure this issue in the next section by introducing additional sign statistical factors to atoms in the crystal. This will modify the signs of matrix elements in such a way that final generators $e^{(v)}(z)$, $f^{(v)}(z)$, $\psi^{(v)}(z)$ will satisfy relations of $Y(\widehat{\fg\fl}_{1|1})$.

%%%%%%%---------------------------------------------------------------------------------------------------------------------------------------------------------------------------------------------
\subsection{\texorpdfstring{Generalized Conifold $xy=z^mw^n$---$Y(\widehat{\fg\fl}_{m|n})$ Modules}{Generalized Conifold xy=z**m w**n---Y(gl(m|n)) Modules}}

A toric Calabi-Yau resolution of a singular algebraic curve $xy=z^mw^n$ in $\IC^4$ is called a  generalized conifold geometry, and it is encoded in a choice of a vacuum state in a $m+n$ periodic $\mathbb{Z}_2$-spin chain. We will call a signature $\Sigma_{m,n}$ an arrangement of $\mathbb{Z}_2$-spins $\sigma_a$ satisfying the following condition:\footnote{A default choice for $\widehat{\fg\fl}_{m|n}$ 
	$$
	\Sigma=\{\underbrace{ +1,\ldots,+1}_{m\;\mbox{times}},\underbrace{-1,\ldots,-1}_{n\;\mbox{times}} \}
	$$ 
	is known in math literature \cite{Bezerra:2019dmp} as the standard choice of Cartan matrix and parity.}
\be
\Sigma_{m,n}:\; \{1,2,\ldots,m+n \}\to \{+1,-1\}, \;\mbox{so that}\; \#(+1)=m,\;\#(-1)=n \;.
\ee
In this section we understand the spin index $(v)$ periodic modulo $m+n$. As a default example we will choose
\be\label{ex_sign}
\Sigma_{8,4}=\{1, 1, 1, 1, -1, -1, 1, 1, -1, 1, -1, 1\}  \;.
\ee

The quiver representation of the generalized conifold is constructed according to the following procedure: the quiver $\CQ(\Sigma_{m,n})$ has $m+n$ vertices, and one assigns to a quiver node $v$ \emph{even} or \emph{odd} parity depending on whether the product $\sigma_v\sigma_{v+1}$ is either $+1$ or $-1$. 

The quiver has the following set of maps. To each even node one assigns a map:
$$
A_v:\; V_v\longrightarrow V_v  \;.
$$ 
To each pair $(v,v+1)$ of nodes regardless their parity one assigns a pair of maps:
$$
B_v:\; V_v\longrightarrow V_{v+1} \;,\quad C_v:\; V_{v+1}\longrightarrow V_{v}  \;.
$$
Eventually, we choose some quiver node to connect it to the framing node. For instance, for $\CQ(\Sigma_{8,4})$ we have:
$$
\begin{array}{c}
\begin{tikzpicture}
\draw[fill=red] (0,0) circle (0.15);
\draw[fill=green] (1.5,0) circle (0.15);
\draw[fill=blue] (3,0) circle (0.15);
\draw[fill=purple] (4.5,0) circle (0.15);
\draw[fill=orange] (6,0) circle (0.15);
\draw[fill=cyan] (7.5,0) circle (0.15);
\node[below] at (0,-0.15) {$1$};
\node[below] at (1.5,-0.15) {$2$};
\node[below] at (3,-0.15) {$3$};
\node[below] at (4.5,-0.15) {$4$};
\node[below] at (6,-0.15) {$5$};
\node[below] at (7.5,-0.15) {$6$};
\draw[->] (0.129904, 0.075) -- (1.3701, 0.075);
\draw[<-] (0.129904, -0.075) -- (1.3701, -0.075);
\node[above] at (0.75,0.075) {$B_1$};
\node[below] at (0.75,-0.075) {$C_1$};
\begin{scope}[shift={(1.5,0)}]
\draw[->] (0.129904, 0.075) -- (1.3701, 0.075);
\draw[<-] (0.129904, -0.075) -- (1.3701, -0.075);
\node[above] at (0.75,0.075) {$B_2$};
\node[below] at (0.75,-0.075) {$C_2$};
\end{scope}
\begin{scope}[shift={(3,0)}]
\draw[->] (0.129904, 0.075) -- (1.3701, 0.075);
\draw[<-] (0.129904, -0.075) -- (1.3701, -0.075);
\node[above] at (0.75,0.075) {$B_3$};
\node[below] at (0.75,-0.075) {$C_3$};
\end{scope}
\begin{scope}[shift={(4.5,0)}]
\draw[->] (0.129904, 0.075) -- (1.3701, 0.075);
\draw[<-] (0.129904, -0.075) -- (1.3701, -0.075);
\node[above] at (0.75,0.075) {$B_4$};
\node[below] at (0.75,-0.075) {$C_4$};
\end{scope}
\begin{scope}[shift={(6,0)}]
\draw[->] (0.129904, 0.075) -- (1.3701, 0.075);
\draw[<-] (0.129904, -0.075) -- (1.3701, -0.075);
\node[above] at (0.75,0.075) {$B_5$};
\node[below] at (0.75,-0.075) {$C_5$};
\end{scope}
\begin{scope}[shift={(0,-2)}]
\draw[fill=gray] (0,0) circle (0.15);
\draw[fill=white] (1.5,0) circle (0.15);
\draw[fill=pink] (3,0) circle (0.15);
\draw[fill=brown] (4.5,0) circle (0.15);
\draw[fill=yellow] (6,0) circle (0.15);
\draw[fill=magenta] (7.5,0) circle (0.15);
\node[above] at (0,0.15) {$12$};
\node[above] at (1.5,0.15) {$11$};
\node[above] at (3,0.15) {$10$};
\node[above] at (4.5,0.15) {$9$};
\node[above] at (6,0.15) {$8$};
\node[above] at (7.5,0.15) {$7$};
\draw[->] (0.129904, 0.075) -- (1.3701, 0.075);
\draw[<-] (0.129904, -0.075) -- (1.3701, -0.075);
\node[above] at (0.75,0.075) {$C_{11}$};
\node[below] at (0.75,-0.075) {$B_{11}$};
\begin{scope}[shift={(1.5,0)}]
\draw[->] (0.129904, 0.075) -- (1.3701, 0.075);
\draw[<-] (0.129904, -0.075) -- (1.3701, -0.075);
\node[above] at (0.75,0.075) {$C_{10}$};
\node[below] at (0.75,-0.075) {$B_{10}$};
\end{scope}
\begin{scope}[shift={(3,0)}]
\draw[->] (0.129904, 0.075) -- (1.3701, 0.075);
\draw[<-] (0.129904, -0.075) -- (1.3701, -0.075);
\node[above] at (0.75,0.075) {$C_9$};
\node[below] at (0.75,-0.075) {$B_9$};
\end{scope}
\begin{scope}[shift={(4.5,0)}]
\draw[->] (0.129904, 0.075) -- (1.3701, 0.075);
\draw[<-] (0.129904, -0.075) -- (1.3701, -0.075);
\node[above] at (0.75,0.075) {$C_8$};
\node[below] at (0.75,-0.075) {$B_8$};
\end{scope}
\begin{scope}[shift={(6,0)}]
\draw[->] (0.129904, 0.075) -- (1.3701, 0.075);
\draw[<-] (0.129904, -0.075) -- (1.3701, -0.075);
\node[above] at (0.75,0.075) {$C_7$};
\node[below] at (0.75,-0.075) {$B_7$};
\end{scope}
\end{scope}
\draw[->] (-0.129904, -2.075) to[out=180,in=270] (-2.075,-1) to[out=90,in=180] (-0.129904,0.075);
\draw[<-] (-0.129904, -1.925) to[out=180,in=270] (-1.925,-1) to[out=90,in=180] (-0.129904,-0.075);
\node[left] at (-2.075,-1) {$B_{12}$};
\node[right] at (-1.925,-1) {$C_{12}$};
\begin{scope}[shift={(7.5,0)}]
\begin{scope}[xscale=-1]
\draw[->] (-0.129904, -2.075) to[out=180,in=270] (-2.075,-1) to[out=90,in=180] (-0.129904,0.075);
\draw[<-] (-0.129904, -1.925) to[out=180,in=270] (-1.925,-1) to[out=90,in=180] (-0.129904,-0.075);
\node[right] at (-2.075,-1) {$B_{6}$};
\node[left] at (-1.925,-1) {$C_{6}$};
\end{scope}
\end{scope}
\begin{scope}[shift={(3.75,-1)}]
\draw[fill=white] (-0.1,-0.1) -- (-0.1,0.1) -- (0.1,0.1) -- (0.1,-0.1) -- cycle;
\end{scope}
\draw[->] (3.65,-1) to (1,-1) to[out=180,in=300] (0.075,-0.129904);
\node at (0.4,-1) {$I$};
\draw[->] (-0.075,0.129904) to[out=120,in=180] (0,1) to[out=0,in=60] (0.075,0.129904);
\node[above] at (0,1) {$A_1$};
\begin{scope}[shift={(1.5,0)}]
\draw[->] (-0.075,0.129904) to[out=120,in=180] (0,1) to[out=0,in=60] (0.075,0.129904);
\node[above] at (0,1) {$A_2$};
\end{scope}
\begin{scope}[shift={(3,0)}]
\draw[->] (-0.075,0.129904) to[out=120,in=180] (0,1) to[out=0,in=60] (0.075,0.129904);
\node[above] at (0,1) {$A_3$};
\end{scope}
\begin{scope}[shift={(6,0)}]
\draw[->] (-0.075,0.129904) to[out=120,in=180] (0,1) to[out=0,in=60] (0.075,0.129904);
\node[above] at (0,1) {$A_5$};
\end{scope}
\begin{scope}[shift={(0,-2)},yscale=-1]
\draw[->] (-0.075,0.129904) to[out=120,in=180] (0,1) to[out=0,in=60] (0.075,0.129904);
\node[below] at (0,1) {$A_{12}$};
\end{scope}
\begin{scope}[shift={(7.5,-2)},yscale=-1]
\draw[->] (-0.075,0.129904) to[out=120,in=180] (0,1) to[out=0,in=60] (0.075,0.129904);
\node[below] at (0,1) {$A_{7}$};
\end{scope}
\end{tikzpicture}
\end{array}
$$

Let us remind a construction of periodic quiver $\hat\CQ$.

Consider all the consequent triples of nodes $(v-1,v,v+1)$. To each such triple based on values of $(\sigma_v,\sigma_{v+1})$ one assigns the following type of a tile and a superpotnetial term:
\be
\begin{array}{|l|c|l|}
	\hline
	(\sigma_v,\sigma_{v+1}) & \mbox{Tile} & W_{(v,v+1)}\\
	\hline
	(+1,+1)&
	\begin{array}{c}
		\begin{tikzpicture}
		\draw[fill=black] (-0.7,0) circle (0.06) (0.7,0) circle (0.06)  (0,0.7) circle (0.06)  (0,-0.7) circle (0.06);
		\draw[->] (-0.657574, 0.0424264) -- (-0.0424264, 0.657574);
		\draw[->] (0.657574, 0.0424264) -- (0.0424264, 0.657574);
		\draw[<-] (-0.657574, -0.0424264) -- (-0.0424264, -0.657574);
		\draw[<-] (0.657574, -0.0424264) -- (0.0424264, -0.657574);
		\draw[->] (0,0.64) -- (0,-0.64);
		\node[left] at (-0.7,0) {$v-1$};
		\node[right] at (0.7,0) {$v+1$};
		\node[above] at (0,0.7) {$v$};
		\node[below] at (0,-0.7) {$v$};
		\node[above left] at (-0.35,0.35) {$B_{v-1}$}; 
		\node[above right] at (0.35,0.35) {$C_v$};
		\node[below left] at (-0.35,-0.35) {$C_{v-1}$}; 
		\node[below right] at (0.35,-0.35) {$B_v$};
		\node[right] at (-0.1,0) {$A_v$};
		\end{tikzpicture}
	\end{array}& \Tr\left[C_vB_vA_v-A_vB_{v-1}C_{v-1}\right]\\
	\hline
	(-1,-1)&
	\begin{array}{c}
		\begin{tikzpicture}
		\draw[fill=black] (-0.7,0) circle (0.06) (0.7,0) circle (0.06)  (0,0.7) circle (0.06)  (0,-0.7) circle (0.06);
		\draw[<-] (-0.657574, 0.0424264) -- (-0.0424264, 0.657574);
		\draw[<-] (0.657574, 0.0424264) -- (0.0424264, 0.657574);
		\draw[->] (-0.657574, -0.0424264) -- (-0.0424264, -0.657574);
		\draw[->] (0.657574, -0.0424264) -- (0.0424264, -0.657574);
		\draw[<-] (0,0.64) -- (0,-0.64);
		\node[left] at (-0.7,0) {$v-1$};
		\node[right] at (0.7,0) {$v+1$};
		\node[above] at (0,0.7) {$v$};
		\node[below] at (0,-0.7) {$v$};
		\node[above left] at (-0.35,0.35) {$C_{v-1}$}; 
		\node[above right] at (0.35,0.35) {$B_v$};
		\node[below left] at (-0.35,-0.35) {$B_{v-1}$}; 
		\node[below right] at (0.35,-0.35) {$C_v$};
		\node[right] at (-0.1,0) {$A_v$};
		\end{tikzpicture}
	\end{array}&  \Tr\left[-C_vB_vA_v+A_vB_{v-1}C_{v-1}\right]\\
	\hline
	(+1,-1)&
	\begin{array}{c}
		\begin{tikzpicture}
		\draw[fill=black] (-0.7,0) circle (0.06) (0.7,0) circle (0.06)  (0,0.7) circle (0.06)  (0,-0.7) circle (0.06);
		\draw[->] (-0.657574, 0.0424264) -- (-0.0424264, 0.657574);
		\draw[<-] (0.657574, 0.0424264) -- (0.0424264, 0.657574);
		\draw[<-] (-0.657574, -0.0424264) -- (-0.0424264, -0.657574);
		\draw[->] (0.657574, -0.0424264) -- (0.0424264, -0.657574);
		\node[left] at (-0.7,0) {$v-1$};
		\node[right] at (0.7,0) {$v+1$};
		\node[above] at (0,0.7) {$v$};
		\node[below] at (0,-0.7) {$v$};
		\node[above left] at (-0.35,0.35) {$B_{v-1}$}; 
		\node[above right] at (0.35,0.35) {$B_v$};
		\node[below left] at (-0.35,-0.35) {$C_{v-1}$}; 
		\node[below right] at (0.35,-0.35) {$C_v$};
		\end{tikzpicture}
	\end{array}& \Tr\left[-C_vB_vB_{v-1}C_{v-1}\right]\\
	\hline
	(-1,+1)&
	\begin{array}{c}
		\begin{tikzpicture}
		\draw[fill=black] (-0.7,0) circle (0.06) (0.7,0) circle (0.06)  (0,0.7) circle (0.06)  (0,-0.7) circle (0.06);
		\draw[<-] (-0.657574, 0.0424264) -- (-0.0424264, 0.657574);
		\draw[->] (0.657574, 0.0424264) -- (0.0424264, 0.657574);
		\draw[->] (-0.657574, -0.0424264) -- (-0.0424264, -0.657574);
		\draw[<-] (0.657574, -0.0424264) -- (0.0424264, -0.657574);
		\node[left] at (-0.7,0) {$v-1$};
		\node[right] at (0.7,0) {$v+1$};
		\node[above] at (0,0.7) {$v$};
		\node[below] at (0,-0.7) {$v$};
		\node[above left] at (-0.35,0.35) {$C_{v-1}$}; 
		\node[above right] at (0.35,0.35) {$C_v$};
		\node[below left] at (-0.35,-0.35) {$B_{v-1}$}; 
		\node[below right] at (0.35,-0.35) {$B_v$};
		\end{tikzpicture}
	\end{array}& \Tr\left[C_vB_vB_{v-1}C_{v-1}\right]\\
	\hline
\end{array}
\ee
The quiver superpotential $\W$ is constructed from tile superpotentials:
\be
\W=\sum\lm_{v=1}^{m+n}W_{(v,v+1)} \;.
\ee

We combine tiles in a parquet flooring of a torus universal cover $\IR^2$ according to a ``stacking" rule:
\bea\nn
\begin{array}{c}
	\begin{tikzpicture}
	\draw[fill=gray] (-0.5,0) -- (0,0.5) -- (0.5,0) -- (0,-0.5) -- cycle;
	\draw[fill=black] (-0.5,0) circle (0.05) (0,0.5) circle (0.05) (0.5,0) circle (0.05) (0,-0.5) circle (0.05);
	\node[left] at (-0.5,0) {$v-1$};
	\node[above] at (0,0.5) {$v$};
	\node[right] at (0.5,0) {$v+1$};
	\node[below] at (0,-0.5) {$v$};
	\begin{scope}[shift={(4,0.5)}]
	\draw[fill=gray] (-0.5,0) -- (0,0.5) -- (0.5,0) -- (0,-0.5) -- cycle;
	\draw[fill=black] (-0.5,0) circle (0.05) (0,0.5) circle (0.05) (0.5,0) circle (0.05) (0,-0.5) circle (0.05);
	\node[left] at (-0.5,0) {$v$};
	\node[above] at (0,0.5) {$v+1$};
	\node[right] at (0.5,0) {$v+2$};
	\node[right] at (0,-0.5) {$v+1$};
	\end{scope}
	\begin{scope}[shift={(4,-0.5)}]
	\draw[fill=gray] (-0.5,0) -- (0,0.5) -- (0.5,0) -- (0,-0.5) -- cycle;
	\draw[fill=black] (-0.5,0) circle (0.05) (0,0.5) circle (0.05) (0.5,0) circle (0.05) (0,-0.5) circle (0.05);
	\node[left] at (-0.5,0) {$v$};
	\node[right] at (0.5,0) {$v+2$};
	\node[below] at (0,-0.5) {$v+1$};
	\end{scope}
	\draw[->] (2,0) -- (3.3,0);
	\end{tikzpicture}
\end{array}
\eea
Obviously, under such tile matching the edge orientations will match as well. The parquet pattern is evidently periodic with respect to vertical and horizontal shifts, therefore the quiver we draw is a periodic quiver on a torus.

\begin{figure}[ht!]
	\begin{center}
		\scalebox{0.9}{
		\begin{tikzpicture}
		\node at (-2.2,5){\includegraphics[trim=0 0 0 0,clip,scale=0.3]{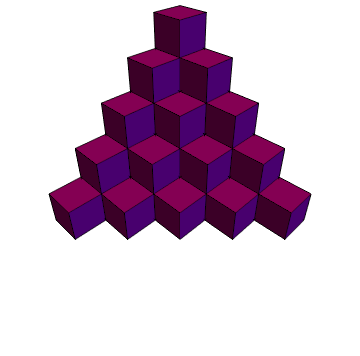}};
		\node at (1.8,5){\includegraphics[trim=0 0 0 0,clip,scale=0.4]{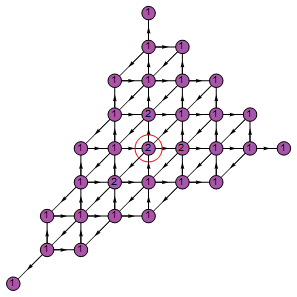}};
		\node at (7,5){\includegraphics[trim=0 0 0 0,clip,scale=0.2]{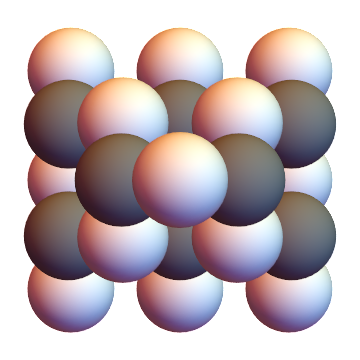}};
		\node at (10,5){\includegraphics[trim=0 0 0 0,clip,scale=0.5]{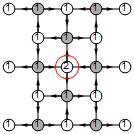}};
		\node{\includegraphics[trim=0 50 0 140,clip,scale=0.4]{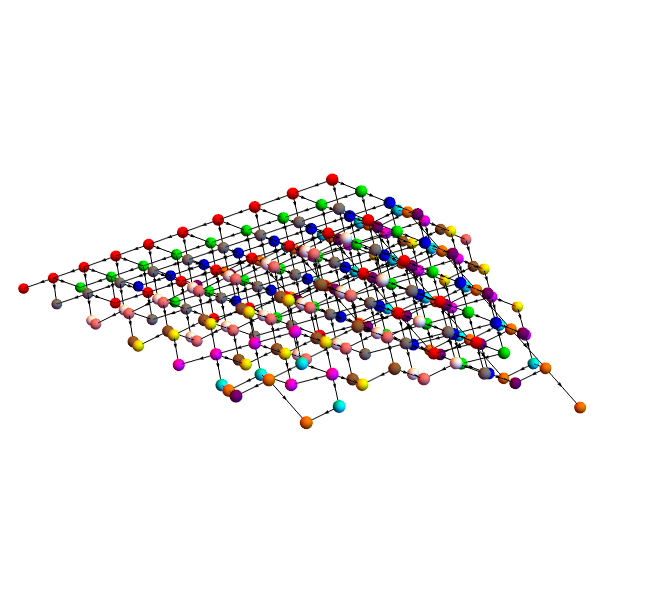}};
		\node at (7.5,0) {\includegraphics[trim=0 0 0 0,clip,scale=0.4]{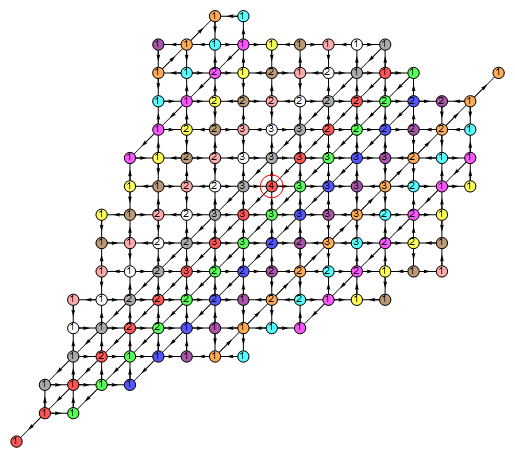}};
		\end{tikzpicture}}
	\end{center}
	\caption{Examples of crystals for $\Sigma_{1,0}$, $\Sigma_{1,1}$, $\Sigma_{8,4}$: 3d arrangement and 2d schematic depiction.}\label{fig:cry_ex}
\end{figure}

One constructs corresponding 3d crystals as it is described in Section \ref{s:CY3}.  We will also adopt notations for crystals as a 2d crystal projection with integer numbers within atoms. Such an atom diagram in position $\vec r$ with integer number $k$ denotes that all vacant atom positions with coordinates:
$$
(\vec r,0) \;,\quad (\vec r,1) \;, \quad \dots  \quad (\vec r, k)
$$
are occupied. Some examples of plane partition lattice, 
pyramid partition lattice and more general crystal drawings are presented in Fig. \ref{fig:cry_ex}.

Let us choose a parameterization for the equivariant weights of of chiral fields satisfying the gauge invariance condition for the superpotential in the following way:
\be
\epsilon_{A_v}=\frac{\sigma_v+\sigma_{v+1}}{2}\hbar_{m+n+1}\;,
\quad 
\epsilon_{B_v}=\hbar_v\;,
\quad 
\epsilon_{C_v}=-\sigma_{v+1}\hbar_{m+n+1}-\hbar_v \;.
\ee

A set of arrows flowing from a quiver vertex $v$ to a quiver vertex $w$ we denoted as $(a \colon v\to w)$. Then the total number of arrows flowing from $v$ to $w$ reads $|a\colon v\to w|$.

For a pair of atoms $\Box_1$ and $\Box_2$ we introduce two statistical factors, functions $\nu_{\pm}$ satisfying \eqref{stat}.
A particular choice of statistical factor functions $\nu_{\pm}$ does not matter. For practical use we can specify it in the following way. Let us order all the atoms in the crystal lattice. We can do this since all the atoms in the lattice represent words in an alphabet formed by letters associated to quiver maps $q_a$. Therefore we can order all the atoms in the crystal as words in a lexicographical order. Then introduce a function
$$
{\rm Sign}(\Box_1,\Box_2)
$$
that returns value $+1$ if a sequence $\{ \Box_1,\Box_2 \}$ has the straight order, and $-1$ otherwise. Then define the statistical factor as:
\be\label{sgn}
\nu_+(\Box_1,\Box_2)\coloneqq (-1)^{\left(|\hat \Box_1||\hat \Box_2|+\delta_{\hat \Box_1,\hat \Box_2}+|a:\;\hat\Box_2\to \hat\Box_1|\right)\frac{1+{\rm Sign}(\Box_1,\Box_2)}{2}}  \;.
\ee

Define Hecke shift generators:
\bea
\begin{split}
	e^{(v)}(z)|\Lambda\rangle=\sum\lm_{\Box\in \Lambda^{+(v)}}\frac{E(\Lambda\to \Lambda+\Box)}{z-\bphi_{\Box}}|\Lambda+\Box\rangle \;,\\
	f^{(v)}(z)|\Lambda\rangle=\sum\lm_{\Box\in \Lambda^{-(v)}}\frac{F( \Lambda\to \Lambda-\Box)}{z-\bphi_{\Box}}|\Lambda-\Box\rangle \;,
\end{split}
\eea
with additional sign shifts:
\bea
\begin{split}
	E(\Lambda\to \Lambda+\Box)=\frac{\widetilde{\rm Eul}(\Lambda)}{\widetilde{\rm Eul}(\Lambda,\Lambda+\Box)}\prod\lm_{\Box'\in\Lambda}\nu_+(\Box',\Box) \;,\\  F( \Lambda+\Box\to \Lambda)=\frac{\widetilde{\rm Eul}(\Lambda)}{\widetilde{\rm Eul}(\Lambda+\Box,\Lambda)}\prod\lm_{\Box'\in\Lambda}\nu_-(\Box',\Box) \;.
\end{split}
\eea
Using this definition we can show in a series of numerical experiments these generators satisfy the following set of relations \cite{Li:2020rij}:
\bea\label{EF}
\begin{split}
&E(\Lambda\to\Lambda+\Box)F(\Lambda+\Box\to \Lambda)=\mathop{\rm res}\lm_{z=p_{\Box}}\psi_{\Lambda}^{(\hat \Box)}(z)=-(-1)^{|\hat \Box|}\mathop{\rm res}\lm_{z=p_{\Box}}\psi_{\Lambda+\Box}^{(\hat \Box)}(z) \;, \\
&\frac{E(\Lambda+\Box_1\to\Lambda+\Box_1+\Box_2)F(\Lambda+\Box_1+\Box_2\to\Lambda+\Box_1)}{F(\Lambda+\Box_1\to \Lambda)E(\Lambda\to\Lambda+\Box_2)}=(-1)^{|\hat\Box_1||\hat\Box_2|} \;, \\
	&\frac{E(\Lambda\to\Lambda+\Box_1)E(\Lambda+\Box_1\to\Lambda+\Box_1+\Box_2)}{E(\Lambda\to\Lambda+\Box_2)E(\Lambda+\Box_2\to\Lambda+\Box_1+\Box_2)}\varphi_{\hat\Box_1,\hat\Box_2}\left(\bphi_{\Box_1}-\bphi_{\Box_2}\right)=(-1)^{|\hat\Box_1||\hat\Box_2|} \;, \\
	&\frac{F(\Lambda+\Box_1+\Box_2\to\Lambda+\Box_1)F(\Lambda+\Box_1\to\Lambda)}{F(\Lambda+\Box_1+\Box_2\to\Lambda+\Box_2)F(\Lambda+\Box_2\to\Lambda)}\varphi_{\hat\Box_1,\hat\Box_2}\left(\bphi_{\Box_1}-\bphi_{\Box_2}\right)=(-1)^{|\hat\Box_1||\hat\Box_2|} \;,
\end{split}
\eea
where (f.n.\ denotes the framed node)
\begin{align}
\psi_{\Lambda}^{(v)}(z)& \coloneqq \left(\frac{1}{z}\right)^{\delta_{{\rm f.n.},v}}\left(\prod\lm_{(a:\;v\to v)\in\CA}-\frac{1}{\epsilon_a}\right)\prod\lm_{\Box\in\Lambda}\varphi_{v,\hat\Box}(z-\bphi_{\Box})  \;, \\
\varphi_{v,w}(z) & \coloneqq \frac{\prod\lm_{(a\colon v\to w)\in\CA}(z+\epsilon_a)}{\prod\lm_{(b:\;w\to v)\in\CA}(z-\epsilon_b)}  \;.
\end{align}

Then we can restore the algebra OPE satisfied by generators $e^{(a)}$, $f^{(a)}$ and $\psi^{(a)}$ \cite{Li:2020rij}:
\bea
\begin{split}
	\left[ e^{(a)}(z),f^{(b)}(w) \right\}&\sim\delta_{a,b}\frac{\psi^{(a)}(z)-\psi^{(a)}(w)}{z-w}  \;,\\
	\psi^{(a)}(z)e^{(b)}(w)&\simeq\varphi_{a,b}(z-w)e^{(b)}(w)\psi^{(a)}(z)\;,\\\
	\psi^{(a)}(z)f^{(b)}(w)&\simeq\left[\varphi_{a,b}(z-w)\right]^{-1}f^{(b)}(w)\psi^{(a)}(z)\;,\\\
	e^{(a)}(z)e^{(b)}(w)&\sim(-1)^{|a||b|}\varphi_{a,b}(z-w)e^{(b)}(w)\psi^{(a)}(z)\;,\\\
	f^{(a)}(z)f^{(b)}(w)&\sim(-1)^{|a||b|}\left[\varphi_{a,b}(z-w)\right]^{-1}f^{(b)}(w)f^{(a)}(z) \;,\\
\end{split}
\eea
where $\sim$ ($\simeq$) imply that both sides coincide in expansion in $z$ and $w$ up to monomials $z^j w^{k\ge 0}$ and $z^{j\ge 0} w^{k}$  ($z^{j\ge 0} w^{k\ge 0}$).

For $m+n\geq 3$, $mn\neq 2$ we check numerically generators satisfy higher order Serre relations:
\bea
\begin{split}
	{\rm Sym}_{z_1,z_2}\left[e^{(a)}(z_1),\left[e^{(a)}(z_2),e^{(a+1)}(w)\right]\right]&=0,\;{\rm if}\; |a|=0\;,\\\
	{\rm Sym}_{z_1,z_2}\left[f^{(a)}(z_1),\left[f^{(a)}(z_2),f^{(a+1)}(w)\right]\right]&=0,\;{\rm if}\; |a|=0\;,\\\
	{\rm Sym}_{z_1,z_2}\left[e^{(a)}(z_1),\left[e^{(a+1)}(w_1),\left[e^{(a)}(z_2),e^{(a-1)}(w_2)\right\} \right\} \right\}&=0,\; {\rm if}\; |a|=1\;,\\\
	{\rm Sym}_{z_1,z_2}\left[f^{(a)}(z_1),\left[f^{(a+1)}(w_1),\left[f^{(a)}(z_2),f^{(a-1)}(w_2)\right\} \right\} \right\}&=0,\; {\rm if}\; |a|=1\;.\\
\end{split}
\eea

These OPE and Serre relations are for  $Y(\widehat{\fg\fl}_{m|n})$ and are a $q\to 1$ reduction of analogous relations for toroidal $\fg\fl_{m|n}$ \cite{Bezerra:2019dmp}. If we did not introduce sign shifts by setting $\nu_{\pm}=1$ the resulting generators $\hat e$, $\hat f$ and $\hat \psi$ will also form an algebra \eqref{true_OPE}, however this algebra will not be isomorphic to $Y(\widehat{\fg\fl}_{m|n})$ for generic $m$ and $n$. In particular, fermionic and bosonic generators will lose their parity specifications. This have already happened in the case of $Y(\widehat{\fg\fl}_{m|n})$ \eqref{e_f_gl_1_1} where all generators are fermionic, however the Cartan generators appear in the commutator of $e$ and $f$ rather than the anti-commutator. We consider an example of parity correction effect of $\nu_{\pm}$ in Appendix \ref{s:parity}.

%%%%%%%---------------------------------------------------------------------------------------------------------------------------------------------------------------------------------------------
\subsection{Manifolds with a Compact 4-cycle}

We next turn to examples with compact $4$-cycles. While our method works for 
a general toric Calabi-Yau geometry, let us here work out two illustrative examples,
canonical bundles over $\mathbb{P}^2$ and $\mathbb{P}^1\times \mathbb{P}^1$; both 
geometries have one compact $4$-cycle.

\subsubsection{Canonical Bundle over $\IP^2$}

The geometry of the canonical bundle $K_{\IP^2}$, which coincides with an orbifold $\mathbb{C}^3/\mathbb{Z}_3$, 
is summarized in the following quiver diagram with a superpotential:%\cite{Li:2020rij}:
\be
\begin{array}{c}
\begin{tikzpicture}
\draw[fill=red] (-1,0) circle (0.15); 
\draw[fill=blue] (1,0) circle (0.15);
\draw[fill=green] (0,1.73205) circle (0.15);
\draw[->] (0.85,0) -- (-0.85,0);
\draw[->] (0.85,0) -- (-0.7,0);
\draw[->] (0.85,0) -- (-0.55,0);
\begin{scope}[shift={(0,0.57735)}]
\begin{scope}[rotate=120]
\begin{scope}[shift={(0,-0.57735)}]
\draw[->] (0.85,0) -- (-0.85,0);
\draw[->] (0.85,0) -- (-0.7,0);
\draw[->] (0.85,0) -- (-0.55,0);
\end{scope}
\end{scope}
\end{scope}
\begin{scope}[shift={(0,0.57735)}]
\begin{scope}[rotate=240]
\begin{scope}[shift={(0,-0.57735)}]
\draw[->] (0.85,0) -- (-0.85,0);
\draw[->] (0.85,0) -- (-0.7,0);
\draw[->] (0.85,0) -- (-0.55,0);
\end{scope}
\end{scope}
\end{scope}
\node[below left] at (-1.10607, -0.106066) {$1$};
\node[below right] at (1.10607, -0.106066) {$3$};
\node[above] at (0,1.88205) {$2$};
\node[below] at (0,0) {$(A_i^{(3)},\alpha_i^{(3)})$};
\node[above left] at (-0.5,0.866025) {$(A_i^{(1)},\alpha_i^{(1)})$};
\node[above right] at (0.5,0.866025) {$(A_i^{(2)},\alpha_i^{(2)})$};
\draw[->] (-2,0) -- (-1.15,0);
\begin{scope}[shift={(-2,0)}]
\draw[fill=white] (-0.1,-0.1) -- (-0.1,0.1) -- (0.1,0.1) -- (0.1,-0.1) -- cycle;
\end{scope}
\end{tikzpicture}
\end{array},\quad W=\sum\lm_{i,j,k=1}^{3}\varepsilon^{ijk}\Tr\left(A_i^{(1)}A_j^{(2)}A_k^{(3)}\right) \;.
\ee
To each triple arrow a triplet of maps $A_i^{(v)}$, $i=1,2,3$ and a triplet of equivariant weights $\alpha^{(v)}_i$, $i=1,2,3$ is associated. 

The constraint \eqref{W-inv} in this case is translated in the following one:
\be
\alpha_i^{(1)}+\alpha_j^{(2)}+\alpha_k^{(3)}=0, \quad \mbox{for any choice }\{i,j,k\}\in\{1,2,3\}\;.
\ee

\begin{figure}[!ht]
	\begin{center}
	\includegraphics[scale=0.7]{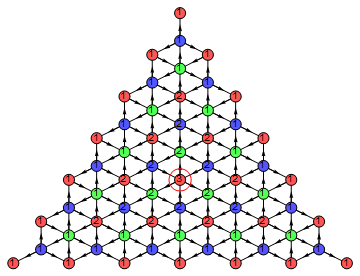};
	\end{center}
	\caption{Example of a crystal for $K_{\IP^2}$}\label{fig:K_P_2_cry}
\end{figure}

To construct a periodic quiver one assigns to maps $A_i^{(v)}$ the following 2d vectors:
$$
A^{(v)}_1\rightsquigarrow\vec x\;,\quad A^{(v)}_2\rightsquigarrow\vec y \;,\quad A^{(v)}_3\rightsquigarrow\vec x +\vec y \;.
$$
Crystal lattice in this case is a cubic lattice corresponding to plane partitions with colored planes of atoms, and its 2d projection is a parquet tessellation by triangular tiles that is easy to depict by hexagonal Bravais lattice, see an example of a crystal in Fig.~\ref{fig:K_P_2_cry}.

\subsubsection{Canonical Bundle over $\IP^1\times \IP^1$.}

The geometry $K_{\IP^1\times\IP^1}$ is summarized in the following quiver diagram with a superpotential: %\cite{Li:2020rij}:
\be
\begin{array}{c}
\begin{tikzpicture}
\draw[fill=red] (-0.8,-0.8) circle (0.15);
\draw[fill=green] (-0.8,0.8) circle (0.15);
\draw[fill=blue] (0.8,0.8) circle (0.15);
\draw[fill=orange] (0.8,-0.8) circle (0.15);
\draw[->] (0.670096, -0.725) to[out=175,in=5] (-0.670096, -0.725);
\draw[->] (0.670096, -0.875) to[out=185,in=355] (-0.670096, -0.875);
\begin{scope}[rotate=90]
\draw[->] (0.670096, -0.725) to[out=175,in=5] (-0.670096, -0.725);
\draw[->] (0.670096, -0.875) to[out=185,in=355] (-0.670096, -0.875);
\end{scope}
\begin{scope}[rotate=180]
\draw[->] (0.670096, -0.725) to[out=175,in=5] (-0.670096, -0.725);
\draw[->] (0.670096, -0.875) to[out=185,in=355] (-0.670096, -0.875);
\end{scope}
\begin{scope}[rotate=270]
\draw[->] (0.670096, -0.725) to[out=175,in=5] (-0.670096, -0.725);
\draw[->] (0.670096, -0.875) to[out=185,in=355] (-0.670096, -0.875);
\end{scope}
\node[below] at (0,-0.9) {$(A_i^{(4)},\alpha_i^{(4)})$};
\node[left] at (-0.9,0) {$(A_i^{(1)},\alpha_i^{(1)})$};
\node[above] at (0,0.9) {$(A_i^{(2)},\alpha_i^{(2)})$};
\node[right] at (0.9,0) {$(A_i^{(3)},\alpha_i^{(3)})$};
\node[left] at (-0.95,-0.8) {$1$};
\node[left] at (-0.95,0.8) {$2$};
\node[right] at (0.95,0.8) {$3$};
\node[right] at (0.95,-0.8) {$4$};
\end{tikzpicture}
\end{array},\begin{array}{r}
W=\Tr\Big( A^{(4)}_1 A^{(3)}_1 A^{(2)}_1 A^{(1)}_1+\\
+A^{(4)}_2 A^{(3)}_2 A^{(2)}_2 A^{(1)}_2-\\
-A^{(4)}_1 A^{(3)}_2 A^{(2)}_1 A^{(1)}_2-\\
-A^{(4)}_2 A^{(3)}_1 A^{(2)}_2 A^{(1)}_1\Big)
\end{array}
\ee

\begin{figure}[ht!]
	\begin{center}
		\includegraphics[scale=0.7]{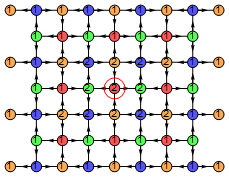}
	\end{center}
	\caption{Example of a crystal for $K_{\IP^1\times \IP^1}$}
	\label{fig:K_P_1_P_1_cry}
\end{figure}

To two arrows flowing from node $v\in\CV$ one associates two maps $A_i^{(v)}$, $i=1,2$ and two equivariant weights $\alpha_i^{(v)}$, $i=1,2$. The 3d crystal represents just a 4-colored pyramid partition, see an example of a crystal in Fig.~\ref{fig:K_P_1_P_1_cry}.

The constraints for the equivariant weights \eqref{W-inv} are the following:
\be
\alpha_1^{(1)}+\alpha_1^{(3)}= \alpha_2^{(1)}+\alpha_2^{(3)}=-\left(\alpha_1^{(2)}+\alpha_1^{(4)}\right)=-\left(\alpha_2^{(2)}+\alpha_2^{(4)}\right)\;.
\ee

\subsubsection{OPE Relations}

Again BPS Hecke operators are constructed in the same fashion as explained in Section~\ref{s:Hecke} with regularized Euler characters \eqref{res_Eul}:
\bea
\begin{split}
	\hat e^{(v)}(z)|\Lambda\rangle&=\sum\lm_{\Box\in \Lambda^{+(v)}}\frac{1}{z-\bphi_{\Box}}\times\frac{\widetilde{\rm Eul}(\Lambda)}{\widetilde{\rm Eul}(\Lambda,\Lambda+\Box)}|\Lambda+\Box\rangle \;,\\
	\hat f^{(v)}(z)|\Lambda\rangle&=\sum\lm_{\Box\in \Lambda^{-(v)}}\frac{1}{z-\bphi_{\Box}}\times\frac{\widetilde{\rm Eul}(\Lambda)}{\widetilde{\rm Eul}(\Lambda-\Box,\Lambda)}|\Lambda-\Box\rangle \;.
\end{split}
\eea

In both cases the resulting generators fall illustratively in a set of generic examples \eqref{OPE0} \cite{Li:2020rij}:
\bea\label{true_OPE}
\begin{split}
\left[\hat e^{(v)}(x),\hat f^{(w)}(y)\right]&\sim\delta_{vw}\frac{\hat \psi^{(v)}(x)-\hat\psi^{(w)}(y)}{x-y} \;,\\
\hat\psi^{(v)}(x)\hat e^{(w)}(y)&\simeq\hat \varphi_{v,w}(x-y)\hat e^{(w)}(y)\hat \psi^{(v)}(x)\;,\\
\hat \psi^{(v)}(x)\hat f^{(w)}(y)&\simeq\left[\hat\varphi_{v,w}(x-y)\right]^{-1}\hat f^{(w)}(y)\hat \psi^{(v)}(x)\;,\\
\hat e^{(v)}(x)\hat e^{(w)}(y)&\sim\hat\varphi_{v,w}(x-y)\hat e^{(w)}(y)\hat e^{(v)}(x)\;,\\
\hat f^{(v)}(x)\hat f^{(w)}(y)&\sim\left[\hat\varphi_{v,w}(x-y)\right]^{-1}\hat f^{(w)}(y)\hat f^{(v)}(x) \;,\\
\end{split}
\eea
with the binding factor $\hat\varphi$ given by \eqref{psihat}.

For a crystal representation the eigenvalues of the Cartan elements read \cite{Li:2020rij}:
\be\label{psi}
\hat\psi^{(v)}_{\Lambda}(z)=\left(\frac{1}{z}\right)^{\delta_{{\rm f.n.},v}}\left(\prod\lm_{(a:\;v\to v)\in\CA}-\frac{1}{\epsilon_a}\right)\prod\lm_{\Box\in\Lambda}\hat\varphi_{v,\hat\Box}(z-\bphi_{\Box}) \;.
\ee

%%%%%%%%%%%%%%%%%%%%%%%%%%%%%%%%%%%%%%%%%%%%%%%%%%%%%%%%%%%%%%%%%%%%%%%%%%%%%%%%%%%%
\section{Summary and Discussion}
%%%%%%%%%%%%%%%%%%%%%%%%%%%%%%%%%%%%%%%%%%%%%%%%%%%%%%%%%%%%%%%%%%%%%%%%%%%%%%%%%%%%

In this paper we physically identified the BPS state algebra
associated with D-branes wrapping holomorphic cycles in a non-compact toric Calabi-Yau three-fold.
Starting with the effective SQM on the D-branes,
we applied supersymmetric localization onto the Higgs branch,
and extracted the representations of the BPS operators by 
working out the expressions for the Hecke shift operators,
and hence the representations of the BPS state algebra.
We have checked (by explicit computations for many examples)
that the representations of the BPS algebras in this paper coincide with the 
representations of the quiver Yangian introduced previously in \cite{Li:2020rij}.

There are clearly many open questions. Many of the questions raised in Section 10 of \cite{Li:2020rij} are relevant here.
For example, while in this paper we focused on a particular chamber of the moduli space,
we could discuss wall crossing phenomena by varying the stability parameters 
(see \cite{Young:2008hn,MR2836398,MR2999994,Jafferis:2008uf,Chuang:2008aw,Nagao:2009ky,Nagao:2009rq,Sulkowski:2009rw,Aganagic:2010qr,Yamazaki:2010fz} for discussion of wall crossing phenomena in 
terms of crystal melting).  We should note, however, that our Higgs branch localization requires a large value of the stability parameter as in \eqref{hierarchy}
and does not necessarily apply in a general chamber.

We can also discuss truncations of the algebra. Such truncations was discussed for the affine Yangian of $\mathfrak{gl}_1$ in \cite{Fukuda:2015ura, Prochazka:2015deb, Gaiotto:2017euk, Harada:2018bkb},
and related truncations of $\mathcal{W}$ algebras appear in \cite{Prochazka:2017qum, Rapcak:2019wzw}. The truncations of general quiver Yangians was
discussed in \cite{Li:2020rij}, and the corresponding discussion in the spirit of the paper would be 
an analogue for the discussion of $\mathbb{C}^2$ in Section \ref{s:gl2} (which is a truncation of $\mathbb{C}^3$ discussed in Section \ref{s:C3}).

One of the most pressing questions is to prove in general the 
consistency with the results of this paper with the representation of the quiver Yangian 
constructed in \cite{Li:2020rij}. While (as we reported in Section \ref{s:Examples})
we have done extensive checks of this consistency with 
the help of a computer program, we do not have a general
proof of applicable to a general toric Calabi-Yau manifold.
In practice this amounts to proving expressions 
for the matrix elements (as in \eqref{e_z_Fock} for $\mathbb{C}^2$),
and proving associated combinatorial identities as demonstrated 
in Appendix \ref{s:calc}.

Finally, one speculates that the methods of this paper is applicable to 
more general geometries beyond non-compact toric Calabi-Yau manifolds.
One hopes that a suitable SQM will still be an effective theory on the D-brane worldvolume,
from which one should in principle be able to extract the BPS state algebra. 

%%%%%%%%%%%%%%%%%%%%%%%%%%%%%%%%%%%%%%%%%%%%%%%%%%%%%%%%%%%%%%%%%%%%%%%%%%%%%%%%%%%%
\section*{Acknowledgements}
%%%%%%%%%%%%%%%%%%%%%%%%%%%%%%%%%%%%%%%%%%%%%%%%%%%%%%%%%%%%%%%%%%%%%%%%%%%%%%%%%%%%

We would like to thank Wei Li, Andrei Mironov, Dinakar Muthiah, Hiraku Nakajima, Boris Pioline, Alexei Sleptsov, Andrei Smirnov, Yaping Yang, Gufang Zhao and Yegor Zenkevich for stimulating discussion. DG would like to thank Yaping Yang and Gufang Zhao for sharing a draft of their work \cite{RSYZ} prior to publication. 
This research was supported in part by WPI Research Center Initiative, MEXT, Japan. MY was also supported by the JSPS Grant-in-Aid for Scientific Research (No.\ 17KK0087, No.\ 19K03820 and No.\ 19H00689). DG would like to thank  Eidgen\"ossische Technische Hochschule Z\"urich, Switzerland; Laboratoire de Physique Th\'eorique et Hautes Energies, Sorbonne Universit\'e et CNRS, France; Moscow Institute of Physics and Technology (MIPT), Russia, for generous hospitality during the work on this project, the research activity at MIPT was supported in part by RFBR grants 18-31-20046 mol-a-ved, 19-51-53014 GFEN-a, 19-01-00680 A.

%%%%%%%%%%%%%%%%%%%%%%%%%%%%%%%%%%%%%%%%%%%%%%%%%%%%%%%%%%%%%%%%%%%%%%%%%%%%%%%%%%%%
%%%%%%%%%%%%%%%%%%%%%%%%%%%%%%%%%%%%%%%%%%%%%%%%%%%%%%%%%%%%%%%%%%%%%%%%%%%%%%%%%%%%
\appendix
%%%%%%%%%%%%%%%%%%%%%%%%%%%%%%%%%%%%%%%%%%%%%%%%%%%%%%%%%%%%%%%%%%%%%%%%%%%%%%%%%%%%
%%%%%%%%%%%%%%%%%%%%%%%%%%%%%%%%%%%%%%%%%%%%%%%%%%%%%%%%%%%%%%%%%%%%%%%%%%%%%%%%%%%%

%%%%%%%%%%%%%%%%%%%%%%%%%%%%%%%%%%%%%%%%%%%%%%%%%%%%%%%%%%%%%%%%%%%%%%%%%%%%%%%%%%%%
\section{\texorpdfstring{$\CN=4$ 1d Gauged Linear Sigma Model}{N=4 1d Gauged Linear Sigma Model}}
\label{s:SQM}
%%%%%%%%%%%%%%%%%%%%%%%%%%%%%%%%%%%%%%%%%%%%%%%%%%%%%%%%%%%%%%%%%%%%%%%%%%%%%%%%%%%%

In this Appendix we summarize some basic results on $\CN=4$ one-dimensional gauged SQM following \cite{Denef:2002ru,Ohta:2014ria}.

The Lagrangian for the $U(N)$ $\CN=4$ SQM reads:
\be
L_{G}=\Tr\left[\frac{1}{2}(\calD_0X^i)^2+\frac{1}{4}\left[X^i, X^j \right]^2-\I \bar\lambda\bar\sigma^0\calD_0 \lambda+\bar\lambda\bar\sigma^i\left[X^i,\lambda \right] +\frac{1}{2}D^2-\theta D  \right]  \;,
\ee
where the covariant derivative is defined as $\calD_0 X=\p_0X+\I\left[A,X \right]$
%\be
%\calD_0 X=\p_0X+\I\left[A,X \right]  \;,
%\ee
and 
we denoted an $\mathcal{N}=4$ vector multiplet as $(A, X^{1,2,3}, \lambda, D)$.
In the following we combine two of the adjoint scalars into a complex field
\be
\Phi \coloneqq X^1+\I X^2 \;.
\ee

This Lagrangian is invariant up to total derivatives under the following SUSY transformations:
\bea\label{SUSY}
\begin{split}
	\delta A&=-\I \xi\sigma^0 \bar\lambda+\I  \lambda\sigma^0\bar\xi \;, \\
	\delta X^i&=\I \xi\sigma^i \bar\lambda-\I  \lambda\sigma^i\bar\xi \;, \\
	\delta \lambda&=\I\xi D+2\sigma^{0i}\xi \calD_0 X^i+\I \sigma^{ij} \xi\left[X^i,X^j \right] \;, \\
	\delta D&=-\xi\sigma^0\calD_0\bar\lambda-\I\xi\sigma^i\left[X^i,\bar\lambda \right] -\calD_0\lambda\sigma^0\bar\xi-\I \left[X^i,\lambda \right] \sigma^i \bar\xi 
  \;.
  \end{split}
\eea

The Lagrangian for a chiral multiplet $(q_a, \psi_a, F_a)$ in an irreducible representation $\rho$ of the gauge group reads:
\bea
\begin{split}
	L_{\chi}=\Tr\left[ |\calD_0 q_a|^2-|X^i\cdot q_a|^2 - \I \bar\psi_a\bar\sigma^0 \calD_0   \psi_a + \bar\psi_a \bar\sigma^i X^i\cdot \psi_a+|F_a|^2+\right.\\
	\left. +\I\sqrt{2}\bar q_a\lambda\cdot \psi_a+\bar q_a D \cdot q_a \right] \;.
\end{split}
\eea
The action of the gauge algebra element $\varphi\in\mathfrak{u}(n)$ on chiral fields we denoted as $\varphi \cdot q$.
The covariant derivative in this case corresponds to the following expression:
\be
\calD_0 q_a=\p_0 q_a+\I A\cdot q_a  \;,
\ee
and the SUSY transformations of the fields are:
\bea
\begin{split}
	\delta q_a&=\sqrt{2}\xi\psi_a \;,\\
	\delta\psi_a&=\I\sqrt{2}\left( \sigma^0 \bar\xi \calD_0 q_a+\I \sigma^i \bar\xi X^i\cdot  q_a \right)+\sqrt{2}\xi F_a\;,\\
	\delta F_a&=\I\sqrt{2}\left( \bar\xi\bar\sigma^0 \calD_0 \psi_a+\I \bar\xi \bar\sigma^i X^i\cdot \psi_a  \right)-2\I\bar\xi (\bar\lambda \cdot)^T q_a  \;.
\end{split}
\eea

The superpotential Lagrangian reads:
\be
L_W=\left(F_a \p_a W-\frac{1}{2}\p^2_{ab}W\psi_a\psi_b\right)+{\rm (c.c.)}  \;.
\ee

The total Lagrangian is a sum of all these Lagrangian pieces:
\be
L=L_G+L_{\chi}+L_W \;.
\ee

From the variation $\delta_G \CO=\frac{\I}{\hbar}\left[\hat G,\CO\right]$ we derive the following expressions for the supercharges generating the action of $\delta=\xi Q+\bar \xi \bar Q$:
\bea
\begin{split}
	Q_1={\rm Tr}\Bigg[\bar\lambda^{\dot 1}\left(D+\I \calD_0 X^3-\frac{1}{2}\left[\Phi,\bar \Phi\right]\right)+\bar\lambda^{\dot 2}\left(\I \calD_0 \bar \Phi-\left[\bar \Phi,X^3\right]\right)+\\
	+ \sqrt{2}\left(\calD_0\bar q+\I\bar q X^3\cdot \right)\psi_1+\I\sqrt{2}\bar q \bar \Phi\cdot \psi_2+\I\sqrt{2}\bar\psi_{\dot 2} F\Bigg]  \;,\\
	Q_2={\rm Tr}\Bigg[ \bar\lambda^{\dot 2}\left(D-\I\calD_0X^3+\frac{1}{2}\left[ \Phi,\bar \Phi \right] \right)+\bar\lambda^{\dot 1}\left(\I\calD_0 \Phi+\left[\Phi,X^3\right]\right) +\\
	+\sqrt{2}\left(\calD_0\bar q-\I\bar q X^3\cdot \right)\psi_2+\I\sqrt{2}\bar q \Phi\cdot  \psi_1-\I\sqrt{2}\bar\psi_{\dot 1} F  \Bigg]\;,\\
	\bar Q_{\dot 1}={\rm Tr}\Bigg[\left(D-\I \calD_0 X^3-\frac{1}{2}\left[\Phi,\bar \Phi\right]\right)\lambda^{1}-\left(\I \calD_0 \Phi-\left[\Phi,X^3\right]\right)\lambda^{2}+\\
	+ \sqrt{2}\bar\psi_{\dot 1}\left(\calD_0q-\I X^3\cdot q\right)-\I\sqrt{2}\bar\psi_{\dot 2} \Phi\cdot  q-\I\sqrt{2}\bar F \psi_2\Bigg]\;,\\
	\bar Q_{\dot 2}={\rm Tr}\Bigg[ \left(D+\I\calD_0X^3+\frac{1}{2}\left[ \Phi,\bar \Phi \right] \right)\lambda^{2}-\left(\I\calD_0\bar \Phi+\left[\bar \Phi,X^3\right]\right)\lambda^{1} +\\
	+\sqrt{2}\bar\psi_{\dot 2}\left(\calD_0 q+\I X^3\cdot q\right)-\I\sqrt{2}\bar\psi_{\dot 1}\bar \Phi \cdot q +\I\sqrt{2}\bar F \psi_1 \Bigg] \;,
\end{split}
\eea
where fields $D$ and $F$ are set to their expectation values:
\bea
\begin{split}
D=\theta\; \mathds{1}_{n\times n}-\sum\lm_a T_a \; q_a\bar q_a,\quad F_a=-\overline{\p_a W}\; ,
\end{split}
\eea
where $T_a$ is the charge of the field $q_a$ under the gauge algebra $\fu(n)$.

We can also derive the gauge current:
\bea
\begin{split}
	\hat G(\varphi)={\rm Tr}\Big(\I\left[\varphi,X^i \right]\calD_0 X^i+\bar\lambda\bar\sigma^0\left[\varphi, \lambda\right] 
	+\bar\psi_a\bar\sigma^0 \cdot \varphi \cdot \psi_a+\\
	+\I\left(\calD_0\bar q_a\cdot\varphi\cdot q_a-\bar q_a\cdot\varphi\cdot \calD_0q_a\right) \Big)  \;.
\end{split}
\eea

These supercharges satisfy the following commutation relations:
\begin{equation}\label{superalgebra}
\begin{split}
	\left\{Q_{\alpha},\bar Q_{\dot{\alpha}} \right\}&=-2\sigma^0_{\alpha\dot{\alpha}}H-2\I\sigma^{\mu}_{\alpha\dot \alpha}\hat G(X_{\mu}) \;,\\
	\left\{Q_{\alpha},Q_{\beta} \right\}&=8 \bar q_a \left[\left(\sigma^{0i}\right)_{\alpha}{}\!^{\gamma}\epsilon_{\gamma\beta}X^i\right]F_a  \;,
\end{split}
\end{equation}
where $H$ is the Hamiltonian of the system and we denoted $X^{\mu}=(A,X^1,X^2,X^3)$.

We would like to rewrite supercharges as differential forms on the SQM target space. Let us choose one of the four supercharges, say, $\bar Q_{\dot 1}$. One identifies fermion operators with operators acting on the space of differential forms on the target space:
\bea\label{dict}
\begin{split}
	&-\lambda^1\rightsquigarrow dX^3\wedge \;,\quad -\lambda^2\rightsquigarrow d\bar \Phi\wedge \;,\quad -\I\sqrt{2}\bar\psi_{\dot 1}\rightsquigarrow d\bar q\wedge \;,\quad \I\sqrt{2}\psi_2\rightsquigarrow dq\wedge \:,\\
	&\I\sqrt{2}\psi_1=\I\sqrt{2}\bar\psi_{\dot 1}^{\dagger}\rightsquigarrow \iota_{\p/\p \bar q}=\star \,d\bar q\wedge \star \;, \quad -\I\sqrt{2}\bar\psi_{\dot 2}=-\I\sqrt{2}\psi_2^{\dagger}\rightsquigarrow \iota_{\p/\p q}=\star \,d q\wedge\star \;,
\end{split}
\eea
where $\star$ is the Hodge star operator. In addition we have to assign some fermion vacuum state to a particular form, so that all others forms can be created from it by generator actions. Conventionally, one chooses a canonical fermion vacuum  to be annihilated by all fermion annihilation operators, and creation operators produce new states:
\bea\nn
\langle 0|0\rangle =1,\quad \psi_{1}|0\rangle=\psi_{ 2}|0\rangle=0\;.
\eea
However the form $1$ of degree 0 satisfies the following condition $\iota_V 1=0$. Using the above dictionary we find that the fermion vacuum assigned to $1$ is not the canonical one:
\be
\bar\psi_{\dot 2}|0\rangle \rightsquigarrow 1\;.
\ee
In these terms we can represent the supercharge $\bar Q_{\dot 1}$ as a conjugated differential operator:
\be
\bar Q_{\dot 1}=e^{-\fH}\left(d_{X^3}+\bar\p_{ \Phi,q}+\iota_V+dW\wedge\right)e^{\fH}  \;,
\ee
where we define the height function
\be
\fH \coloneqq \Tr \; X^3\left(\frac{1}{2}\left[\Phi,\bar \Phi\right]-D\right) \;,
\ee
and a vector field on chiral fields:
\be\label{VF}
V(q) \coloneqq \Phi\cdot q\; \frac{\p}{\p q}  \;.
\ee
In these terms the ground states are identified with gauge-invariant harmonic forms.

The SQM has two global symmetries: $\SU(2)_J$ covering $\SO(3)$ rotating compactified direction obtained from reduction of $\CN=1$ 4d SYM to $\CN=4$ 1d SQM, and $R$-symmetry of supercharges. SQM fields form representations of these two symmetries summarized in the following table:\footnote{$R$-charges of the chiral fields should be chosen in such a way that the superpotential is $R$-invariant.}
\be
\begin{array}{c|c|c|c|c|c|c|c}
	& A & X_i & \lambda_{\alpha} & D & q & \psi_{\alpha} & F \\
	\hline
	\SU(2)_J & {\bf 1} & {\bf 3} & {\bf 2} & {\bf 1} & {\bf 1} & {\bf 2} & {\bf 1}\\
	\hline
	\U(1)_R & 0 & 0 & 1/2 & 0& r & r-1/2 & r-1\\
\end{array}\; .
\ee
Generators of these symmetries are given by corresponding Noether currents:
\begin{align}
\begin{split}
	J_i&=\Tr\left[-\I\epsilon_{ijk}X_j\frac{\p}{\p X_k}+\bar\lambda\bar\sigma^0\sigma^{0i}\lambda+\sum\lm_a\bar\psi_a\bar\sigma^0\sigma^{0i}\psi_a\right],\\
	R&=\Tr\left[\frac{1}{2}\bar\lambda\bar\sigma^0\lambda+r\sum\lm_a\left(q_a\frac{\p}{\p q_a}-\bar q_a\frac{\p}{\p \bar q_a}\right)+\left(r-\frac{1}{2}\right)\sum\lm_a\bar\psi_a\bar\sigma^0\psi_a\right]
	\;.
\end{split}
\end{align}

Supercharges commute with the angular momentum operator in the following way:
\be
\left[J_k, Q_{\alpha} \right]=(\sigma^{k0}){}_{\alpha}{}^{\beta}Q_\beta\;,\quad 
\left[J_k, \bar Q_{\dot\alpha} \right]=\bar Q_{\dot\beta}(\bar\sigma^{k0}){}^{\dot\beta}{}_{\dot\alpha}\;.
\ee
In particular we have
$$
\left[J_3,Q_1 \right]=-\frac{1}{2}Q_1\;,\quad \left[J_3,Q_2 \right]=\frac{1}{2}Q_2\;,\quad \left[J_3,\bar Q_{\dot 1} \right]=\frac{1}{2}\bar Q_{\dot 1}\;,\quad \left[J_3,\bar Q_{\dot 2} \right]=-\frac{1}{2}\bar Q_{\dot 2}\; ;
$$
and with $R$-charge:
\be
\left[R, Q_{\alpha} \right]=\frac{1}{2} Q_{\alpha}\;,\quad \left[R, \bar Q_{\dot\alpha} \right]=-\frac{1}{2} \bar Q_{\dot\alpha}\; .
\ee

Using these relations it is simple to construct two non-perturbative operators:
\be\label{angular}
\CJ_+=J_3+R\;,\quad \CJ_-=J_3-R\;,
\ee
commuting with one half of the complete $\CN=4$ 1d supersymmetry:
$$
\left[Q_1,\CJ_+\right]=\left[\bar Q_{\dot 1},\CJ_+\right]=0,\quad \left[Q_2,\CJ_-\right]=\left[\bar Q_{\dot 2},\CJ_-\right]=0\;.
$$

%%%%%%%---------------------------------------------------------------------------------------------------------------------------------------------------------------------------------------------
\section{Example of SQM Wave-Function}
\label{s:exapmle}

In this Appendix we present an example for the computation of the SQM wavefunction.
We choose the example of the ADHM quiver with dimension 2 and for the fixed point $\{2\}$.
We follow the notations of Section \ref{s:gl2}.

The supercharge we localize with respect to reads:
\be
\scalebox{0.8}{$\begin{aligned}
	\bar Q_{\dot 1}={\rm Tr}\Bigg\{ 
	&\lambda^1\left(\left(\left[B_1,B_1^{\dag}\right]+\left[B_2,B_2^{\dag}\right]+II^{\dag}-J^{\dag}J-\theta \, 1_{N\times N}\right)-\right.\\ 
	&\left.-\p_{X^3}-\frac{1}{2}\left[\Phi,\Phi^{\dag}\right]+\left[B_3,B_3^{\dag}\right] \right)-\lambda^2\left(2\p_{\Phi^{\dag}}-\left[\Phi,X^3\right]\right)-\\
	&-\I\sqrt{2}\bar{\psi}_{\dot 1}(B_1)\left(\p_{B_1^{\dag}}+\left[X^3,B_1\right]\right)-\I\sqrt{2}\bar{\psi}_{\dot 2}(B_1)\left(\left[\Phi,B_1\right]-\epsilon_1B_1\right)+\\
	&+\I\sqrt{2}\psi_2(B_1)\left[B_2,B_3\right]-\I\sqrt{2}\bar{\psi}_{\dot 1}(B_2)\left(\p_{B_2^{\dag}}+\left[X^3,B_2\right]\right)-\I\sqrt{2}\bar{\psi}_{\dot 2}(B_2)\left(\left[\Phi,B_2\right]-\epsilon_2B_2\right)+\\
	&+\I\sqrt{2}\psi_2(B_2)\left[B_3,B_1\right]-\I\sqrt{2}\bar{\psi}_{\dot 1}(I)\left(\p_{I^{\dag}}+X^3I\right)-\I\sqrt{2}\bar{\psi}_{\dot 2}(I)\left(\Phi I-a I\right)+\I\sqrt{2}\psi_2(I)JB_3-\\
	&-\I\sqrt{2}\bar{\psi}_{\dot 1}(J)\left(\p_{J^{\dag}}-JX^3\right)-\I\sqrt{2}\bar{\psi}_{\dot 2}(J)\left(-J\Phi+(a-\epsilon_1-\epsilon_2)J\right)+\I\sqrt{2}\psi_2(J)B_3 I-\\
	&-\I\sqrt{2}\bar{\psi}_{\dot 1}(B_3)\left(\p_{B_3^{\dag}}+\left[X^3,B_3\right]\right)-\I\sqrt{2}\bar{\psi}_{\dot 2}(B_3)\left(\left[\Phi,B_3\right]+(\epsilon_1+\epsilon_2)B_3\right)+\\
	&+\I\sqrt{2}\psi_2(B_3)\left(\left[B_1,B_2\right]+IJ\right)
	\Bigg\} \;.
\end{aligned}$
}
\ee

Consider a fixed critical point defined by a partition $\{2\}$. We will denote corresponding expectation values by bars:
\bea
\begin{split}
&\bar X^3=0\;,\quad \bar \Phi=\left(\begin{array}{cc}
	a & 0\\
	0 & a+\epsilon_1\\
\end{array}\right)\;,\quad 
\bar B_1=\left(\begin{array}{cc}
	0 & 0\\
	\sqrt{\theta} & 0\\
\end{array}\right)\;,\quad 
\bar B_2=0\;,\\ 
&\bar I=\left(\begin{array}{c}
	\sqrt{2\theta} \\ 0\\
\end{array}\right)\;,\quad
\bar J=0 \;,\quad \bar B_3=0\;.
\end{split}
\eea

We introduce the following notations for deviations of the matrices values from expectation values:
\bea
\begin{split}
	&\delta X^3=\left(\begin{array}{cc}
		x_{11}& x_{12}\\
		x_{21}& x_{22}\\
	\end{array}\right),\quad \delta \Phi=\left(\begin{array}{cc}
		\phi_{11}& \phi_{12}\\
		\phi_{21}& \phi_{22}\\
	\end{array}\right),\\ 
	&\delta I=\left(\begin{array}{c}
		\iota_1\\
		\iota_2
	\end{array}\right),\quad \delta J=\left(\begin{array}{cc}
		\upsilon_1& \upsilon_2\\
	\end{array}\right),\\
	&\delta B_1=\left(\begin{array}{cc}
		\alpha_{11}& \alpha_{12}\\
		\alpha_{21}& \alpha_{22}\\
	\end{array}\right),\quad \delta B_2=\left(\begin{array}{cc}
		\beta_{11}& \beta_{12}\\
		\beta_{21}& \beta_{22}\\
	\end{array}\right),\quad \delta B_3=\left(\begin{array}{cc}
		\gamma_{11}& \gamma_{12}\\
		\gamma_{21}& \gamma_{22}\\
	\end{array}\right)\;.
\end{split}
\eea

The weights induced by the $\Omega$-background on the deformation parameters for arrows are given by eigenvalues of operator $\bar \Phi\cdot (*)-\epsilon_a (*)$. It is easy to calculate these weights (we define weights in terms of 2-vectors $(a,b)$ implying that the eigenvalue for such a vector is $a\epsilon_1+b\epsilon_2$):
\be
\begin{array}{c|c|c|c|c}
	\mbox{d.o.f}& \iota_1&\iota_2 & \upsilon_1 & \upsilon_2  \\
	\hline
	\mbox{wt.} & (0,0) & (1,0)&(-1,-1) &(-2,-1)\\
	\hline
	\mbox{d.o.f}&\alpha_{11}&\alpha_{12}& \alpha_{21}& \alpha_{22}\\
	\hline
	\mbox{wt.}&(-1,0) & (-2,0)&(0,0) &(-1,0)\\
	\hline
	\mbox{d.o.f}& \beta_{11}&\beta_{12}& \beta_{21}& \beta_{22} \\
	\hline
	\mbox{wt.} & (0,-1)&(-1,-1)&(1,-1)& (0,-1)\\
	\hline
	\mbox{d.o.f}&\gamma_{11}&\gamma_{12}& \gamma_{21}& \gamma_{22}\\
	\hline
	\mbox{wt.} & (1,1)& (0,1)& (2,1)&(1,1)\\
\end{array}
\ee

We divide all the degrees of freedom in the proposed set of four groups.

\paragraph{Group 1:} $x_{11}$, $x_{22}$, $\phi_{11}$, $\phi_{22}$, $\alpha_{21}$, $\iota_1$

These terms contribute only to the height function and vector field:
\begin{align}
&\fH=x_{11}\left(\left|\sqrt{2\theta}+\iota_1\right|^2-\left|\sqrt{\theta}+\alpha_{21}\right|^2-\theta\right)+x_{22}\left(\left|\sqrt{\theta}+\alpha_{21}\right|^2-\theta\right)\;,\\
&V(B_{1,21})=\sqrt{\theta}\left(\phi_{22}-\phi_{11}\right)\;,\quad V(I_1)= \sqrt{2\theta}\phi_{11} \;.
\end{align}
Frequencies of all these degrees of freedom are of order $\sqrt{\theta}$.

\paragraph{Group 2:} $\phi_{12}$, $\phi_{21}$, $\alpha_{11}-\alpha_{22}$, $\iota_2$

We call these gauge degrees of freedom for the following reason. Let us expand the action of the vector field up to the first order in the fields. For a field $q_a$ associated to arrow $a$ this expansion reads:
\be
\delta q_a=\Phi\cdot q_a-\epsilon_a q_a=\left(\bar\Phi\cdot -\epsilon_a\right) q_a+\Phi\cdot\bar  q_a={\bf wt}\cdot q_a+\Phi\cdot\bar q_a \;,
\ee
where by {\bf wt} we denote the weight function that acts on the arrow fields diagonally. In these terms an infinitesimal action of a \emph{complexification} $\GL(n,\IC)$ of the initial gauge group $U(n)$ on the fields reads:
\be
\delta_G q_a=g\cdot \bar q_a \;, \quad g\in\fg\fl(n,\IC) \;.
\ee
In these terms we see that the vector field $V$ mimics the action of the gauge group complexification parameterized by $\Phi\in\fg\fl(n,\IC)$.

For the mentioned fields we find:
\bea
\begin{split}
	&\fH=x_{12}\left(-\frac{1}{2}\epsilon_1^* \phi _{12}+\frac{1}{2} \epsilon_1 \phi_{21}^*+\sqrt{2\theta } \iota_2^*+\sqrt{\theta } \alpha _{11}-\sqrt{\theta } \alpha _{22}\right)+\\
	&\qquad\qquad+x_{21}\left(\sqrt{\theta } \alpha _{11}^*-\sqrt{\theta } \alpha _{22}^*+\frac{1}{2} \epsilon_1^* \phi _{21}-\frac{1}{2} \epsilon_1 \phi _{12}^*+\sqrt{2\theta } \iota _2\right) \;,\\
	&\delta \alpha_{11}=\sqrt{\theta } \phi _{12}-\epsilon_1 \alpha _{11} \;,\quad \delta\alpha_{22}=-\epsilon_1 \alpha _{22}-\sqrt{\theta } \phi _{12} \;,\quad \delta \iota_2=\sqrt{2\theta } \phi _{21}+\epsilon_1\iota _2\;.
\end{split}
\eea
Let us concentrate on fields $\iota_2$ and $\phi_{21}$, since the same we will be able to conclude about remaining degrees of freedom. The metric reads:
\be
ds^2=\left|d\iota_2\right|^2+\frac{1}{2}\left|d\phi_{21}\right|^2 \;.
\ee

Let us make an orthonormal change of coordinates:
\be\label{mixing}
\iota_2=-\frac{2 \sqrt{2\theta } t_1-\epsilon_1^*t_2 }{\sqrt{2 \left(4 \theta+|\epsilon_1|^2 \right)}} \;, \quad \phi_{21}=\frac{2 \left(\sqrt{2\theta } t_2+t_1 \epsilon_1 \right)}{\sqrt{2 \left(4 \theta +|\epsilon_1|^2\right)}} \;.
\ee

For metric, a height function contribution and vector fields this change of coordinates gives the following expressions:
\be
ds^2=|dt_1|^2+\frac{1}{2}|dt_2|^2 \;, \,\,\, \fH=\sqrt{2\theta+\frac{|\epsilon_1|^2}{2}}x_{21}t_1+{\rm (c.c.)} \;, \,\,\, \delta \iota_2=\sqrt{2\theta+\frac{|\epsilon_1|^2}{2}}t_2 \;.
\ee
Clearly, these degrees of freedom have weights of order $\sqrt{\theta}$.

\paragraph{Group 3:} $\gamma_{ij}$, $\upsilon_a$, $\beta_{12}$, $\beta_{11}-\beta_{22}$

These degrees of freedom give a non-trivial contribution to the first order of the superpotential expansion:
$$
W=\sqrt{\theta } \left(\sqrt{2}\upsilon _1 -\beta _{12}\right) \gamma _{11}+\sqrt{\theta } \left(\beta _{11} -\beta _{22}\right) \gamma _{12}+\sqrt{\theta } \beta _{12} \gamma _{22}+\sqrt{2\theta } \upsilon _2 \gamma _{21} \;.
$$

The superpotential is quadratic in fields. Consider pairs of fields entering the superpotential expression. Denote such a pair $x$ and $y$, so that the superpotential contribution reads:
$$
W=\sqrt{\theta}xy \;.
$$
The weights of fields entering this pair are opposite in order to satisfy the gauge invariance condition for the superpotential. In this case the differential has the following form:
\bea
\begin{split}
&\bar Q_{\dot 1}\sim \bar\psi_{\dot 1}(x)\p_{\bar x}+\bar\psi_{\dot 2}(x)\left[w x\right]+\psi_2(x)\left[\sqrt{\theta}y\right]+\\
&\qquad +\bar\psi_{\dot 1}(y)\p_{\bar y}+\bar\psi_{\dot 2}(y)\left[-w y\right]+\psi_2(y)\left[\sqrt{\theta}x\right]\;.
\end{split}
\eea
The weight $w$ of field $x$ is a linear function in $\hbar_1$ and $\hbar_2$. Let us change fermionic variables as:
\be
\bar \chi_{\dot 1}=\frac{w\bar\psi_{\dot 2}(x)+\sqrt{\theta}\psi_2(y)}{\sqrt{\theta+|w|^2}} \;,\quad \bar \chi_{\dot 2}=\frac{\sqrt{\theta}\psi_{2}(x)-w\bar\psi_{\dot 2}(y)}{\sqrt{\theta+|w|^2}} \;.
\ee
These fermions satisfy the standard anti-commutation relations:
$$
\left\{\chi_\alpha,\chi_\beta \right\}=\left\{\bar\chi_{\dot \alpha},\bar\chi_{\dot\beta} \right\}=0 \;,\quad\left\{\chi_{\alpha},\bar\chi_{\dot\beta} \right\}=\delta_{\alpha\beta} \;.
$$
In these terms the supercharge reads:
\be
\bar Q_{\dot 1}\sim \bar\psi_{\dot 1}(x)\p_{\bar x}+\bar\chi_{\dot 1}\sqrt{\theta+|w|^2}x+\bar\psi_{\dot 1}(y)\p_{\bar y}+\bar\chi_{\dot 2}\sqrt{\theta+|w|^2}y \;.
\ee
Eventually, we find that the weights of both degrees of freedom are of order $\sqrt{\theta}$.
\paragraph{Group 4:} $\alpha_{11}+\alpha_{22}$, $\alpha_{12}$, $\beta_{11}+\beta_{22}$, $\beta_{21}$

The corresponding weights read:
\be
\begin{array}{c|c|c|c|c}
	\mbox{d.o.f}& \alpha_{11}+\alpha_{22}& \alpha_{12} & \beta_{11}+\beta_{22}& \beta_{21}\\
	\hline
	\mbox{wt.}& -\epsilon_1& -2\epsilon_1& -\epsilon_2 & \epsilon_1-\epsilon_2\\
\end{array} \;.
\ee
These are the only degrees of freedom contributing to $\Psi_{|\omega|<\Lambda}$.

A product of these weights gives the correct expression for the corresponding Euler character (see \eqref{Eul}):
\be
{\rm Eul}_{\{2 \} }=2\epsilon_1^2\epsilon_2(\epsilon_2-\epsilon_1) \;.
\ee

%%%%%%%%%%%%%%%%%%%%%%%%%%%%%%%%%%%%%%%%%%%%%%%%%%%%%%%%%%%%%%%%%%%%%%%%%%%%%%%%%%%%
\section{Examples of Explicit Calculations}
\label{s:calc}
%%%%%%%%%%%%%%%%%%%%%%%%%%%%%%%%%%%%%%%%%%%%%%%%%%%%%%%%%%%%%%%%%%%%%%%%%%%%%%%%%%%%

In this Appendix we include some examples of explicit computations for 
relations of the BPS state algebra. We find rather intricate cancellations
are required for our consistency. While we have done similar computations
for many more complicated examples with computer programs,  we restrict here to discuss some 
simple examples.

%%%%%%%---------------------------------------------------------------------------------------------------------------------------------------------------------------------------------------------
\subsection{Explicit Example of Calculation for Hilbert Scheme Fixed Points}\label{s:ideals}

Consider two partitions:
\be
\Lambda_1=\begin{array}{c}
	\begin{tikzpicture}
	\begin{scope}[scale=0.6]
	\begin{scope}[shift={(0,0)}]
	\draw[thick, fill=gray!20!white] (0,0) -- (1,0) -- (1,1) -- (0,1) -- cycle;
	\node at (0.5,0.5) {$1$};
	\end{scope}
	\begin{scope}[shift={(1,0)}]
	\draw[thick, fill=gray!20!white] (0,0) -- (1,0) -- (1,1) -- (0,1) -- cycle;
	\node at (0.5,0.5) {$x$};
	\end{scope}
	\begin{scope}[shift={(2,0)}]
	\draw[thick, fill=gray!20!white] (0,0) -- (1,0) -- (1,1) -- (0,1) -- cycle;
	\node at (0.5,0.5) {$x^2$};
	\end{scope}
	\begin{scope}[shift={(0,1)}]
	\draw[thick, fill=gray!20!white] (0,0) -- (1,0) -- (1,1) -- (0,1) -- cycle;
	\node at (0.5,0.5) {$y$};
	\end{scope}
	\begin{scope}[shift={(1,1)}]
	\draw[thick, fill=gray!20!white] (0,0) -- (1,0) -- (1,1) -- (0,1) -- cycle;
	\node at (0.5,0.5) {$x y$};
	\end{scope}
	\begin{scope}[shift={(3,0)}]
	\draw[dashed] (0,0) -- (1,0) -- (1,1) -- (0,1) -- cycle;
	\node at (0.5,0.5) {$x^3$};
	\end{scope}
	\begin{scope}[shift={(2,1)}]
	\draw[dashed] (0,0) -- (1,0) -- (1,1) -- (0,1) -- cycle;
	\node at (0.5,0.5) {$x^2 y$};
	\end{scope}
	\begin{scope}[shift={(0,2)}]
	\draw[dashed] (0,0) -- (1,0) -- (1,1) -- (0,1) -- cycle;
	\node at (0.5,0.5) {$y^2$};
	\end{scope}
	\end{scope}
	\end{tikzpicture}
\end{array},\quad \Lambda_{2}=\begin{array}{c}
	\begin{tikzpicture}
	\begin{scope}[scale=0.6]
	\begin{scope}[shift={(0,0)}]
	\draw[thick, fill=gray!20!white] (0,0) -- (1,0) -- (1,1) -- (0,1) -- cycle;
	\node at (0.5,0.5) {$1$};
	\end{scope}
	\begin{scope}[shift={(1,0)}]
	\draw[thick, fill=gray!20!white] (0,0) -- (1,0) -- (1,1) -- (0,1) -- cycle;
	\node at (0.5,0.5) {$x$};
	\end{scope}
	\begin{scope}[shift={(2,0)}]
	\draw[thick, fill=gray!20!white] (0,0) -- (1,0) -- (1,1) -- (0,1) -- cycle;
	\node at (0.5,0.5) {$x^2$};
	\end{scope}
	\begin{scope}[shift={(0,1)}]
	\draw[thick, fill=gray!20!white] (0,0) -- (1,0) -- (1,1) -- (0,1) -- cycle;
	\node at (0.5,0.5) {$y$};
	\end{scope}
	\begin{scope}[shift={(1,1)}]
	\draw[thick, fill=gray!20!white] (0,0) -- (1,0) -- (1,1) -- (0,1) -- cycle;
	\node at (0.5,0.5) {$x y$};
	\end{scope}
	\begin{scope}[shift={(0,2)}]
	\draw[thick, fill=gray!20!white] (0,0) -- (1,0) -- (1,1) -- (0,1) -- cycle;
	\node at (0.5,0.5) {$y^2$};
	\end{scope}
	\begin{scope}[shift={(3,0)}]
	\draw[dashed] (0,0) -- (1,0) -- (1,1) -- (0,1) -- cycle;
	\node at (0.5,0.5) {$x^3$};
	\end{scope}
	\begin{scope}[shift={(2,1)}]
	\draw[dashed] (0,0) -- (1,0) -- (1,1) -- (0,1) -- cycle;
	\node at (0.5,0.5) {$x^2 y$};
	\end{scope}
	\begin{scope}[shift={(0,3)}]
	\draw[dashed] (0,0) -- (1,0) -- (1,1) -- (0,1) -- cycle;
	\node at (0.5,0.5) {$y^3$};
	\end{scope}
	\begin{scope}[shift={(1,2)}]
	\draw[dashed] (0,0) -- (1,0) -- (1,1) -- (0,1) -- cycle;
	\node at (0.5,0.5) {$x y^2$};
	\end{scope}
	\end{scope}
	\end{tikzpicture}
\end{array}
\ee

In this section we will calculate explicitly the tangent spaces for these two fixed points,  the tangent space to the incidence locus and corresponding Euler characters.

Both fixed points represent ideals in $\IC[x,y]$ generated by monomials corresponding to boxes located in corners of diagrams:
\be
\CI_1=\{x^3,\; x^2 y,\; y^2 \} \;,\quad \CI_2=\{x^3,\; x^2 y,\; x y^2,\; y^3 \} \;.
\ee

Let us define tangent spaces to these two fixed points. We start with constructing deformations to ideals. The monomial generators of the ideal for a chosen diagram can be deformed to polynomials with the help of monomials corresponding to the entries of the diagram. For example, we consider an \emph{initially deformed} generators of the ideals:
\bea\label{def}
\begin{split}
	\CI_1^{({\rm in.d.})}&=\left\{ \begin{array}{c}
		x^3-t \alpha _{1,1}-t  \alpha _{1,2}x-t \alpha _{1,3}x^2 -t  \alpha _{1,4}y-t  \alpha _{1,5}x y\\
		x^2 y-t \alpha _{2,1}-t \alpha _{2,2}x -t \alpha _{2,3}x^2 -t  \alpha _{2,4}y -t \alpha _{2,5}x y \\
		y^2-t \alpha _{3,1}-t \alpha _{3,2}x -t \alpha _{3,3}x^2-t \alpha _{3,4} y-t  \alpha _{3,5}x y
	\end{array} \right\} \;,\\
	\CI_2^{({\rm in.d.})}&=\left\{ \begin{array}{c}
		x^3-t \beta _{1,1}-t \beta _{1,2}x -t \beta _{1,3}x^2 -t \beta _{1,4}y -t \beta _{1,5}x y -t \beta _{1,6}y^2 \\
		x^2 y-t \beta _{2,1}-t \beta _{2,2}x -t \beta _{2,3}x^2 -t \beta _{2,4}y -t \beta _{2,5}x y -t \beta _{2,6}y^2 \\
		x y^2-t \beta _{3,1}-t \beta _{3,2}x -t \beta _{3,3}x^2 -t \beta _{3,4}y -t \beta _{3,5}x y -t \beta _{3,6}y^2\\
		y^3-t \beta _{4,1}-t \beta _{4,2}x -t \beta _{4,3}x^2 -t \beta _{4,4}y -t \beta _{4,5}x y -t \beta _{4,6} y^2
	\end{array} \right\} \;.
\end{split}
\eea

Here $t$ is a small parameter to measure deviation from the fixed point therefore we will work up to $O(t^2)$ order. Indices of coefficients $\alpha_{i,j}$ and $\beta_{i,j}$ are chosen in such a way that the first index represents the order number of the ideal generator we are deforming, and the second one indexes the order number of deforming monomial in chosen ordered monomial bases:
\bea
\begin{split}
	V_1&\coloneqq \IC[x,y]/\CI_1={\rm Span}\left\{1, x, x^2, y, x y\right\} \;,\\
	V_2& \coloneqq \IC[x,y]/\CI_2={\rm Span}\left\{1,x,x^2,y,x y,y^2\right\} \;.
\end{split}
\eea

Unconstrained polynomials \eqref{def} do not form ideals in $\IC[x,y]$. Consider operators $B_1^{(i)}$ and $B_2^{(i)}$ defined on corresponding $V_i$ as multiplication by $x$ and $y$ modulo $\CI_i$, in particular:
\be
B_1^{(1)}=\left(
\begin{array}{ccccc}
	0 & 0 & t \alpha _{1,1} & 0 & t \alpha _{2,1} \\
	1 & 0 & t \alpha _{1,2} & 0 & t \alpha _{2,2} \\
	0 & 1 & t \alpha _{1,3} & 0 & t \alpha _{2,3} \\
	0 & 0 & t \alpha _{1,4} & 0 & t \alpha _{2,4} \\
	0 & 0 & t \alpha _{1,5} & 1 & t \alpha _{2,5} \\
\end{array}
\right),\quad B_2^{(1)}=\left(
\begin{array}{ccccc}
	0 & 0 & t \alpha _{2,1} & t \alpha _{3,1} & 0 \\
	0 & 0 & t \alpha _{2,2} & t \alpha _{3,2} & t \alpha _{3,1} \\
	0 & 0 & t \alpha _{2,3} & t \alpha _{3,3} & t \alpha _{3,2} \\
	1 & 0 & t \alpha _{2,4} & t \alpha _{3,4} & 0 \\
	0 & 1 & t \alpha _{2,5} & t \alpha _{3,5} & t \alpha _{3,4} \\
\end{array}
\right) \;.
\ee

A constraint on these matrices is a commutativity of $x\cdot$ and $y\cdot$, or, in a representation $[B_1^{(i)},B_2^{(i)}]=0$:
\be
\left[B_1^{(1)},B_2^{(1)}\right]=t \left(
\begin{array}{ccccc}
	0 & 0 & 0 & 0 & 0 \\
	0 & 0 & \alpha _{2,1} & 0 & 0 \\
	0 & 0 & \alpha _{2,2} & 0 & \alpha _{3,1} \\
	0 & 0 & -\alpha _{1,1} & 0 & -\alpha _{2,1} \\
	0 & 0 & \alpha _{2,4}-\alpha _{1,2} & 0 & -\alpha _{2,2} \\
\end{array}
\right)+O(t^2) \;.
\ee

From this commutator we read off the constraints on $\CI^{({\rm in.d.})}$ to become deformed ideal generators $\CI^{(d)}$ in $\IC[x,y]$, and we do the same for $\CI_2$ and $\beta$. As the result we have the following ideals deformed by tangent directions in both cases of fixed points:
\bea
\begin{split}
	\CI^{(d)}_1&=\left\{\begin{array}{c}
		x^3-t\alpha _1 x-t \alpha _2 x^2-t \alpha _3 y-t \alpha _4 x y\\
		x^2 y-t\alpha _1 y-t\alpha _5 x^2-t\alpha _6 x y\\
		y^2-t\alpha _7 x-t\alpha _8 x^2-t\alpha _9 y-t\alpha _{10} x y
	\end{array} \right\}\;,\\
	\CI^{(d)}_2&=\left\{\begin{array}{c}
		x^3-t\beta _1 x^2-t\beta _2 x y-t\beta _3 y^2\\
		x^2 y-t\beta _4 x^2-t\beta _5 x y-t\beta _6y^2\\
		x y^2-t\beta _7 x^2-t\beta _8 x y-t\beta _9y^2\\
		y^3-t\beta _{10} x^2-t\beta _{11} x y-t\beta _{12} y^2
	\end{array} \right\} \;.
\end{split}
\eea

These polynomials belong to eigenspaces for the torus action. Assuming that weights of the $x$ and $y$ are $\epsilon_1$ and $\epsilon_2$ we easily calculate corresponding weights of tangent vectors:
\bea
\begin{split}
	&w_{\alpha _1}=2 \epsilon _1,\;w_{\alpha _2}=\epsilon _1,\;w_{\alpha _3}=3 \epsilon _1-\epsilon _2,\;w_{\alpha _4}=2 \epsilon _1-\epsilon _2,\;w_{\alpha _5}=\epsilon _2,\\
	&\qquad w_{\alpha _6}=\epsilon _1,\;w_{\alpha _7}=2 \epsilon _2-\epsilon _1,\;w_{\alpha _8}=2 \epsilon _2-2 \epsilon _1,\;w_{\alpha _9}=\epsilon _2,\;w_{\alpha _{10}}=\epsilon _2-\epsilon _1;\\
	&w_{\beta _1}=\epsilon _1,\; w_{\beta _2}=2 \epsilon _1-\epsilon _2,\; w_{\beta _3}=3 \epsilon _1-2 \epsilon _2,\; w_{\beta _4}=\epsilon _2,\; w_{\beta _5}=\epsilon _1,\\
	&\qquad w_{\beta _6}=2 \epsilon _1-\epsilon _2,\;
	w_{\beta _7}=2 \epsilon _2-\epsilon _1,\; w_{\beta _8}=\epsilon _2,\; w_{\beta _9}=\epsilon _1,\\ 
	&\qquad w_{\beta _{10}}=3 \epsilon _2-2 \epsilon _1,\; w_{\beta _{11}}=2 \epsilon _2-\epsilon _1,\; w_{\beta _{12}}=\epsilon _2.
\end{split}
\eea

We calculate corresponding Euler characters as products over weights.
The result is in agreement with the hook formula \cite{Macdonald}:
\bea
\begin{split}
{\rm Eul}_{\Lambda_1}&=\prod\lm_{i}w_{\alpha_i}=-4 \epsilon _1^3 \left(\epsilon _1-2 \epsilon _2\right) \left(\epsilon _1-\epsilon _2\right){}^2 \left(2 \epsilon _1-\epsilon _2\right) \left(3 \epsilon _1-\epsilon _2\right) \epsilon _2^2 \;,\\
{\rm Eul}_{\Lambda_2}&=\prod\lm_{i}w_{\beta_i}=-\epsilon _1^3 \left(2 \epsilon _1-3 \epsilon _2\right) \left(\epsilon _1-2 \epsilon _2\right){}^2 \left(3 \epsilon _1-2 \epsilon _2\right) \left(2 \epsilon _1-\epsilon _2\right){}^2 \epsilon _2^3 \;.
\end{split}
\eea

According to the definition of the Heisenberg algebra we have to calculate an incidence locus:
$$
\CJ\subset \CI\;,\quad {\rm Supp}(\CI/\CJ)=\{ {\rm pt} \} \;.
$$

In our particular case when we consider fixed points corresponding to two partitions whose diagrams can be embedded into each other the support condition is fulfilled automatically, therefore the tangent space inside the incidence locus to the fixed point given by a pair of diagrams $(\Lambda_1,\Lambda_2)$ is defined by the following constraint:
$$
\CI_2^{(d)}\subset \CI_{1}^{(d)} \;.
$$

Algebraically, it is very easy to translate it to the following constraint: $\CI_{2}^{(d)}$ restricted on $V_1$ as a multiplication modulo $\CI_1^{(d)}$ gives a zero. In this case for generators we have:
\be
\CI_{2,k}^{(d)}=\sum\lm_l c_{k,l}\CI_{1,l}^{(d)},\quad c_{k,l}\in \IC[x,y] \;.
\ee

In our particular case we find:
\be
\Big.\CI_{2}^{(d)}\Big|_{V_1}=t\left(
\begin{array}{ccccc}
	0 & \alpha _1 & \alpha _2-\beta _1 & \alpha _3 & \alpha _4-\beta _2 \\
	0 & 0 & \alpha _5-\beta _4 & \alpha _1 & \alpha _6-\beta _5 \\
	0 & 0 & \alpha _7-\beta _7 & 0 & \alpha _9-\beta _8 \\
	0 & 0 & -\beta _{10} & 0 & \alpha _7-\beta _{11} \\
\end{array}
\right)\cdot V_1+O(t^2) \;.
\ee

There is an obvious solution to these linear equations:
\bea
\begin{split}
	&\alpha_1= \alpha_3=\beta_{10}=0 \;, \quad \gamma_1\coloneqq\alpha_2=\beta_1 \;,\quad \gamma_2\coloneqq \alpha_4=\beta_2 \;,\\
	&\gamma_3\coloneqq\alpha_5=\beta_4,\;\quad \gamma_4\coloneqq\alpha_6=\beta_5\;,\quad \gamma_5\coloneqq \alpha_7=\beta_7=\beta_{11}\;,\quad \gamma_6=\alpha_9=\beta_8\;.
\end{split}
\eea
Notice that parameters $\alpha_8$, $\alpha_{10}$, $\beta_3$, $\beta_6$, $\beta_9$, $\beta_{12}$ and all $\gamma_i$ are unconstrained and span the incidence locus hypersurface. Aslo it is useful to change variables as $\beta_{11}\to \beta_7+\beta_{11}$, then we will have an equation $\beta_{11}=0$ instead. Referring to decomposition \eqref{decomp} we could regroup bases in tangent spaces as:
\bea\label{VUW}
\begin{split}
	&V_1={\rm Span}\left\{\underline{\alpha}_1,\;\underline{\alpha}_3\right\},\; V_2={\rm Span}\left\{\underline{\beta}_{10},\; \underline{\beta}_{11} \right\},\\
	&W_1={\rm Span}\left\{ \alpha_8,\;\alpha_{10}\right\},\; 
	W_2={\rm Span}\left\{\beta_3,\; \beta_6,\;\beta_9,\;\beta_{12} \right\},\\
	&U={\rm Span}\left\{\gamma_1,\;\gamma_2,\;\gamma_3,\;\gamma_4,\;\gamma_5,\;\gamma_6 \right\}\;.
\end{split}
\eea
To distinguish between degrees of freedom belonging to $V_i$ and $W_i$ we will underline the variables representing the former ones. For equivariant weights of $\gamma$'s we derive:
\be
w_{\gamma_1}=\epsilon _1,\; w_{\gamma_2}=2 \epsilon _1-\epsilon _2,\;
w_{\gamma_3}=\epsilon _2,\;w_{\gamma_4}=\epsilon _1,\;
w_{\gamma_5}=2 \epsilon _2-\epsilon _1,\; w_{\gamma_6}=\epsilon _2\;.
\ee

The corresponding Euler character for the incidence locus reads:
\bea
\begin{split}
{\rm Eul}_{(\Lambda_1,\Lambda_2)}= w_{\alpha_8}w_{\alpha_{10}}w_{\beta_3} w_{\beta_6}w_{\beta_9}w_{\beta_{12}}w_{\gamma_1}w_{\gamma_2}w_{\gamma_3}w_{\gamma_4}w_{\gamma_5}w_{\gamma_6} =\\
=-2 \epsilon _1^3 \left(\epsilon _1-2 \epsilon _2\right) \left(3 \epsilon _1-2 \epsilon _2\right) \left(\epsilon _1-\epsilon _2\right){}^2 \left(2 \epsilon _1-\epsilon _2\right){}^2 \epsilon _2^3\;.
\end{split}
\eea

The corresponding coefficient in the product of Jack polynomials reads:
\bea
\begin{split}
	&p_1J_{\Lambda_1}^{(\beta)}(p_1,p_2,\ldots)=-\frac{2 (\beta +3)}{\beta  (\beta +2) (2 \beta +3)}J_{\Lambda_2}^{(\beta)}(p_1,p_2,\ldots)+\ldots\\
	&\frac{{\rm Eul}_{\Lambda_1}(-1,\beta)}{{\rm Eul}_{(\Lambda_1,\Lambda_2)}(-1,\beta)}=-\frac{2 (\beta +3)}{\beta  (\beta +2) (2 \beta +3)} \;.
\end{split}
\eea
Morphism $\tau$ defines a central extension of quiver representations. So for quiver representations we expect to derive the following short exact sequence of homomorphisms:
\be
0\longrightarrow \CE_{\CI_1/\CI_2}\mathop{\longrightarrow}^{\eta} \IC[x,y]/\CI_2 \mathop{\longrightarrow}^{\tau} \IC[x,y]/\CI_1 \longrightarrow 0 \;.
\ee
Homomorphisms $\tau$ and $\eta$ satisfy the following set of relations:
\be
\Big.B_a\Big|_{\IC[x,y]/\CI_1}\cdot \eta=\eta\cdot \Big.B_a\Big|_{\IC[x,y]/\CI_2},\quad 
\Big.B_a\Big|_{\IC[x,y]/\CI_2}\cdot \tau=\tau\cdot \Big.B_a\Big|_{\CE_{\CI_1/\CI_2}} \;.
\ee

For map restrictions we have:
\bea
\begin{split}
B_1^{(1)}&=\left.B_1\right|_{V_1}=\left(
\begin{array}{ccccc}
	0 & 0 & 0 & 0 & 0 \\
	1 & 0 & t \underline{\alpha }_1 & 0 & 0 \\
	0 & 1 & \gamma _1 t & 0 & \gamma _3 t \\
	0 & 0 & t \underline{\alpha }_3 & 0 & t \underline{\alpha }_1 \\
	0 & 0 & \gamma _2 t & 1 & \gamma _4 t \\
\end{array}
\right)\;,\\ 
B_2^{(1)}&=\left.B_2\right|_{V_1}=\left(
\begin{array}{ccccc}
	0 & 0 & 0 & 0 & 0 \\
	0 & 0 & 0 & \gamma _5 t & 0 \\
	0 & 0 & \gamma _3 t & \alpha _8 t & \gamma _5 t \\
	1 & 0 & t \underline{\alpha }_1 & \gamma _6 t & 0 \\
	0 & 1 & \gamma _4 t & \alpha _{10} t & \gamma _6 t \\
\end{array}
\right)\;,
\end{split}
\eea
\bea
\begin{split}
B_1^{(2)}&=\left.B_1\right|_{V_2}=\left(
\begin{array}{cccccc}
	0 & 0 & 0 & 0 & 0 & 0 \\
	1 & 0 & 0 & 0 & 0 & 0 \\
	0 & 1 & \gamma _1 t & 0 & \gamma _3 t & \gamma _5 t \\
	0 & 0 & 0 & 0 & 0 & 0 \\
	0 & 0 & \gamma _2 t & 1 & \gamma _4 t & \gamma _6 t \\
	0 & 0 & \beta _3 t & 0 & \beta _6 t & \beta _9 t \\
\end{array}
\right)\;,\\ 
B_2^{(2)}&=\left.B_2\right|_{V_2}=\left(
\begin{array}{cccccc}
	0 & 0 & 0 & 0 & 0 & 0 \\
	0 & 0 & 0 & 0 & 0 & 0 \\
	0 & 0 & \gamma _3 t & 0 & \gamma _5 t & t \underline{\beta }_{10} \\
	1 & 0 & 0 & 0 & 0 & 0 \\
	0 & 1 & \gamma _4 t & 0 & \gamma _6 t & t \underline{\beta }_{11}+\gamma _5 t \\
	0 & 0 & \beta _6 t & 1 & \beta _9 t & \beta _{12} t \\
\end{array}
\right)\;.
\end{split}
\eea

On $V_1$ polynomials $\CI_{1,k}$ are annihilated, therefore we determine homomorphism $\tau$ by equations:
\be
\tau\cdot \CI_{1,k}\left(\left.B_1\right|_{V_2},\left.B_2\right|_{V_2}\right)= \CI_{1,k}\left(\left.B_1\right|_{V_1},\left.B_2\right|_{V_1}\right)\cdot \tau=0 \;.
\ee
Thus we find (that works only if all underlined variables are set to zero):
\be
\tau=\left(
\begin{array}{cccccc}
	1 & 0 & 0 & 0 & 0 & 0 \\
	0 & 1 & 0 & 0 & 0 & \gamma _5 t \\
	0 & 0 & 1 & 0 & 0 & \alpha _8 t \\
	0 & 0 & 0 & 1 & 0 & \gamma _6 t \\
	0 & 0 & 0 & 0 & 1 & \alpha _{10} t \\
\end{array}
\right) \;.
\ee

And the second homomorphism $\eta$ is defined as ${\rm ker}\;\tau$:
\be
\eta=\left(\begin{array}{c}
0 \\
-\gamma _5 t \\
-\alpha _8 t \\
-\gamma _6 t \\
-\alpha _{10} t \\
1\\
\end{array}\right) \;.
\ee

It is easy to find that restricted to $\CE_{\CI_1/\CI_2}$ matrices $B_{1,2}$ have specific eigenvalues:
\be\label{eigen}
\left.B_1\right|_{\CE_{\CI_1/\CI_2}}=t \beta_9\cdot {\bf Id}\;,\quad \left.B_2\right|_{\CE_{\CI_1/\CI_2}}=t (\beta _{12}-\gamma _6)\cdot {\bf Id} \;.
\ee

These eigenvalues parameterize exactly coordinates of a point that constitutes ${\rm Supp}(\CI_1^{(d)}/\CI_2^{(d)})$. This point parameterizes a physical position of a D0 brane in $\IC^2$ split from the larger crystal.

Finally, consider a complex gauge transform $g^{(2)}$ acting on $V_2$:
\be
g^{(2)}=t\left(
\begin{array}{cccccc}
	0 & 0 & 0 & 0 & 0 & 0 \\
	0 & 0 & 0 & 0 & 0 & \gamma _5 \\
	0 & 0 & 0 & 0 & 0 & \alpha _8 \\
	0 & 0 & 0 & 0 & 0 & \gamma _6 \\
	0 & 0 & 0 & 0 & 0 & \alpha _{10} \\
	0 & 0 & 0 & 0 & 0 & 0 \\
\end{array}
\right)\;.
\ee
For gauge transformed quantities we have:
\be\nn
\tilde \tau= \left(
\begin{array}{cccccc}
	1 & 0 & 0 & 0 & 0 & 0 \\
	0 & 1 & 0 & 0 & 0 & 0 \\
	0 & 0 & 1 & 0 & 0 & 0 \\
	0 & 0 & 0 & 1 & 0 & 0 \\
	0 & 0 & 0 & 0 & 1 & 0 \\
\end{array}
\right)\;,
\ee
\be\label{block_decomp}
\begin{split}
\tilde B^{(2)}_1&=\left(
\begin{array}{cccccc}
	0 & 0 & 0 & 0 & \multicolumn{1}{c|}{0} & 0 \\
	1 & 0 & 0 & 0 & \multicolumn{1}{c|}{0} & 0 \\
	0 & 1 & \gamma _1 t & 0 &\multicolumn{1}{c|}{ \gamma _3 t} & 0 \\
	0 & 0 & 0 & 0 & \multicolumn{1}{c|}{0} & 0 \\
	0 & 0 & \gamma _2 t & 1 &\multicolumn{1}{c|}{ \gamma _4 t} & 0 \\
	\hline
	0 & 0 & \beta _3 t & 0 &\beta _6 t & \beta _9 t\\
\end{array} 
\right) \;,\\
\tilde B^{(2)}_2&=\left(
\begin{array}{cccccc}
	0 & 0 & 0 & 0 &\multicolumn{1}{c|}{0} & 0 \\
	0 & 0 & 0 & \gamma _5 t & \multicolumn{1}{c|}{0} & 0 \\
	0 & 0 & \gamma _3 t & \alpha _8 t & \multicolumn{1}{c|}{\gamma _5 t} & t \underline{\beta }_{10} \\
	1 & 0 & 0 & \gamma _6 t & \multicolumn{1}{c|}{0} & 0 \\
	0 & 1 & \gamma _4 t & \alpha _{10} t &\multicolumn{1}{c|}{ \gamma _6 t} & t \underline{\beta }_{11} \\
	\hline
	0 & 0 & \beta _6 t & 1 & \beta _9 t & \beta _{12} t-\gamma _6 t \\
\end{array}
\right)\;.
\end{split}
\ee
For $\tilde B^{(2)}_i$ the block structure proposed in \eqref{block} becomes transparent.

%%%%%%%---------------------------------------------------------------------------------------------------------------------------------------------------------------------------------------------
\subsection{\texorpdfstring{Example of $e$ -- $f$ Relation Calculation for $Y(\widehat{\mathfrak{gl}}_1)$}{Example of e -- f Relation Calculation for Y(gl1)}}
\label{s:ex_gl_1}
Let us consider two crystals:  
\be
\Lambda_1=\left[\begin{array}{c}
	\begin{tikzpicture}
	\draw[fill=white!60!purple] (0,0) circle (0.2) (0,1) circle (0.2) (1,0) circle (0.2) (-1,-1) circle (0.2);
	\node at (0,0) {$1$};
	\node at (0,1) {$1$};
	\node at (1,0) {$1$};
	\node at (-1,-1) {$1$};
	\draw[->] (0,0.2) -- (0,0.8);
	\draw[->] (0.2,0) -- (0.8,0);
	\draw[->] (-0.141421, -0.141421) -- (-0.858579, -0.858579);
	\draw[thick,red] (0,0) circle (0.3);
	\end{tikzpicture}
\end{array}\right]\;,\quad 
\Lambda_2=\left[\begin{array}{c}
	\begin{tikzpicture}
	\draw[fill=white!60!purple] (0,0) circle (0.2) (0,1) circle (0.2) (1,0) circle (0.2) (-1,-1) circle (0.2) (0,-1) circle (0.2);
	\node at (0,0) {$1$};
	\node at (0,1) {$1$};
	\node at (1,0) {$1$};
	\node at (-1,-1) {$1$};
	\node at (0,-1) {$1$};
	\draw[->] (0,0.2) -- (0,0.8);
	\draw[->] (0.2,0) -- (0.8,0);
	\draw[->] (-0.8,-1) -- (-0.2,-1);
	\draw[->] (-0.141421, -0.141421) -- (-0.858579, -0.858579);
	\draw[->] (0.858579, -0.141421) -- (0.141421, -0.858579);
	\draw[thick,red] (0,0) circle (0.3);
	\draw[thick,blue] (0,-1) circle (0.3);
	\end{tikzpicture}
\end{array}\right]\;.
\ee

A crystal $\Lambda_2$ differs from $\Lambda_1$ by a single atom emphasized by a blue circle. In both cases we parameterize bases in vector spaces of quiver representations by vectors associated to crystal atoms. We will denote a basis vector as $\nu_{\vec r}$, where $\vec r$ is an atom position in the crystal. Thus we have for representation vector spaces:
\begin{align}
\begin{split}
	V_1&={\rm Span}\{ \nu_{(0,0,0)},\nu_{(1,0,0)},\nu_{(0,1,0)},\nu_{(0,0,1)} \}\;,\\
	V_2&={\rm Span}\{ \nu_{(0,0,0)},\nu_{(1,0,0)},\nu_{(0,1,0)},\nu_{(0,0,1)},\nu_{(1,0,1)} \} \;.
\end{split}
\end{align}
In such a basis expectation values of chiral fields $\bar B_i$ are maps that just shift corresponding coordinate in $\vec r$ by a unit and $\bar I$ maps to the root atom $\nu_{(0,0,0)}$:
\bea
\begin{split}
	&\bar B_1^{(1)}=\left(
	\begin{array}{cccc}
		0 & 0 & 0 & 0 \\
		1 & 0 & 0 & 0 \\
		0 & 0 & 0 & 0 \\
		0 & 0 & 0 & 0 \\
	\end{array}
	\right),\; \quad \bar B_2^{(1)}=\left(
	\begin{array}{cccc}
		0 & 0 & 0 & 0 \\
		0 & 0 & 0 & 0 \\
		1 & 0 & 0 & 0 \\
		0 & 0 & 0 & 0 \\
	\end{array}
	\right),\\ 
	&\bar B_3^{(1)}=\left(
	\begin{array}{cccc}
		0 & 0 & 0 & 0 \\
		0 & 0 & 0 & 0 \\
		0 & 0 & 0 & 0 \\
		1 & 0 & 0 & 0 \\
	\end{array}
	\right),\; \quad \bar I^{(1)}=\left(\begin{array}{c}
		1\\ 0\\ 0\\ 0\\
	\end{array}\right)\;,  \\
	&\bar B_1^{(2)}=\left(
	\begin{array}{ccccc}
		0 & 0 & 0 & 0 & 0 \\
		1 & 0 & 0 & 0 & 0 \\
		0 & 0 & 0 & 0 & 0 \\
		0 & 0 & 0 & 0 & 0 \\
		0 & 0 & 0 & 1 & 0 \\
	\end{array}
	\right),\;\quad \bar  B_2^{(2)}=\left(
	\begin{array}{ccccc}
		0 & 0 & 0 & 0 & 0 \\
		0 & 0 & 0 & 0 & 0 \\
		1 & 0 & 0 & 0 & 0 \\
		0 & 0 & 0 & 0 & 0 \\
		0 & 0 & 0 & 0 & 0 \\
	\end{array}
	\right),\\ 
	&\bar B_3^{(2)}=\left(
	\begin{array}{ccccc}
		0 & 0 & 0 & 0 & 0 \\
		0 & 0 & 0 & 0 & 0 \\
		0 & 0 & 0 & 0 & 0 \\
		1 & 0 & 0 & 0 & 0 \\
		0 & 1 & 0 & 0 & 0 \\
	\end{array}
	\right),\; \quad \bar I^{(2)}=\left(\begin{array}{c}
		1\\ 0\\ 0\\ 0\\ 0\\
	\end{array}\right) \;.
\end{split}
\eea
The tangent spaces are spanned by all deformation degrees of freedom  $\delta B_i$, $\delta I$ modulo complexified gauge degrees of freedom. Gauge degrees of freedom are parameterized by elements of $G\in \GL({\rm dim}\; V_r,\IC)$. The linearized chiral fields transform under complexified gauge transformations according to the following rule:
\be
\delta_G B_i=\delta B_i+\left[G\;, \bar B_i\right], \quad \delta_G I=\delta I+G\bar I \;.
\ee

We will denote chiral degrees of freedom in the following way:
\be
\delta B_i^{(r)}=\left(b_{i,jk}^{(r)} \right)_{j,k=1}^{{\rm dim}\; V_r} \;,
\quad \delta I^{(r)}=\left(\xi_{j}^{(r)} \right)_{j=1}^{{\rm dim}\; V_r} \;.
\ee
We fix the gauge by identifying the gauge degrees of freedom with the following chiral degrees of freedom:
\begin{equation}
\scalebox{0.9}{$\begin{aligned}
	G^{(1)}&=\left(
	\begin{array}{cccc}
		-\xi^{(1)}_1 & -b^{(1)}_{1,11} & b^{(1)}_{1,23} & b^{(1)}_{1,24} \\
		-\xi^{(1)}_{2} & -b^{(1)}_{1,21}-\xi^{(1)}_{1} & -b^{(1)}_{2,21} & -b^{(1)}_{3,21} \\
		-\xi^{(1)}_{3} & -b^{(1)}_{1,31} & -b^{(1)}_{2,31}-\xi^{(1)}_{1} & -b^{(1)}_{3,31} \\
		-\xi^{(1)}_{4} & -b^{(1)}_{1,41} & -b^{(1)}_{2,41} & -b^{(1)}_{3,41}-\xi^{(1)}_{1} \\
	\end{array}
	\right)\;,\\
	G^{(2)}&=\left(
	\begin{array}{ccccc}
		-\xi^{(2)}_{1} & -b^{(2)}_{1,11} & b^{(2)}_{1,23} & b^{(2)}_{2,34} & -b^{(2)}_{1,14} \\
		-\xi^{(2)}_{2} & -b^{(2)}_{1,21}-\xi^{(2)}_{1} & -b^{(2)}_{2,21} & -b^{(2)}_{3,21} & b^{(2)}_{2,34}-b^{(2)}_{1,24} \\
		-\xi^{(2)}_{3} & -b^{(2)}_{1,31} & -b^{(2)}_{2,31}-\xi^{(2)}_{1} & -b^{(2)}_{3,31} & -b^{(2)}_{1,34} \\
		-\xi^{(2)}_{4} & -b^{(2)}_{1,41} & b^{(2)}_{1,53} & -b^{(2)}_{3,41}-\xi^{(2)}_{1} & -b^{(2)}_{1,44} \\
		-\xi^{(2)}_{5} & -b^{(2)}_{1,51}-\xi^{(2)}_{4} & -b^{(2)}_{2,51} & -b^{(2)}_{3,51}-\xi^{(2)}_{2} & -b^{(2)}_{1,54}-b^{(2)}_{3,41}-\xi^{(2)}_{1} \\
	\end{array}
	\right)\;.
\end{aligned}$}
\end{equation}
This gauge choice cancels some tangent chiral direction contributions leaving us only with meson  degrees of freedom. In particular, we have:
\bea
\begin{split}
	&\delta B_1^{(1)}=\left(
	\begin{array}{cccc}
		0 & b^{(1)}_{1,12} & b^{(1)}_{1,13} & b^{(1)}_{1,14} \\
		0 & b^{(1)}_{1,22} & 0 & 0 \\
		0 & b^{(1)}_{1,32} & b^{(1)}_{1,33} & b^{(1)}_{1,34} \\
		0 & b^{(1)}_{1,42} & b^{(1)}_{1,43} & b^{(1)}_{1,44} \\
	\end{array}
	\right),\;
	\delta B_2^{(1)}=\left(
	\begin{array}{cccc}
		b^{(1)}_{2,11} & b^{(1)}_{2,12} & b^{(1)}_{2,13} & b^{(1)}_{2,14}\\
		0 & b^{(1)}_{2,22}& b^{(1)}_{2,23} & b^{(1)}_{2,24} \\
		0 & b^{(1)}_{2,32} & b^{(1)}_{2,33} & b^{(1)}_{2,34} \\
		0 & b^{(1)}_{2,42} & b^{(1)}_{2,43} & b^{(1)}_{2,44} \\
	\end{array}
	\right),\\ 
	&\delta B_3^{(1)}=\left(
	\begin{array}{cccc}
		b^{(1)}_{3,11}  & b^{(1)}_{3,12} & b^{(1)}_{3,13} & b^{(1)}_{3,14} \\
		0 & b^{(1)}_{3,22} & b^{(1)}_{3,23} & b^{(1)}_{3,24} \\
		0 & b^{(1)}_{3,32} & b^{(1)}_{3,33} & b^{(1)}_{3,34} \\
		0 & b^{(1)}_{3,42} & b^{(1)}_{3,43} & b^{(1)}_{3,44} \\
	\end{array}
	\right)\;, \quad \delta I^{(1)}=0 \;,
\end{split}
\eea
and
\bea
\begin{split}
&\delta B_1^{(2)}=\left(
\begin{array}{ccccc}
	0 & b^{(2)}_{1,12} & b^{(2)}_{1,13} & 0 & b^{(2)}_{1,15} \\
	0 & b^{(2)}_{1,22} & 0 & 0 & b^{(2)}_{1,25} \\
	0 & b^{(2)}_{1,32} & b^{(2)}_{1,33} & 0 & b^{(2)}_{1,35} \\
	0 & b^{(2)}_{1,42} & b^{(2)}_{1,43} & 0 & b^{(2)}_{1,45} \\
	0 & b^{(2)}_{1,52} & 0 & 0 & b^{(2)}_{1,55} \\
\end{array}
\right),\\
&\delta B_2^{(2)}=\left(
\begin{array}{ccccc}
	b^{(2)}_{2,11} & b^{(2)}_{2,12} & b^{(2)}_{2,13} & b^{(2)}_{2,14} & b^{(2)}_{2,15}\\
	0 & b^{(2)}_{2,22} & b^{(2)}_{2,23} & b^{(2)}_{2,24}& b^{(2)}_{2,25}\\
	0 & b^{(2)}_{2,32} &  b^{(2)}_{2,32} & 0 & b^{(2)}_{2,35} \\
	b^{(2)}_{2,41} & b^{(2)}_{2,42} & b^{(2)}_{2,43} & b^{(2)}_{2,44} & b^{(2)}_{2,45} \\
	0 & b^{(2)}_{2,52} & b^{(2)}_{2,53} & b^{(2)}_{2,54} & b^{(2)}_{2,55} \\
\end{array}
\right),\\
&\delta B_3^{(2)}=\left(
\begin{array}{ccccc}
	b^{(2)}_{3,11} & b^{(2)}_{3,12} & b^{(2)}_{3,13} & b^{(2)}_{3,14} & b^{(2)}_{3,15}\\
	0 & b^{(2)}_{3,22} & b^{(2)}_{3,23} & b^{(2)}_{3,24} & b^{(2)}_{3,25}\\
	0 & b^{(2)}_{3,32} & b^{(2)}_{3,33} & b^{(2)}_{3,34} & b^{(2)}_{3,35} \\
	0 & b^{(2)}_{3,42} & b^{(2)}_{3,43} & b^{(2)}_{3,44} & b^{(2)}_{3,45}\\
	0 & b^{(2)}_{3,52} & b^{(2)}_{3,53} & b^{(2)}_{3,54} & b^{(2)}_{3,55}\\
\end{array}
\right)\;, \qquad\delta I^{(2)}=0 \;.
\end{split}
\eea

Expectation values of the complex fields $\Phi$ on the chosen bases are defined by the weight function $\bphi_{\Box}$, therefore we derive:
\be
\Phi^{(1)}={\rm diag}\,\left(0,\,\epsilon_1,\,\epsilon_3,\,\epsilon_2\right)\;,\quad 
\Phi^{(2)}={\rm diag}\,\left(0,\,\epsilon_1,\,\epsilon_3,\,\epsilon_2,\,\epsilon_1+\epsilon_2\right)\;.
\ee

Equivariant weights $\omega$ for the gauge invariant tangent directions are defined using the action of $\Phi$ (with a parametrization $\epsilon_1=\hbar_1$, $\epsilon_2=\hbar_2$, $\epsilon_3=-\hbar_1-\hbar_2$):
\be
\left(\left[\Phi,B_i\right]-\epsilon_i B_i\right)_{jk}=\omega(b_{i,jk})\cdot b_{i,jk} \;.
\ee

To describe the weights of tangent spaces we will use also characters. In general, weight $\omega$ is a linear function in all $\hbar_i$, let us denote $\omega_{\hbar_i}$ a coefficient $\p_{\hbar_i}\omega$. Then we define character for the equivariant torus action on the vector space $V$ as:
\be
\chi(V)=\sum\lm_{e\in V} \prod\lm_{j}q_j^{\omega_{\hbar_j}(e)} \;.
\ee

For vector spaces at hands we have:
\bea
\begin{split}
	\chi(V_1)&=q_2^2 q_1^3+q_2^3 q_1^2+q_2^2 q_1^2+2 q_1^2+5 q_2 q_1+\frac{q_1}{q_2^2}+2 q_1+2 q_2^2+\frac{q_2}{q_1^2}+2 q_2+\\
	&\qquad +\frac{1}{q_1^2}+\frac{5}{q_2}+\frac{2}{q_1 q_2}
	+\frac{1}{q_1^3 q_2}+\frac{1}{q_2^2}+\frac{2}{q_1^2 q_2^2}+\frac{1}{q_1 q_2^3}+\frac{5}{q_1}\;,\\
	\chi(V_2)&=q_2^3 q_1^3+q_2^2 q_1^3+q_2^3 q_1^2+q_2^2 q_1^2+2 q_2 q_1^2+2 q_1^2+2 q_2^2 q_1+6 q_2 q_1+\\
	&\qquad+\frac{q_1}{q_2}+\frac{q_1}{q_2^2}+3 q_1+2 q_2^2
	+\frac{q_2}{q_1^2}+3 q_2+\frac{2}{q_1^2}+\frac{5}{q_2}+\frac{4}{q_1 q_2}+\frac{1}{q_1^2 q_2}+\frac{1}{q_1^3 q_2}+\\
	&\qquad+\frac{2}{q_2^2}+\frac{1}{q_1 q_2^2}+\frac{1}{q_1^2 q_2^2}+\frac{1}{q_1^3 q_2^2}
	+\frac{1}{q_1 q_2^3}+\frac{1}{q_1^2 q_2^3}+\frac{q_2}{q_1}+\frac{5}{q_1}+2 \;.
\end{split}
\eea

Representation $V_1$ is a subrepresentation of $V_2$  if there is a homomorphism:
\be
\tau:\quad V_2\to V_1
\ee
satisfying a set of equations:
\be\label{homom}
\tau\cdot B_i^{(2)}=B_i^{(1)}\cdot \tau\;, \quad i=1,2,3 \;, \quad \tau\cdot I^{(2)}=I^{(1)} \;.
\ee
Homomorphism $\tau$ also can be decomposed as an expectation value $\bar\tau$ and tangent degrees of freedom $\delta\tau$. It is pretty easy to calculate expectation value $\bar\tau$. It is just a projection of $V_2$ to $V_1$ mapping a vector $\nu_{(1,0,1)}\in V_2/V_1$ to zero:
\be
\bar\tau=\left(\begin{array}{ccccc}
	1& 0& 0& 0& 0\\
	0& 1& 0& 0& 0\\
	0& 0& 1& 0& 0\\
	0& 0& 0& 1& 0\\
\end{array}\right)\;.
\ee
The set of linearized equations for the homomorphism reads:
\bea
\begin{split}
&\bar\tau\cdot \delta B_i^{(2)}+\delta\tau\cdot \bar B_i^{(2)}=\delta B_i^{(1)}\cdot \bar\tau+\bar B_i^{(1)}\cdot \delta\tau\;, \quad i=1,2,3 \;,\\
&\delta \tau\cdot \bar I^{(2)}+\bar \tau\cdot \delta I^{(2)}=\delta I^{(1)} \;.
\end{split}
\eea
These equations define a hyperplane in a vector space spanned by $\delta B_i^{(r)}$, $i=1,2,3$, $r=1,2$ and $\delta\tau$. All $\delta\tau$ are fixed. The incidence locus hyperplane is spanned by the following vectors:
\bea
\begin{split}
	V_3& ={\rm Span}\left(b^{(1)}_{3,32},\,b^{(1)}_{3,42},\,b^{(1)}_{3,44},\,b^{(2)}_{1,12},\,b^{(2)}_{1,13},\,b^{(2)}_{1,22},\,b^{(2)}_{1,32},\,b^{(2)}_{1,33},\,b^{(2)}_{1,42},\,b^{(2)}_{1,43},\right.\\ 
	& \left.b^{(2)}_{1,52},\,b^{(2)}_{1,55},\,b^{(2)}_{2,11},\,b^{(2)}_{2,12},\, b^{(2)}_{2,13},\,b^{(2)}_{2,14},\,b^{(2)}_{2,22},\,b^{(2)}_{2,23},\,b^{(2)}_{2,24},\,b^{(2)}_{2,32},\,b^{(2)}_{2,33},\,b^{(2)}_{2,41},\,b^{(2)}_{2,42},\right.\\ 
	&\left. b^{(2)}_{2,43},\,b^{(2)}_{2,44},\,b^{(2)}_{2,52},\,b^{(2)}_{2,53},\,b^{(2)}_{2,54},\,b^{(2)}_{2,55},\,b^{(2)}_{3,11},\,b^{(2)}_{3,12},\, b^{(2)}_{3,13},\,b^{(2)}_{3,14},\,b^{(2)}_{3,22},\,b^{(2)}_{3,23}, \right.\\ 
	&\left. b^{(2)}_{3,24},\,b^{(2)}_{3,32},\,b^{(2)}_{3,33},\,b^{(2)}_{3,34},\,b^{(2)}_{3,42},\,b^{(2)}_{3,43},\,b^{(2)}_{3,44},\,b^{(2)}_{3,45},\,b^{(2)}_{3,52},\,b^{(2)}_{3,53},\,b^{(2)}_{3,54},\,b^{(2)}_{3,55}\right)\;.
\end{split} 
\eea

Space $V_3$ is the graded tangent space to the incidence locus. We easily calculate its character:
\bea
\begin{split}
	\chi(V_3)=q_2^3 q_1^3+q_2^2 q_1^3+q_2^3 q_1^2+q_2^2 q_1^2+2 q_2 q_1^2+2 q_1^2+2 q_2^2 q_1+6 q_2 q_1+\\
	+\frac{q_1}{q_2}+\frac{q_1}{q_2^2}+2 q_1+2 q_2^2+\frac{q_2}{q_1^2}+2 q_2+\frac{1}{q_1^2}+\frac{6}{q_2}+\frac{2}{q_1 q_2}+\\
	+\frac{1}{q_1^3 q_2}+\frac{1}{q_2^2}+\frac{2}{q_1^2 q_2^2}+\frac{1}{q_1 q_2^3}+\frac{q_2}{q_1}+\frac{6}{q_1}+1 \;.
\end{split}
\eea

It is easy to translate these characters to Euler characters according to equation \eqref{res_Eul}:
\bea
\begin{split}
	&\widetilde{\rm Eul}(V_1) =  -512 \hbar _1^{10} \left(\hbar _1-2 \hbar _2\right) \left(2 \hbar _1-\hbar _2\right) \hbar _2^{10} \left(\hbar _1+\hbar _2\right){}^{10} \left(3 \hbar _1+\hbar _2\right) \times\\ 
	&\quad\times\left(3 \hbar _1+2 \hbar _2\right) \left(\hbar _1+3 \hbar _2\right) \left(2 \hbar _1+3 \hbar _2\right) \;,\\
	&\widetilde{\rm Eul}(V_2)=  3072 \hbar _1^{12} \left(\hbar _1-2 \hbar _2\right) \left(\hbar _1-\hbar _2\right){}^2 \left(2 \hbar _1-\hbar _2\right) \hbar _2^{12} \left(\hbar _1+\hbar _2\right){}^{13} \times\\ 
	&\quad\times \left(2 \hbar _1+\hbar _2\right){}^3 \left(3 \hbar _1+\hbar _2\right) \left(\hbar _1+2 \hbar _2\right){}^3 \left(3 \hbar _1+2 \hbar _2\right){}^2 \left(\hbar _1+3 \hbar _2\right) \left(2 \hbar _1+3 \hbar _2\right){}^2\;,\\
	&\widetilde{\rm Eul}(V_3)=  1536 \hbar _1^{11} \left(\hbar _1-2 \hbar _2\right) \left(\hbar _1-\hbar _2\right){}^2 \left(2 \hbar _1-\hbar _2\right) \hbar _2^{11} \left(\hbar _1+\hbar _2\right){}^{12} \times\\ 
	&\quad \times \left(2 \hbar _1+\hbar _2\right){}^2  \left(3 \hbar _1+\hbar _2\right) \left(\hbar _1+2 \hbar _2\right){}^2 \left(3 \hbar _1+2 \hbar _2\right) \left(\hbar _1+3 \hbar _2\right) \left(2 \hbar _1+3 \hbar _2\right) \;.
\end{split}
\eea

We calculate matrix elements:
\bea
\begin{split}
	&E(\Lambda_1\to\Lambda_2)=-\frac{1}{3 \hbar _1 \left(\hbar _1-\hbar _2\right){}^2 \hbar _2 \left(\hbar _1+\hbar _2\right){}^2 \left(2 \hbar _1+\hbar _2\right){}^2 \left(\hbar _1+2 \hbar _2\right){}^2} \;,\\
	&F(\Lambda_2\to\Lambda_1)=2 \hbar _1 \hbar _2 \left(\hbar _1+\hbar _2\right) \left(2 \hbar _1+\hbar _2\right) \left(\hbar _1+2 \hbar _2\right) \left(3 \hbar _1+2 \hbar _2\right) \left(2 \hbar _1+3 \hbar _2\right) \;.
\end{split}
\eea

For the product we find:
\bea
\begin{split}
&E(\Lambda_1\to\Lambda_2)F(\Lambda_2\to\Lambda_1)=\\
&\qquad =-\frac{2 \left(3 \hbar _1+2 \hbar _2\right) \left(2 \hbar _1+3 \hbar _2\right)}{3 \left(\hbar _1-\hbar _2\right){}^2 \left(\hbar _1+\hbar _2\right) \left(2 \hbar _1+\hbar _2\right) \left(\hbar _1+2 \hbar _2\right)}=\frac{\mathop{\rm res}\lm_{z=\hbar_1+\hbar_2}\psi_{\Lambda_1}(z)}{-\hbar_1\hbar_2(\hbar_1+\hbar_2)} \;,
\end{split}
\eea
where
\bea
\begin{split}
\psi_{\Lambda_1}(z)=-\frac{z^2 \left(z-\hbar _1-2 \hbar _2\right) \left(z-2 \hbar _1-\hbar _2\right) \left(z+\hbar _1-\hbar _2\right) }{\left(z-2 \hbar _1\right) \left(z-\hbar _1\right) \left(z+\hbar _1\right) \left(z-2 \hbar _2\right) \left(z-\hbar _2\right) \left(z-\hbar _1-\hbar _2\right) }\times\\
\times \frac{\left(z-\hbar _1+\hbar _2\right) \left(z+2 \hbar _1+\hbar _2\right) \left(z+\hbar _1+2 \hbar _2\right)}{\left(z+\hbar _2\right) \left(z+\hbar _1+\hbar _2\right) \left(z+2 \hbar _1+2 \hbar _2\right)} \;.
\end{split}
\eea

%%%%%%%---------------------------------------------------------------------------------------------------------------------------------------------------------------------------------------------
\subsection{\texorpdfstring{Example of $e$ -- $f$ Relations for Distinguished Atoms and Statistics}{Example of e -- f Relations for Distinguished Atoms and Statistics}}\label{s:parity}

Consider the following crystal of $Y(\widehat{\fg\fl}_{8|4})$ with signature \eqref{ex_sign}:
\be\label{cry1}
\begin{array}{c}
	\begin{tikzpicture}
	\begin{scope}[shift={(0,0)}]
	\draw[fill=white!40!red] (0,0) circle (0.2);
	\node at (0,0) {$2$};
	\end{scope}
	\begin{scope}[shift={(1,0)}]
	\draw[fill=white!40!green] (0,0) circle (0.2);
	\node at (0,0) {$1$};
	\end{scope}
	\begin{scope}[shift={(2,0)}]
	\draw[fill=white!40!blue] (0,0) circle (0.2);
	\node at (0,0) {$1$};
	\end{scope}
	\begin{scope}[shift={(3,0)}]
	\draw[fill=white!40!purple] (0,0) circle (0.2);
	\node at (0,0) {$1$};
	\end{scope}
	\begin{scope}[shift={(-1,0)}]
	\draw[fill=white!40!gray] (0,0) circle (0.2);
	\node at (0,0) {$1$};
	\end{scope}
	\begin{scope}[shift={(0,1)}]
	\draw[fill=white!40!gray] (0,0) circle (0.2);
	\node at (0,0) {$1$};
	\end{scope}
	\begin{scope}[shift={(1,1)}]
	\draw[fill=white!40!red] (0,0) circle (0.2);
	\node at (0,0) {$1$};
	\end{scope}
	\begin{scope}[shift={(-1,-1)}]
	\draw[fill=white!40!red] (0,0) circle (0.2);
	\node at (0,0) {$1$};
	\end{scope}
	\begin{scope}[shift={(0,-1)}]
	\draw[fill=white!40!green] (0,0) circle (0.2);
	\node at (0,0) {$1$};
	\end{scope}
	\begin{scope}[shift={(0,2)}]
	\draw[fill=white] (0,0) circle (0.2);
	\node at (0,0) {$1$};
	\end{scope}
	\draw[->] (-0.8,0) -- (-0.2,0);
	\draw[->] (0.2,0) -- (0.8,0);
	\draw[->] (1.2,0) -- (1.8,0);
	\draw[->] (2.2,0) -- (2.8,0);
	\draw[->] (0.2,1) -- (0.8,1);
	\draw[->] (-0.8,-1) -- (-0.2,-1);
	\draw[->] (0,0.2) -- (0,0.8);
	\draw[->] (0,1.2) -- (0,1.8);
	\draw[->] (0,-0.8) -- (0,-0.2);
	\draw[->] (1,0.2) -- (1,0.8);
	\draw[->] (-1,-0.8) -- (-1,-0.2);
	\draw[->] (0.858579, 0.858579) -- (0.141421, 0.141421);
	\draw[->] (-0.141421, -0.141421) -- (-0.858579, -0.858579);
	\draw[->] (-0.141421, 0.858579) -- (-0.858579, 0.141421);
	\draw[->] (0.858579, -0.141421) -- (0.141421, -0.858579);
	\draw[red, thick] (0,0) circle (0.3);
	\draw[blue, thick] (3,0) circle (0.3);
	\draw[blue, thick] (0,2) circle (0.3);
	\draw[orange, thick] (-1,-1.5) -- (0,-1.5) to[out=0,in=270] (2.5,0) to[out=90,in=0] (1,1.5) -- (0,1.5) to[out=180,in=90] (-1.5,0) -- (-1.5,-1) to[out=270,in=180] cycle;
	\node[left, blue] at (-0.3,2) {$1$};
	\node[right, blue] at (3.3,0) {$2$};
	\end{tikzpicture}\\
	\mbox{(a)}
\end{array}\quad 
\begin{array}{c}
	\begin{tikzpicture}
	\begin{scope}[shift={(0,0)}]
	\draw[fill=white!40!red] (0,0) circle (0.2);
	\end{scope}
	\begin{scope}[shift={(1,0)}]
	\draw[fill=white!40!green] (0,0) circle (0.2);
	\end{scope}
	\begin{scope}[shift={(2,0)}]
	\draw[fill=white!40!blue] (0,0) circle (0.2);
	\end{scope}
	\begin{scope}[shift={(3,0)}]
	\draw[fill=white!40!purple] (0,0) circle (0.2);
	\end{scope}
	\begin{scope}[shift={(-1,0)}]
	\draw[fill=white!40!gray] (0,0) circle (0.2);
	\end{scope}
	\begin{scope}[shift={(0,1)}]
	\draw[fill=white!40!gray] (0,0) circle (0.2);
	\end{scope}
	\begin{scope}[shift={(1,1)}]
	\draw[fill=white!40!red] (0,0) circle (0.2);
	\end{scope}
	\begin{scope}[shift={(-1,-1)}]
	\draw[fill=white!40!red] (0,0) circle (0.2);
	\end{scope}
	\begin{scope}[shift={(0,-1)}]
	\draw[fill=white!40!green] (0,0) circle (0.2);
	\end{scope}
	\begin{scope}[shift={(0,2)}]
	\draw[fill=white] (0,0) circle (0.2);
	\end{scope}
	\draw[->] (-0.8,0) -- (-0.2,0);
	\draw[->] (0.2,0) -- (0.8,0);
	\draw[->] (1.2,0) -- (1.8,0);
	\draw[->] (2.2,0) -- (2.8,0);
	\draw[->] (0.2,1) -- (0.8,1);
	\draw[->] (-0.8,-1) -- (-0.2,-1);
	\draw[->] (0,0.2) -- (0,0.8);
	\draw[->] (0,1.2) -- (0,1.8);
	\draw[->] (0,-0.8) -- (0,-0.2);
	\draw[->] (1,0.2) -- (1,0.8);
	\draw[->] (-1,-0.8) -- (-1,-0.2);
	\draw[->] (0.858579, 0.858579) -- (0.141421, 0.141421);
	\draw[->] (-0.141421, -0.141421) -- (-0.858579, -0.858579);
	\draw[->] (-0.141421, 0.858579) -- (-0.858579, 0.141421);
	\draw[->] (0.858579, -0.141421) -- (0.141421, -0.858579);
	\node at (-1,-1) {$3$};
	\node at (0,-1) {$6$};
	\node at (-1,0) {$11$};
	\node at (0,0) {$4$};
	\node at (1,0) {$5$};
	\node at (2,0) {$7$};
	\node at (3,0) {$8$};
	\node at (0,1) {$10$};
	\node at (1,1) {$2$};
	\node at (0,2) {$9$};
	\end{tikzpicture}\\
	\mbox{(b)}
\end{array}
\ee

This crystal has a subcrystal $\Lambda$ encircled by an orange line in diagram \eqref{cry1} (a). Two additional atoms $\Box_1$ and $\Box_2$ are encircled by blue circles with assigned indices. We would like to check $E$ -- $F$ relations in \eqref{EF} for this crystal combination. Both atom colors $\hat \Box_1$ and $\hat \Box_2$ correspond to nodes we claimed as odd ones. Therefore for the RHS of the second ratio in \eqref{EF} we expect $-1$. Let us first consider matrix elements without sign shifts $\nu_{\pm}$:
\bea
\begin{split}
	\hat E(\Lambda\to \Lambda+\Box)=\frac{\widetilde{\rm Eul}(\Lambda)}{\widetilde{\rm Eul}(\Lambda,\Lambda+\Box)}\;,\\ \hat F( \Lambda+\Box\to \Lambda)=\frac{\widetilde{\rm Eul}(\Lambda)}{\widetilde{\rm Eul}(\Lambda+\Box,\Lambda)} \;.
\end{split}
\eea

Using techniques discussed in the previous section we find:
\bea
\begin{split}
	&\hat F(\Lambda+\Box_1\to \Lambda)=\hat F(\Lambda+\Box_1+\Box_2\to \Lambda+\Box_2)=2 \hbar _{13}^2 \;,\\
	&\hat E(\Lambda\to\Lambda+\Box_2)=\hat E(\Lambda+\Box_1\to\Lambda+\Box_1+\Box_2)=1 \;.\\
\end{split}
\eea

Then as a result we derive:
\be
\frac{\hat E(\Lambda+\Box_1\to\Lambda+\Box_1+\Box_2)\hat F(\Lambda+\Box_1+\Box_2\to\Lambda+\Box_1)}{\hat F(\Lambda+\Box_1\to \Lambda)\hat E(\Lambda\to\Lambda+\Box_2)}=1\;.
\ee

 We choose an ordering of crystal atoms as it is depicted in diagram \eqref{cry1} (b) starting with the root atom that is stored in a 2-atom row in a position marked by a red circle \eqref{cry1} (a). Thus we get for sign shift functions \eqref{stat}, \eqref{sgn}:
\begin{align}
\nu_+(\Box_i,\Box_j)&=\left(
\begin{array}{ccccccccccc}
	-1 & 1 & 1 & 1 & -1 & -1 & 1 & 1 & 1 & -1 & -1 \\
	1 & -1 & 1 & 1 & -1 & -1 & 1 & 1 & 1 & -1 & -1 \\
	1 & 1 & -1 & 1 & -1 & -1 & 1 & 1 & 1 & -1 & -1 \\
	1 & 1 & 1 & -1 & -1 & -1 & 1 & 1 & 1 & -1 & -1 \\
	1 & 1 & 1 & 1 & -1 & 1 & -1 & 1 & 1 & 1 & 1 \\
	1 & 1 & 1 & 1 & 1 & -1 & -1 & 1 & 1 & 1 & 1 \\
	1 & 1 & 1 & 1 & 1 & 1 & -1 & -1 & 1 & 1 & 1 \\
	1 & 1 & 1 & 1 & 1 & 1 & 1 & -1 & -1 & 1 & 1 \\
	1 & 1 & 1 & 1 & 1 & 1 & 1 & 1 & -1 & -1 & -1 \\
	1 & 1 & 1 & 1 & 1 & 1 & 1 & 1 & 1 & -1 & 1 \\
	1 & 1 & 1 & 1 & 1 & 1 & 1 & 1 & 1 & 1 & -1 \\
\end{array}
\right) \;,\\
\nu_-(\Box_i,\Box_j)&=\left(
\begin{array}{ccccccccccc}
	-1 & 1 & 1 & 1 & 1 & 1 & 1 & 1 & 1 & 1 & 1 \\
	1 & -1 & 1 & 1 & 1 & 1 & 1 & 1 & 1 & 1 & 1 \\
	1 & 1 & -1 & 1 & 1 & 1 & 1 & 1 & 1 & 1 & 1 \\
	1 & 1 & 1 & -1 & 1 & 1 & 1 & 1 & 1 & 1 & 1 \\
	-1 & -1 & -1 & -1 & -1 & 1 & 1 & 1 & 1 & 1 & 1 \\
	-1 & -1 & -1 & -1 & 1 & -1 & 1 & 1 & 1 & 1 & 1 \\
	1 & 1 & 1 & 1 & -1 & -1 & -1 & 1 & 1 & 1 & 1 \\
	1 & 1 & 1 & 1 & 1 & 1 & -1 & 1 & -1 & 1 & 1 \\
	1 & 1 & 1 & 1 & 1 & 1 & 1 & 1 & 1 & 1 & 1 \\
	-1 & -1 & -1 & -1 & 1 & 1 & 1 & 1 & -1 & -1 & 1 \\
	-1 & -1 & -1 & -1 & 1 & 1 & 1 & 1 & -1 & 1 & -1 \\
\end{array}
\right) \;.
\end{align}

Applying these expressions to matrix elements we derive:
\begin{align}
\begin{split}
	&F(\Lambda+\Box_1\to \Lambda)=-F(\Lambda+\Box_1+\Box_2\to \Lambda+\Box_2)=2 \hbar _{13}^2 \;,\\
	&E(\Lambda\to\Lambda+\Box_2)= E(\Lambda+\Box_1\to\Lambda+\Box_1+\Box_2)=-1 \;,\\
\end{split}
\end{align}
giving the desired result:
\be
\frac{ E(\Lambda+\Box_1\to\Lambda+\Box_1+\Box_2) F(\Lambda+\Box_1+\Box_2\to\Lambda+\Box_1)}{ F(\Lambda+\Box_1\to \Lambda) E(\Lambda\to\Lambda+\Box_2)}=-1\;.
\ee

%%%%%%%---------------------------------------------------------------------------------------------------------------------------------------------------------------------------------------------
\subsection{Example of Serre Relation Calculation}

Consider the case of $Y(\widehat{\fg\fl}_{3|1})$ with the following parity choice $\Sigma_{3,1}=\{1,-1,1,1\}$, the quiver diagram has the following form:
\be
\begin{array}{c}
	\begin{tikzpicture}
	\draw[fill=white!40!red] (0,0) circle (0.2);
	\draw[fill=white!40!green] (2,0) circle (0.2);
	\draw[fill=white!40!orange] (0,-1) circle (0.2);
	\draw[fill=white!40!blue] (2,-1) circle (0.2);
	\draw[->] (0.173205, 0.1) to[out=30,in=150] (1.82679, 0.1);
	\draw[<-] (0.173205, -0.1) to[out=330,in=210] (1.82679, -0.1);
	\draw[->] (0.173205, -0.9) to[out=30,in=150] (1.82679, -0.9);
	\draw[<-] (0.173205, -1.1) to[out=330,in=210] (1.82679, -1.1);
	\draw[->] (0.1, -0.173205) to[out=300,in=60] (0.1, -0.826795);
	\draw[<-] (-0.1, -0.173205) to[out=240,in=120] (-0.1, -0.826795);
	\draw[->] (2.1, -0.173205) to[out=300,in=60] (2.1, -0.826795);
	\draw[<-] (1.9, -0.173205) to[out=240,in=120] (1.9, -0.826795);
	\draw[->] (0,-1.2) to[out=270,in=315] (-0.5,-1.5) to[out=135,in=180] (-0.2,-1);
	\draw[->] (2,-1.2) to[out=270,in=225] (2.5,-1.5) to[out=45,in=0] (2.2,-1); 
	\draw[->] (-1,0) -- (-0.2,0);
	\begin{scope}[shift={(-1,0)}]
	\draw[fill=black] (-0.1,-0.1) -- (-0.1,0.1) -- (0.1,0.1) -- (0.1,-0.1) -- cycle;
	\end{scope}
	\end{tikzpicture}
\end{array}
\ee
Without loss off generality we use the gauge freedom of choosing the weight space, to obtain the following parameterization:
\bea
\begin{split}
	&w_{B_1}=\hbar _1,\; w_{C_1}= \hbar _2-\hbar _1,\; w_{B_2}= \hbar _2,\; w_{C_2}= -\hbar _2-\hbar _1,\\
	&w_{B_3}=\hbar _1,\; w_{C_3}= -\hbar _1-\hbar _2,\; w_{A_3}=\hbar _2,\; w_{B_4}=\hbar _1,\;w_{C_4}= -\hbar _1-\hbar _2,\; w_{A_4}= \hbar _2 \;.
\end{split}
\eea
Consider a particular 9-atom crystal:
$$
\begin{array}{c}
\begin{tikzpicture}
\begin{scope}[shift={(-1,1)}]
\draw[fill=white!40!green] (0,0) circle (0.2);
\node at (0,0) {$1$}; 
\end{scope}
\begin{scope}[shift={(0,1)}]
\draw[fill=white!40!red] (0,0) circle (0.2);
\node at (0,0) {$1$}; 
\end{scope}
\begin{scope}[shift={(0,0)}]
\draw[fill=white!40!green] (0,0) circle (0.2);
\node at (0,0) {$1$}; 
\end{scope}
\begin{scope}[shift={(0,-1)}]
\draw[fill=white!40!blue] (0,0) circle (0.2);
\node at (0,0) {$1$}; 
\end{scope}
\begin{scope}[shift={(1,0)}]
\draw[fill=white!40!red] (0,0) circle (0.2);
\node at (0,0) {$1$}; 
\end{scope}
\begin{scope}[shift={(2,0)}]
\draw[fill=white!40!orange] (0,0) circle (0.2);
\node at (0,0) {$1$}; 
\end{scope}
\begin{scope}[shift={(2,-1)}]
\draw[fill=white!40!red] (0,0) circle (0.2);
\node at (0,0) {$1$}; 
\end{scope}
\begin{scope}[shift={(3,0)}]
\draw[fill=white!40!blue] (0,0) circle (0.2);
\node at (0,0) {$1$}; 
\end{scope}
\begin{scope}[shift={(4,0)}]
\draw[fill=white!40!green] (0,0) circle (0.2);
\node at (0,0) {$1$}; 
\end{scope}
\draw[->] (-0.2,1) -- (-0.8,1);
\draw[->] (0,0.2) -- (0,0.8);
\draw[->] (0,-0.2) -- (0,-0.8);
\draw[->] (0.8,0) -- (0.2,0);
\draw[->] (1.2,0) -- (1.8,0);
\draw[->] (2.2,0) -- (2.8,0);
\draw[->] (3.2,0) -- (3.8,0);
\draw[->] (2,-0.2) -- (2,-0.8);
\draw[thick,red] (1,0) circle (0.3);
\draw[thick,blue] (-1,1) circle (0.3);
\draw[thick,blue] (0,-1) circle (0.3);
\draw[thick,blue] (2,-1) circle (0.3);
\draw[thick,blue] (4,0) circle (0.3);
\draw[thick,orange] (0,-0.5) -- (3,-0.5) to[out=0,in=270] (3.5,0) to[out=90,in=0] (0,1.5) to[out=180,in=90] (-0.5,0) to[out=270,in=180] cycle;
\node[left,blue] at (-1.3,1) {$2$};
\node[left,blue] at (-0.3,-1) {$4$};
\node[left,blue] at (1.7,-1) {$1$};
\node[right,blue] at (4.3,0) {$3$};
\end{tikzpicture}
\end{array}
$$

We have enumerated marked atoms. It turns out these atoms can be added to the root crystal marked by an orange circle in an arbitrary order, and still on all intermediate steps a derived configuration of atoms is a valid crystal.

Atoms labeled by numbers 2 and 3 have the same color, and moreover they have a flavor corresponding to the color of an odd node. Therefore the raising generators adding these atoms are fermionic. Therefore this is the case exactly suitable for checking quartic Serre relations in $Y(\widehat{\fg\fl}_{3|1})$, where we have a relation between two odd generators of the same color, and two other generators corresponding to neighboring nodes. In this particular case, the supercommutator should be decomposed as follows:
\begin{align}
\begin{split}
	A_1&={\rm Sym}_{z_1,z_2}\left[e^{(2)}(z_1),\left[e^{(3)}(w_1),\left[e^{(2)}(z_2),e^{(1)}(w_2)\right\} \right\} \right\}=\\
	&={\rm Sym}_{z_1,z_2}\left[e^{(2)}(z_1),\left[e^{(3)}(w_1),\left\{e^{(2)}(z_2),e^{(1)}(w_2)\right\} \right] \right]=\\
	&=e^{(1)}(w_2)e^{(2)}(z_1)e^{(3)}(w_1)e^{(2)}(z_2)+e^{(1)}(w_2)e^{(2)}(z_2)e^{(3)}(w_1)e^{(2)}(z_1)-\\
	&-e^{(2)}(z_1)e^{(1)}(w_2)e^{(2)}(z_2)e^{(3)}(w_1)+e^{(2)}(z_1)e^{(1)}(w_2)e^{(3)}(w_1)e^{(2)}(z_2)-\\
	&-e^{(2)}(z_1)e^{(2)}(z_2)e^{(1)}(w_2)e^{(3)}(w_1)+e^{(2)}(z_1)e^{(3)}(w_1)e^{(1)}(w_2)e^{(2)}(z_2)+\\
	&+e^{(2)}(z_1)e^{(3)}(w_1)e^{(2)}(z_2)e^{(1)}(w_2)-e^{(2)}(z_2)e^{(1)}(w_2)e^{(2)}(z_1)e^{(3)}(w_1)+\\
	&+e^{(2)}(z_2)e^{(1)}(w_2)e^{(3)}(w_1)e^{(2)}(z_1)-e^{(2)}(z_2)e^{(2)}(z_1)e^{(1)}(w_2)e^{(3)}(w_1)+\\
	&+e^{(2)}(z_2)e^{(3)}(w_1)e^{(1)}(w_2)e^{(2)}(z_1)+e^{(2)}(z_2)e^{(3)}(w_1)e^{(2)}(z_1)e^{(1)}(w_2)-\\
	&-e^{(3)}(w_1)e^{(1)}(w_2)e^{(2)}(z_1)e^{(2)}(z_2)-e^{(3)}(w_1)e^{(1)}(w_2)e^{(2)}(z_2)e^{(2)}(z_1)-\\
	&-e^{(3)}(w_1)e^{(2)}(z_1)e^{(1)}(w_2)e^{(2)}(z_2)-e^{(3)}(w_1)e^{(2)}(z_2)e^{(1)}(w_2)e^{(2)}(z_1) \;.
\end{split}
\end{align}
We apply this operator to the root crystal $\Lambda_0$. Operators $e^{(a)}(z)$ have poles in points corresponding to atom weights:
\be
1)\; w_2= -\hbar _2 \;, \quad  2)\; z_1=\hbar _1+\hbar _2 \;,\quad 3)\; z_2=-3\hbar _1-3 \hbar _2 \;,\quad 4)\; w_1=\hbar _1+\hbar _2 \;.
\ee
The expression $A_1(z_1,z_2,w_1,w_2)|\Lambda_0\rangle$ will  produce the whole crystal $\Lambda$, and the residue will consist of a sum over matrix elements $E(\Lambda_1\to\Lambda_2)$ in various combinations. Each term in this sum will add consequently 4 atoms to $\Lambda_0$ to get $\Lambda$ in the result. For example, this type of term 
\bea\nn
\begin{split}
E(\Lambda_0\to \Lambda_0+a)E(\Lambda_0+a\to \Lambda_0+a+b)E(\Lambda_0+a+b\to \Lambda_0+a+b+c)\times\\ \times E(\Lambda_0+a+b+c\to \Lambda_0+a+b+c+d)
\end{split}
\eea
adds atoms $a$, $b$, $c$, $d$ in a sequence $[a,b,c,d]$. For the sake of brevity let us denote such quintic $E$-terms by such sequences. The residue of our interest acquires the following form:
\bea
\begin{split}
	A_2& \coloneqq \mathop{\rm Res}\lm_{z_1,z_2,w_1,w_2}\langle\Lambda|A_1|\Lambda_0\rangle=\\
	&=[1,2,4,3]+[1,3,4,2]-[2,1,3,4]+[2,1,4,3]-[2,3,1,4]+[2,4,1,3]+\\
	&+[2,4,3,1]-[3,1,2,4]
	+[3,1,4,2]-[3,2,1,4]+[3,4,1,2]+[3,4,2,1]-\\
	&-[4,1,2,3]-[4,1,3,2]-[4,2,1,3]-[4,3,1,2] \;.
\end{split}
\eea

Using methods discussed above we calculate the necessary coefficients in multipliers of $$\left(\hbar _2^3 \left(\hbar _1+\hbar _2\right) \left(2 \hbar _1+\hbar _2\right){}^2 \left(4 \hbar _1+\hbar _2\right)\right)^{-1}$$ for brevity of expressions. The corresponding expressions read
\bea
\begin{split}
	&[2,4,1,3\text{]=}-\frac{1}{48}\;, \quad [4,2,1,3\text{]=}-\frac{1}{96}\;, \quad [2,1,4,3\text{]=}-\frac{1}{48}\;, \quad [1,2,4,3\text{]=}\frac{1}{32}\;, \\
	&[4,1,2,3\text{]=}\frac{1}{64}\;, \quad
	[1,4,2,3\text{]=}\frac{1}{64}\;, \quad [4,1,3,2\text{]=}-\frac{1}{64}\;, \quad [1,4,3,2\text{]=}-\frac{1}{64}\;, \\ 
	&[2,4,3,1\text{]=}\frac{2 \hbar _1+\hbar _2}{24 \left(4 \hbar _1+\hbar _2\right)}\;, \quad
	[4,2,3,1\text{]=}\frac{2 \hbar _1+\hbar _2}{48 \left(4 \hbar _1+\hbar _2\right)}\;, \\ 
	&[2,3,4,1\text{]=}\frac{\left(2 \hbar _1+\hbar _2\right){}^2}{12 \left(4 \hbar _1+\hbar _2\right) \left(4 \hbar _1+3 \hbar _2\right)}\;, \quad [3,2,4,1\text{]=}-\frac{\left(2 \hbar _1+\hbar _2\right){}^2}{12 \left(4 \hbar _1+\hbar _2\right) \left(4 \hbar _1+3 \hbar _2\right)}\;, \\ 
	&[4,3,2,1\text{]=}-\frac{2 \hbar _1+\hbar _2}{48 \left(4 \hbar _1+\hbar _2\right)}\;, \quad [3,4,2,1\text{]=}-\frac{\left(2 \hbar _1+\hbar _2\right){}^2}{24 \left(4 \hbar _1+\hbar _2\right) \left(4 \hbar _1+3 \hbar _2\right)}\;, \\ 
	&[2,1,3,4\text{]=}-\frac{2 \hbar _1+\hbar _2}{24 \left(4 \hbar _1+3 \hbar _2\right)}\;, \quad
	[1,2,3,4\text{]=}\frac{2 \hbar _1+\hbar _2}{16 \left(4 \hbar _1+3 \hbar _2\right)}\;, \\ 
	&[2,3,1,4\text{]=}\frac{\left(2 \hbar _1+\hbar _2\right){}^2}{12 \left(4 \hbar _1+\hbar _2\right) \left(4 \hbar _1+3 \hbar _2\right)}\;, \quad [3,2,1,4\text{]=}-\frac{\left(2 \hbar _1+\hbar _2\right){}^2}{12 \left(4 \hbar _1+\hbar _2\right) \left(4 \hbar _1+3 \hbar _2\right)}\;, \\
	&[1,3,2,4\text{]=}-\frac{2 \hbar _1+\hbar _2}{16 \left(4 \hbar _1+3 \hbar _2\right)}\;, \;
	[3,1,2,4\text{]=}\frac{\left(2 \hbar _1+\hbar _2\right){}^2}{8 \left(4 \hbar _1+\hbar _2\right) \left(4 \hbar _1+3 \hbar _2\right)}\;, \\ 
	&[4,3,1,2\text{]=}\frac{2 \hbar _1+\hbar _2}{32 \left(4 \hbar _1+\hbar _2\right)}\;, \quad
	[3,4,1,2\text{]=}\frac{\left(2 \hbar _1+\hbar _2\right){}^2}{16 \left(4 \hbar _1+\hbar _2\right) \left(4 \hbar _1+3 \hbar _2\right)}\;, \\ 
	&[1,3,4,2\text{]=}-\frac{2 \hbar _1+\hbar _2}{32 \left(4 \hbar _1+3 \hbar _2\right)}\;, \quad [3,1,4,2\text{]=}\frac{\left(2 \hbar _1+\hbar _2\right){}^2}{16 \left(4 \hbar _1+\hbar _2\right) \left(4 \hbar _1+3 \hbar _2\right)}.
\end{split}
\eea

Summing up these contributions with appropriate signs we conclude:
\be
A_2=0 \;.
\ee
As this example shows, the consistency of the Serre relation requires highly non-trivial cancellations.

\clearpage

%%%%%%%%%%%%%%%%%%%%%%%%%%%%%%%
%%%%%%%%%%%%%%%%%%%%%%%%%%%%%%%
\bibliographystyle{nb}
\bibliography{ref-W-algebra}
%%%%%%%%%%%%%%%%%%%%%%%%%%%%%%%
%%%%%%%%%%%%%%%%%%%%%%%%%%%%%%%
\end{document}